\documentstyle[bezier,11pt]{article} 
\def\tenrm{}
\newcommand{\be}{\begin{equation}}
\newcommand{\ee}{\end{equation}}
\newcommand{\bearray}{\begin{eqnarray}}
\newcommand{\eearray}{\end{eqnarray}}
\newcommand{\nl}{\nonumber \\}

\newenvironment{exercise}{\small\begin{description}
                         \item[{Exercise:}]}{\end{description}}

\newcommand{\eq}[1]{{(\ref{#1})}}

\newcommand{\e}{{\rm e}}
\newcommand{\order}{{\cal O}}

\newcommand{\lder}[1]{\Delta^{(#1)}}

\newcommand{\Lag}{{\cal L}}
\newcommand{\cutoff}{{(a)}}
\newcommand{\ub}{\overline{u}}
\newcommand{\psib}{\overline{\psi}}
\newcommand{\half}{\mbox{$\frac{1}{2}$}}
\newcommand{\third}{\mbox{$\frac{1}{3}$}}
\renewcommand{\Re}{{\rm\,Re\,}}
\newcommand{\Tr}{{\rm\,Tr\,}}
\renewcommand{\P}{{\rm\,P\,}}
\newcommand{\F}[2]{F_{#1#2}}
\newcommand{\Pl}[2]{P_{#1#2}}
\newcommand{\Rt}[2]{R_{#1#2}}
\newcommand{\su}{{${\rm SU}_3$}}
\newcommand{\lat}{{\rm lat}}
\newcommand{\C}{{\cal C}}
\newcommand{\ct}{{\rm c}}

\def\D{{\rm D}}
\def\I{{\rm I}}
\newcommand{\Dv}{{\bf D}}
\newcommand{\xv}{{\bf x}}
\newcommand{\rv}{{\bf r}}
\newcommand{\pv}{{\bf p}}

\newcommand{\dd}{{\rm d}}

\def\U#1{U_{#1}}
\def\Udag#1{{U^\dagger_{#1}}}
\def\A#1{A_{#1}}

\newcommand{\gammav}{\mbox{\boldmath$\gamma$}}
\newcommand{\sigmav}{\mbox{\boldmath$\sigma$}}
\newcommand{\Deltav}{\mbox{\boldmath$\Delta$}}
\title{Redesigning Lattice QCD}
\author{G.\,Peter Lepage\\
\small Newman Laboratory of Nuclear Studies,
Cornell University, Ithaca NY 14853}
\begin{document}
\maketitle
\begin{abstract}
There has been major progress in recent years in the development of
improved discretizations of the QCD action, current operators, etc for
use in numerical simulations that employ very coarse lattices. 
These lectures
review the field theoretic techniques used to design these
discretizations, techniques for testing and tuning the new formalisms
that result, and recent simulation results employing these formalisms.
\end{abstract}
\section{Introduction}

Lattice quantum chromodynamics is {\em the\/} fundamental theory of
low-energy strong interactions. In principle, lattice QCD should tell us
everything we want to know about hadronic spectra and structure, including
such phenomenologically useful things as weak-interaction form factors, decay
rates and deep-inelastic structure functions.  In these 
lectures\footnote{These lectures were 
given at the 1996 Schladming Winter School on
Perturbative and Nonperturbative Aspects of Quantum Field Theory
(Schladming, Austria, March 1996)}
I discuss a revolutionary development  that makes the techniques of
lattice QCD much more widely accessible and greatly extends the range
of problems that are tractable.

The basic approximation in lattice QCD is the replacement of continuous space
and time by a discrete grid.
The nodes or ``sites'' of the grid are separated by
lattice spacing~$a$, and the length of a side of the grid is~$L$:
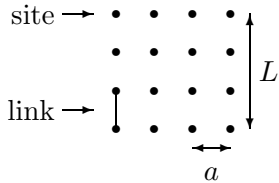
\begin{center}
\setlength{\unitlength}{.02in}
\begin{picture}(75,75)(0,0)
\multiput(25,25)(10,0){4}{\circle*{2}}
\multiput(25,35)(10,0){4}{\circle*{2}}
\multiput(25,45)(10,0){4}{\circle*{2}}
\multiput(25,55)(10,0){4}{\circle*{2}}

\put(60,40){\vector(0,1){15}}
\put(60,40){\vector(0,-1){15}}
\put(65,40){\makebox(0,0){$L$}}

\put(50,20){\vector(1,0){5}}
\put(50,20){\vector(-1,0){5}}
\put(50,15){\makebox(0,0)[t]{$a$}}

\put(11,55){\vector(1,0){8}}
\put(9,55){\makebox(0,0)[r]{site}}

\put(11,30){\vector(1,0){8}}
\put(9,30){\makebox(0,0)[r]{link}}
\put(25,25){\line(0,1){10}}

\end{picture}
\end{center}
The quark
and gluon fields from which the theory is built are specified only on the
sites of the grid, or on the ``links'' joining adjacent sites;
interpolation is used to find the fields  between the sites.
In this lattice approximation, the path integral, from which all quantum
mechanical properties of the theory can be extracted, becomes an ordinary
multidimensional integral where the integration variables are the values of
the fields at each of the grid sites:
\be
\int {\cal D}A_\mu\,\ldots\,\e^{-\int L \dd t}
\longrightarrow \int \prod_{x_j\epsilon\,{\rm grid}}\!\! \dd A_\mu(x_j)\,\ldots
\,\e^{-a\sum L_j}.
\ee
Thus the problem of nonperturbative relativistic quantum field theory is
reduced to one of numerical integration. The integral is over a large
number of variables and so Monte Carlo methods are generally used in its
evaluation. Note that the path integral uses euclidean time
rather than ordinary minkowski time, where $t_{\rm eucl} = \I\,t_{\rm
mink}$; this removes a factor of~$\I$ from the exponent, getting rid of
high-frequency oscillations in the integrand that are hard to integrate.

Early enthusiasm for this approach to nonperturbative QCD gradually
gave way to the sobering
realization that very large computers would be needed for the numerical
integration of the path integral\,---\,computers much larger than those that
existed in the mid~1970's, when lattice QCD was invented. Much
of a lattice theorist's effort in the first twenty years of lattice QCD was
spent in accumulating computing time on the world's largest supercomputers,
or in designing and building computers far larger than the largest
commercially available computers. By the early~1990's, it was widely felt
that teraflop computers costing tens of million of dollars would be essential
to the final numerical solution of full QCD.

The revolutionary development I discuss here was the introduction of new
techniques that allow one to do realistic numerical simulations
of QCD on ordinary desktop workstations or even personal computers.
To understand this development one must understand the factors that govern
the cost of a full QCD simulation. This cost is governed
by a formula like
\be
\mbox{cost} \propto \left(\frac{L}{a}\right)^4\,\left(\frac{1}{a}\right)
\,\left(\frac{1}{m_\pi^2\,a}\right)
\ee
where the first factor is just the number of lattice sites in the grid, and
the remaining factors account for the ``critical slowing down'' of
the algorithms used in the numerical integration. This formula shows that the
single most important determinant of cost is the lattice spacing. The cost
varies as the sixth power of~$1/a$, suggesting that one ought to keep the
lattice spacing as large as possible. Until very recently it was thought
that lattice spacings as small as .05--.1\,fm would be essential for reliable
simulations of QCD. As I describe in these lectures, new simulation results
based on new techniques indicate that spacings as large as .4\,fm give
accurate results. Assuming that the cost is proportional to~$(1/a)^6$,
these new simulations using coarse lattices should cost $10^3$--$10^6$~times
less than traditional simulations on fine lattices.

The computational advantage of coarse lattices is enormous and will
certainly redefine numerical QCD: the simplest calculations can be
done on a personal computer, while problems of unprecedented
difficulty and precision can be tackled with large supercomputers.  In
these lectures I explain why we now think coarse lattices can be made
to work well. I review the techniques from quantum field theory needed
to redesign lattice QCD for coarse lattices. These techniques are
based on ideas from renormalization theory and effective field
theory. I begin in Sect.~2 discussing the field-theoretic implications
of discretizing space and time. Then in Sects.~3, 4~and~5 I discuss in
detail how to discretize the dynamics of gluons, light quarks and
heavy quarks, respectively. I also discuss the crucial issue of
testing and tuning improved discretizations.  These sections present a
mixture of very old and very new results.  I do not discuss in any
detail the numerical techniques used in the simulations; these are
described in standard texts\,\cite{creutz85,montvay94,gpl89}.  As these
are lectures from a school, I include a number of exercises that
illustrate the concepts developed in the lectures.  Finally I
summarize the current situation and prospects for the future in
Sect.~6.
\section{Field Theory on a Lattice}
In this section  I discuss the factors that limit the maximum size of
the lattice spacing. Replacing space-time by a discrete lattice is an
approximation. If we make the lattice spacing too large, our answers will not
be sufficiently accurate. For the purpose of these lectures, I define
a   ``sufficiently accurate'' simulation to be one that reproduces the
low-energy
properties of hadrons to within a few percent, which is very precise for
low-energy strong-interaction physics.  The issue then is: How large can we
make~$a$ while keeping errors of order a few percent or less?

A nonzero lattice spacing results in two types of error: the error that
arises when we replace derivatives in the field equations by
finite-difference approximations, and the error due to the ultraviolet
cutoff imposed by the lattice. I now discuss each of these in turn, and then
focus on the key role played by perturbation theory in correcting for
finite-$a$ errors.

\subsection{Approximate Derivatives}
In the lattice approximation, field values are known only at the sites on
the lattice. Consequently we approximate  derivatives in the field equations
by finite-differences that use only field values at the sites.
This is completely conventional, and very familiar, numerical analysis. For
example, the derivative of a field $\psi$ evaluated at lattice site~$x_j$ is
approximated by
\be
\frac{\partial \psi(x_j)}{\partial x} \approx
\Delta_x\psi(x_j)
\ee
where
\be
\Delta_x\psi(x) \equiv \frac{\psi(x+a)-\psi(x-a)}{2a}.
\ee
It is easy to analyze the error associated with this approximation. Taylor's
Theorem implies that
\bearray
2a\,\Delta_x\psi(x) &\equiv& \psi(x+a) - \psi(x-a) \nl
&=& \left(\e^{a\partial_x} - \e^{-a\partial_x}\right) \psi(x)
\eearray
and therefore
\be \label{delta-psi}
\Delta_x\psi = \left(\partial_x + \frac{a^2}{6}\partial^3_x +
\order(a^4)\right)
\psi.
\ee
Thus the relative error in $\Delta_x\psi$ is of order~$(a/\lambda)^2$ where
$\lambda$ is the typical length scale in~$\psi(x)$.

On coarse lattices we generally need more accurate discretizations than this
one. These are easily constructed. For example from \eq{delta-psi} it is
obvious that
\be \label{deriv-exp}
\frac{\partial\psi}{\partial x} = \Delta_x\psi - \frac{a^2}{6}\Delta^3_x
\psi + \order(a^4),
\ee
which is a more accurate discretization.
When one wishes to reduce the finite-$a$ errors in a simulation,
it is usually far more efficient to improve the discretization of the
derivatives than to reduce
the lattice spacing. For example, with just the first term in the
approximation to~$\partial_x\psi$, cutting the lattice
spacing in half would reduce a 20\%~error to~5\%; but the cost
would increase by a factor of $2^6\!=\!64$ in a simulation where cost
goes like~$1/a^6$.  On
the other hand, including the $a^2$ correction to the derivative, while
working at the larger
lattice spacing, achieves the same reduction in error but with a cost
increase of only a factor of~2.

Equation~\ref{deriv-exp} shows the first two terms
of a  systematic expansion of the continuum derivative in powers of $a^2$.
In principle, higher-order terms can be included to obtain greater
accuracy, but in practice the first couple of terms are sufficiently
accurate for most purposes. Simple numerical experiments with $a^3$-accurate
discretizations like this one (see below) show
that only three or four lattice sites per bump in~$\psi$ are needed to
achieve accuracies of a few percent or less.
Since ordinary hadrons are approximately
1.8\,fm in diameter, these experiments suggest that a lattice spacing of
.4\,fm would suffice for simulating these hadrons. However QCD is a quantum
theory, and, as I discuss in the next section, quantum effects can change
everything.

  \begin{exercise} Show that
  \be \label{deriv2-exp}
  \frac{\partial^2\psi}{\partial x^2} = \lder2_x\psi - \frac{a^2}{12}\left(
  \lder2_x\right)^2\psi + \order(a^4)
  \ee
  where
  \be
  \lder2_x\psi(x) \equiv \frac{\psi(x+a)-2\psi(x)+\psi(x-a)}{a^2}.
  \ee
  Note that $\Delta_x$ and $\lder2_x$ can be used to construct lattice
  approximations for derivatives of any order: for example,
  \bearray
  \lder3_x\psi &\equiv& \Delta_x\lder2_x\psi = \lder2\Delta_x\psi\nl
             &=& \partial_x^3\psi + \order(a^2) \\
  \lder4_x\psi &\equiv& \left(\lder2_x\right)^2\psi \nl
             &=& \partial_x^4\psi + \order(a^2) \\
  &\vdots& \nonumber
  \eearray
  \end{exercise}

  \begin{exercise} Use
  \be \psi(x) = \e^{-x^2/2\sigma^2} \ee
  with $\sigma=2a$ as a sample function to test our discretizations of
  $\partial_x\psi$ and $\partial_x^2\psi$. Evaluate the $a^2$ errors of the
  simplest discretization in each case, and the $a^4$ errors of the more
  accurate discretizations. Show that the $a^2$~errors are of order
  10--20\%, while the $a^4$~errors are only a few percent or less. Note
  that $\lder{2}$ is more substantially more accurate than $\lder{1}$.
		\end{exercise}

\subsection{Ultraviolet Cutoff}
The shortest wavelength oscillation that can be modeled on a lattice is one
with wavelength~$\lambda_{\rm min} = 2a$; for example, the function
$\psi(x)=+1,-1,+1\ldots$ for $x=0,a,2a\ldots$ oscillates with this
wavelength. Thus gluons and quarks  with momenta $k\!=\!2\pi/\lambda$ larger
than~$\pi/a$ are excluded from the lattice theory by the lattice; that is,
the lattice functions as an ultraviolet cutoff. In simple classical field
theories this is often irrelevant: short-wavelength ultraviolet modes are
either unexcited or decouple from the long-wavelength infrared modes of
interest. However, in a noisy nonlinear theory, like an interacting quantum
field theory, ultraviolet modes  strongly affect infrared modes by
renormalizing masses and interactions. Thus we cannot simply discard
all particles with momenta larger than~$\pi/a$; we must somehow mimick their
effects on infrared states. This we can do by modifying our discretized
lagrangian.\footnote{The idea of modifying the lagrangian to
compensate for a finite UV~cutoff is central to chiral field theories
for pions, nonrelativistic QED/QCD and all other
effective field theories. The application to lattice field theories was
pioneered in the form discussed here by Symanzik and is refered to as
``Symanzik improvement'' of lattice operators\,\cite{symanzik83}.}

\begin{figure}
\begin{center}
\setlength{\unitlength}{4pt}
\begin{picture}(35,15)(-5,7)
\put(5,10){\line(1,0){20}}
\put(5,8){\makebox(0,0){$p_2$}}\put(26,8){\makebox(0,0)[r]{$p_4$}}
\put(5,20){\line(1,0){20}}
\put(5,22){\makebox(0,0){$p_1$}}\put(26,22){\makebox(0,0)[r]{$p_3$}}
\put(7,15){\makebox(0,0)[l]{$k$}}
\put(-2,20){\makebox(0,0)[lb]{a)}}
\linethickness{1pt}
\multiput(10,10.5)(0,1){10}{\line(0,1){.25}}
\multiput(20,10.5)(0,1){10}{\line(0,1){.25}}
\end{picture}\qquad
\begin{picture}(40,13)(0,5)
\put(5,10){\line(1,0){30}}
\put(5,8){\makebox(0,0)[r]{$p$}}
\put(20,17){\makebox(0,0)[b]{$k$}}
\put(0,18){\makebox(0,0)[lb]{b)}}
\linethickness{1pt}
\bezier{20}(10,10)(20,20)(30,10)
\end{picture}
\end{center}
\caption{One-loop amplitudes contributing to: a) $qq\!\to\! qq$, and b)
the quark self energy.}
\label{qq-fig}
\end{figure}
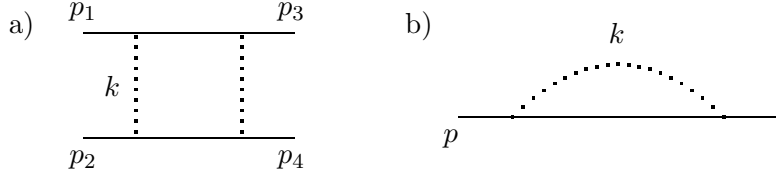

To see how we might mimick the effects of $k\!>\!\pi/a$ states on low
momentum states, consider the scattering amplitude~$T$ for quark-quark
scattering in one-loop perturbation theory (Fig.~\ref{qq-fig}a). The
difference between the correct amplitude~$T$ in the continuum theory and the
cut-off amplitude~$T^\cutoff$ in our lattice theory involves internal
gluons with momentum $k$ of order $\pi/a$ or larger. This is because the
classical theories agree at low momenta, and therefore all propagators and
vertices agree there as well. Thus the loop contributions from low momenta
will be the same in~$T$ and~$T^\cutoff$, and cancel in the difference.
Given that the external  quarks have momenta~$p_i$ that are small relative
to $\pi/a$, we can expand the difference $T-T^\cutoff$ in a Taylor series
in~$a\,p_i$ to obtain
 \bearray
T-T^{(a)} & = &
a^2\,c(a)\,\ub(p_2)\gamma_\mu u(p_1)\,\ub(p_4)\gamma^\mu u(p_3) \nl
&+& a^2\,c_A(a)\,\ub\gamma_\mu\gamma_5 u\,\ub\gamma^\mu\gamma_5 u \nl
&+& a^4\,d(a)\,(p_1-p_2)^2\,\ub\gamma_\mu u\,\ub\gamma^\mu u \nl
&+&\cdots .
 \eearray
where the couplings $c$, $c_A$\,\ldots{}\,are dimensionless functions of the
cutoff. This difference is what is missing from the lattice theory; it is
the contribution of the $k\!>\!\pi/a$~states excluded by the
lattice.\footnote{More precisely, this contribution is due to states
above the cutoff, and to corrections needed to fix up the
states below but sufficiently near the cutoff that they suffer
severe lattice distortion.} The key observation is that
we can reintroduce this high-$k$ contribution into the lattice theory by
adding new interactions to the lattice lagrangian:
\bearray
\delta\Lag_{\rm 4q}^\cutoff &=&
\half\,a^2\,c(a)\,\psib\gamma_\mu\psi\,\psib\gamma^\mu\psi \nl
&+& \half\,a^2\,c_A(a)\,\psib\gamma_\mu\gamma_5\psi\,
					\psib\gamma^\mu\gamma_5\psi \nl
&+& a^4\,d(a)\,\psib D^2\gamma_\mu\psi\,\psib\gamma^\mu\psi\nl
&+&\cdots.
\eearray
These new local interactions give the same contribution to~$T(qq\to qq)$
in the lattice theory as the $k\!>\!\pi/a$ states do in the continuum
theory. Although these correction terms are
nonrenormalizable, they do not cause problems in our lattice theory
because it is cut off at~$\pi/a$. On the contrary, they bring our lattice
theory closer to the continuum without reducing the lattice spacing.

This simple analysis illustrates a general result of renormalization
theory: one can mimick the effects of states excluded by a cutoff with
extra local interactions in the cut-off lagrangian. The
correction terms in $\delta\Lag^\cutoff$ are all local\,---\,that is, they
are polynomials in the fields and in derivatives of the
fields\,---\,since  contributions omitted as a consequence of the cutoff
can be Taylor-expanded in powers of $ap_i$. This is intuitively reasonable
since these terms correct for contributions from intermediate states that
are highly virtual and thus quite local in extent; the locality is a
consequence of the uncertainty principle. The relative importance of the
different interactions is determined by the number of powers of~$a$ in their
coefficients, and that is determined by the dimensions of the operators
involved: each term in $\delta\Lag^\cutoff$ must have total (energy)
dimension four, and so if the operator in a particular term has dimension
$n+4$ then the coefficient must contain a factor~$a^n$. In principle there
are infinitely many correction terms, but we need only include interactions
with operators of dimension~$n+4$ or less to achieve accuracy through
order~$(ap_i)^n$. In practice we only need precision to a given order~$n$
in $ap_i$, and there are only a finite number of local operators with
dimension less than or equal to any given~$n+4$.

Initially the $k\!>\!\pi/a$~contributions appear to be bad news. As we
argued in the previous section, $a^2$~corrections are necessary for
precision when
$a$~is large. However the quantum nature of our theory implies that there
are
$a^2$~terms, due to $k\!>\,\pi/a$~states, which depend in detail on
the nature of the theory. Thus, for example, when we discretize the
derivative in the QCD quark action we replace
\be
\psib\partial\cdot\gamma\psi \to \psib\Delta\cdot\gamma\psi
+ a^2\,d(a)\,\psib\lder3\cdot\gamma\psi + \cdots
\ee
where now $d(a)$ has a part, $-1/6$, from numerical analysis
(equation \eq{deriv-exp}) plus a contribution that mimicks the $k\!>\!\pi/a$
part of the quark self energy (Fig.~\ref{qq-fig}b). The problem is that the
latter contribution is, by its nature, completely theory specific; we
cannot look it up in a book on numerical analysis. We must somehow solve
the $k\!>\!\pi/a$ part of the field theory in order to calculate it;
otherwise we are unable to correct our lagrangian and unable to use large
lattice spacings.

The good news, for lattice QCD, is that $k\!>\!\pi/a$~QCD is perturbative
provided $a$ is small enough (because of asymptotic freedom). Then
correction coefficients like~$d(a)$ can be computed perturbatively, using
Feynman diagrams, in an expansion in powers of~$\alpha_s(\pi/a)$. Thus,
for example, our corrected lattice action for quarks becomes
\bearray
\Lag^\cutoff &=& \psib\left(\Delta\cdot\gamma + m(a)\right)\psi \nl
&+& d(a)\,a^2\,\psib\lder3\cdot\gamma\psi \nl
&+& c(a)\,a^2\,\psib\gamma_\mu\psi\,\psib\gamma^\mu\psi + \ldots
\eearray
where
\bearray
d(a) &=& -\frac{1}{6} + d_1\,\alpha_s(\pi/a) + \cdots \\
c(a) &=& c_2\,\alpha_s^2(\pi/a) + \cdots \\
\cdots \nonumber
\eearray
are computed to whatever order in perturbation theory is necessary.
In this way we use perturbation theory to, in effect, fill in the
gaps between lattice points, allowing us to obtain continuum results without
taking the lattice spacing to zero.

\subsection{Perturbation Theory and Tadpole Improvement\protect\footnotemark}
\footnotetext{This  section is based upon work with Paul Mackenzie
that is described in~\cite{gpl93}.}
Improved discretizations and large lattice spacings are old ideas,
pioneered by Wilson, Symanzik and others\,\cite{wilson83}.
However, perturbation theory is
essential; the lattice spacing~$a$ must be small enough so that
$k\!\approx\!\pi/a$~QCD is perturbative. This was the requirement that
drove lattice QCD towards very costly simulations with tiny lattice
spacings. Traditional perturbation theory for
lattice QCD begins to fail at distances of order $1/20$
to $1/10$\,fm, and therefore lattice spacings must be at least this small
before improved actions are useful. This seems very odd since
phenomenological applications of continuum perturbative QCD suggest that
perturbation theory works well down to energies of order 1\,GeV, which
corresponds to a lattice spacing of~0.6\,fm. The breakthrough, in the early
1990's, was the discovery of a trivial modification of lattice QCD, called
``tadpole improvement,'' that allows perturbation theory to work even at
distances as large as $1/2$\,fm\,\cite{gpl93,gpl94}.

One can readily derive Feynman diagram rules for lattice QCD using the
same techniques as in the continuum, but applied to the lattice
lagrangian\,\cite{kawai81}.
The particle propagators and interaction vertices are usually
complicated functions of the momenta that become identical to their
continuum analogues in the low-momentum limit. All loop momenta are cut
off at $k_\mu=\pm\pi/a$.

Testing perturbation theory is also straightforward. One designs
short-distance quantities that can be computed easily in a simulation (i.e.,
in a Monte Carlo evaluation of the lattice path integral). The Monte Carlo
gives the exact value which can then be compared with the perturbative
expansion for the same quantity. An example of such a quantity is the
expectation value of the Wilson loop operator,
\be
W({\cal C})\equiv\langle 0|\third\Re\Tr\P\e^{-\I g\oint_{\cal C}A\cdot
\dd x}|0\rangle,
\ee
where $A$ is the QCD vector potential, $\P$~denotes path ordering, and
${\cal C}$ is any small, closed path or loop on the lattice. $W({\cal
C})$~is perturbative for sufficiently small loops~$\cal C$.
We can test the utility of perturbation theory over any range of distances
by varying the loop size while comparing
numerical Monte Carlo results for~$W({\cal C})$ with perturbation theory.

Fig.~\ref{creutz-fig} illustrates the highly unsatisfactory state of
traditional lattice-QCD perturbation theory.
There I show the ``Creutz ratio'' of $2a\times2a$, $2a\times a$ and
$a\times a$ Wilson loops,
\be \label{creutz-ratio}
\chi_{2,2} \equiv -\ln\left(\frac{W(2a\times2a)\,W(a\times a)}{W^2(2a\times
a)}\right),
\ee
plotted versus the size~$2a$ of the largest loop. Traditional perturbation
theory (dotted lines)
underestimates the exact result by factors of three or four for
loops of order~$1/2$\,fm; only when the loops are smaller than $1/20$\,fm
does perturbation theory begin to give accurate results.
\begin{figure}
\begin{center}
\setlength{\unitlength}{0.240900pt}
\ifx\plotpoint\undefined\newsavebox{\plotpoint}\fi
\sbox{\plotpoint}{\rule[-0.200pt]{0.400pt}{0.400pt}}%
\begin{picture}(1200,900)(0,0)
\font\gnuplot=cmr10 at 10pt
\gnuplot
\sbox{\plotpoint}{\rule[-0.200pt]{0.400pt}{0.400pt}}%
\put(220.0,113.0){\rule[-0.200pt]{220.664pt}{0.400pt}}
\put(220.0,401.0){\rule[-0.200pt]{4.818pt}{0.400pt}}
\put(198,401){\makebox(0,0)[r]{0.2}}
\put(1116.0,401.0){\rule[-0.200pt]{4.818pt}{0.400pt}}
\put(220.0,688.0){\rule[-0.200pt]{4.818pt}{0.400pt}}
\put(198,688){\makebox(0,0)[r]{0.4}}
\put(1116.0,688.0){\rule[-0.200pt]{4.818pt}{0.400pt}}
\put(449.0,113.0){\rule[-0.200pt]{0.400pt}{4.818pt}}
\put(449,68){\makebox(0,0){0.001}}
\put(449.0,812.0){\rule[-0.200pt]{0.400pt}{4.818pt}}
\put(678.0,113.0){\rule[-0.200pt]{0.400pt}{4.818pt}}
\put(678,68){\makebox(0,0){0.01}}
\put(678.0,812.0){\rule[-0.200pt]{0.400pt}{4.818pt}}
\put(907.0,113.0){\rule[-0.200pt]{0.400pt}{4.818pt}}
\put(907,68){\makebox(0,0){0.1}}
\put(907.0,812.0){\rule[-0.200pt]{0.400pt}{4.818pt}}
\put(220.0,113.0){\rule[-0.200pt]{220.664pt}{0.400pt}}
\put(1136.0,113.0){\rule[-0.200pt]{0.400pt}{173.207pt}}
\put(220.0,832.0){\rule[-0.200pt]{220.664pt}{0.400pt}}
\put(45,472){\makebox(0,0){$\chi_{22}$}}
\put(678,-22){\makebox(0,0){loop size (fm)}}
\put(678,877){\makebox(0,0){$\chi_{22}$ Loop Ratio\,---\,$1^{\rm st}$ Order}}
\put(220.0,113.0){\rule[-0.200pt]{0.400pt}{173.207pt}}
\put(518,688){\makebox(0,0)[r]{exact}}
\put(562,688){\circle{24}}
\put(1029,650){\circle{24}}
\put(975,495){\circle{24}}
\put(946,448){\circle{24}}
\put(605,259){\circle{24}}
\put(254,208){\circle{24}}
\put(518,643){\makebox(0,0)[r]{new P.Th.}}
\put(540.0,643.0){\rule[-0.200pt]{15.899pt}{0.400pt}}
\put(254,209){\usebox{\plotpoint}}
\put(254,209.17){\rule{1.700pt}{0.400pt}}
\multiput(254.00,208.17)(4.472,2.000){2}{\rule{0.850pt}{0.400pt}}
\put(262,210.67){\rule{1.927pt}{0.400pt}}
\multiput(262.00,210.17)(4.000,1.000){2}{\rule{0.964pt}{0.400pt}}
\put(270,211.67){\rule{1.927pt}{0.400pt}}
\multiput(270.00,211.17)(4.000,1.000){2}{\rule{0.964pt}{0.400pt}}
\put(278,212.67){\rule{1.686pt}{0.400pt}}
\multiput(278.00,212.17)(3.500,1.000){2}{\rule{0.843pt}{0.400pt}}
\put(285,213.67){\rule{1.927pt}{0.400pt}}
\multiput(285.00,213.17)(4.000,1.000){2}{\rule{0.964pt}{0.400pt}}
\put(293,215.17){\rule{1.700pt}{0.400pt}}
\multiput(293.00,214.17)(4.472,2.000){2}{\rule{0.850pt}{0.400pt}}
\put(301,216.67){\rule{1.927pt}{0.400pt}}
\multiput(301.00,216.17)(4.000,1.000){2}{\rule{0.964pt}{0.400pt}}
\put(309,217.67){\rule{1.927pt}{0.400pt}}
\multiput(309.00,217.17)(4.000,1.000){2}{\rule{0.964pt}{0.400pt}}
\put(317,218.67){\rule{1.686pt}{0.400pt}}
\multiput(317.00,218.17)(3.500,1.000){2}{\rule{0.843pt}{0.400pt}}
\put(324,219.67){\rule{1.927pt}{0.400pt}}
\multiput(324.00,219.17)(4.000,1.000){2}{\rule{0.964pt}{0.400pt}}
\put(332,221.17){\rule{1.700pt}{0.400pt}}
\multiput(332.00,220.17)(4.472,2.000){2}{\rule{0.850pt}{0.400pt}}
\put(340,222.67){\rule{1.927pt}{0.400pt}}
\multiput(340.00,222.17)(4.000,1.000){2}{\rule{0.964pt}{0.400pt}}
\put(348,223.67){\rule{1.927pt}{0.400pt}}
\multiput(348.00,223.17)(4.000,1.000){2}{\rule{0.964pt}{0.400pt}}
\put(356,224.67){\rule{1.686pt}{0.400pt}}
\multiput(356.00,224.17)(3.500,1.000){2}{\rule{0.843pt}{0.400pt}}
\put(363,226.17){\rule{1.700pt}{0.400pt}}
\multiput(363.00,225.17)(4.472,2.000){2}{\rule{0.850pt}{0.400pt}}
\put(371,227.67){\rule{1.927pt}{0.400pt}}
\multiput(371.00,227.17)(4.000,1.000){2}{\rule{0.964pt}{0.400pt}}
\put(379,228.67){\rule{1.927pt}{0.400pt}}
\multiput(379.00,228.17)(4.000,1.000){2}{\rule{0.964pt}{0.400pt}}
\put(387,229.67){\rule{1.927pt}{0.400pt}}
\multiput(387.00,229.17)(4.000,1.000){2}{\rule{0.964pt}{0.400pt}}
\put(395,230.67){\rule{1.686pt}{0.400pt}}
\multiput(395.00,230.17)(3.500,1.000){2}{\rule{0.843pt}{0.400pt}}
\put(402,232.17){\rule{1.700pt}{0.400pt}}
\multiput(402.00,231.17)(4.472,2.000){2}{\rule{0.850pt}{0.400pt}}
\put(410,233.67){\rule{1.927pt}{0.400pt}}
\multiput(410.00,233.17)(4.000,1.000){2}{\rule{0.964pt}{0.400pt}}
\put(418,234.67){\rule{1.927pt}{0.400pt}}
\multiput(418.00,234.17)(4.000,1.000){2}{\rule{0.964pt}{0.400pt}}
\put(426,235.67){\rule{1.927pt}{0.400pt}}
\multiput(426.00,235.17)(4.000,1.000){2}{\rule{0.964pt}{0.400pt}}
\put(434,236.67){\rule{1.686pt}{0.400pt}}
\multiput(434.00,236.17)(3.500,1.000){2}{\rule{0.843pt}{0.400pt}}
\put(441,238.17){\rule{1.700pt}{0.400pt}}
\multiput(441.00,237.17)(4.472,2.000){2}{\rule{0.850pt}{0.400pt}}
\put(449,239.67){\rule{1.927pt}{0.400pt}}
\multiput(449.00,239.17)(4.000,1.000){2}{\rule{0.964pt}{0.400pt}}
\put(457,240.67){\rule{1.927pt}{0.400pt}}
\multiput(457.00,240.17)(4.000,1.000){2}{\rule{0.964pt}{0.400pt}}
\put(465,241.67){\rule{1.686pt}{0.400pt}}
\multiput(465.00,241.17)(3.500,1.000){2}{\rule{0.843pt}{0.400pt}}
\put(472,242.67){\rule{1.927pt}{0.400pt}}
\multiput(472.00,242.17)(4.000,1.000){2}{\rule{0.964pt}{0.400pt}}
\put(480,244.17){\rule{1.700pt}{0.400pt}}
\multiput(480.00,243.17)(4.472,2.000){2}{\rule{0.850pt}{0.400pt}}
\put(488,245.67){\rule{1.927pt}{0.400pt}}
\multiput(488.00,245.17)(4.000,1.000){2}{\rule{0.964pt}{0.400pt}}
\put(496,246.67){\rule{1.927pt}{0.400pt}}
\multiput(496.00,246.17)(4.000,1.000){2}{\rule{0.964pt}{0.400pt}}
\put(504,247.67){\rule{1.686pt}{0.400pt}}
\multiput(504.00,247.17)(3.500,1.000){2}{\rule{0.843pt}{0.400pt}}
\put(511,249.17){\rule{1.700pt}{0.400pt}}
\multiput(511.00,248.17)(4.472,2.000){2}{\rule{0.850pt}{0.400pt}}
\put(519,250.67){\rule{1.927pt}{0.400pt}}
\multiput(519.00,250.17)(4.000,1.000){2}{\rule{0.964pt}{0.400pt}}
\put(527,251.67){\rule{1.927pt}{0.400pt}}
\multiput(527.00,251.17)(4.000,1.000){2}{\rule{0.964pt}{0.400pt}}
\put(535,252.67){\rule{1.927pt}{0.400pt}}
\multiput(535.00,252.17)(4.000,1.000){2}{\rule{0.964pt}{0.400pt}}
\put(543,253.67){\rule{1.686pt}{0.400pt}}
\multiput(543.00,253.17)(3.500,1.000){2}{\rule{0.843pt}{0.400pt}}
\put(550,255.17){\rule{1.700pt}{0.400pt}}
\multiput(550.00,254.17)(4.472,2.000){2}{\rule{0.850pt}{0.400pt}}
\put(558,256.67){\rule{1.927pt}{0.400pt}}
\multiput(558.00,256.17)(4.000,1.000){2}{\rule{0.964pt}{0.400pt}}
\put(566,257.67){\rule{1.927pt}{0.400pt}}
\multiput(566.00,257.17)(4.000,1.000){2}{\rule{0.964pt}{0.400pt}}
\put(574,258.67){\rule{1.927pt}{0.400pt}}
\multiput(574.00,258.17)(4.000,1.000){2}{\rule{0.964pt}{0.400pt}}
\put(582,259.67){\rule{1.686pt}{0.400pt}}
\multiput(582.00,259.17)(3.500,1.000){2}{\rule{0.843pt}{0.400pt}}
\put(589,261.17){\rule{1.700pt}{0.400pt}}
\multiput(589.00,260.17)(4.472,2.000){2}{\rule{0.850pt}{0.400pt}}
\put(597,262.67){\rule{1.927pt}{0.400pt}}
\multiput(597.00,262.17)(4.000,1.000){2}{\rule{0.964pt}{0.400pt}}
\put(605,263.67){\rule{1.927pt}{0.400pt}}
\multiput(605.00,263.17)(4.000,1.000){2}{\rule{0.964pt}{0.400pt}}
\put(613,264.67){\rule{1.686pt}{0.400pt}}
\multiput(613.00,264.17)(3.500,1.000){2}{\rule{0.843pt}{0.400pt}}
\put(620,266.17){\rule{1.700pt}{0.400pt}}
\multiput(620.00,265.17)(4.472,2.000){2}{\rule{0.850pt}{0.400pt}}
\put(628,267.67){\rule{1.927pt}{0.400pt}}
\multiput(628.00,267.17)(4.000,1.000){2}{\rule{0.964pt}{0.400pt}}
\put(636,268.67){\rule{1.927pt}{0.400pt}}
\multiput(636.00,268.17)(4.000,1.000){2}{\rule{0.964pt}{0.400pt}}
\put(644,269.67){\rule{1.686pt}{0.400pt}}
\multiput(644.00,269.17)(3.500,1.000){2}{\rule{0.843pt}{0.400pt}}
\put(651,271.17){\rule{1.700pt}{0.400pt}}
\multiput(651.00,270.17)(4.472,2.000){2}{\rule{0.850pt}{0.400pt}}
\put(659,272.67){\rule{1.927pt}{0.400pt}}
\multiput(659.00,272.17)(4.000,1.000){2}{\rule{0.964pt}{0.400pt}}
\put(667,274.17){\rule{1.700pt}{0.400pt}}
\multiput(667.00,273.17)(4.472,2.000){2}{\rule{0.850pt}{0.400pt}}
\put(675,276.17){\rule{1.500pt}{0.400pt}}
\multiput(675.00,275.17)(3.887,2.000){2}{\rule{0.750pt}{0.400pt}}
\put(682,278.17){\rule{1.700pt}{0.400pt}}
\multiput(682.00,277.17)(4.472,2.000){2}{\rule{0.850pt}{0.400pt}}
\put(690,279.67){\rule{1.927pt}{0.400pt}}
\multiput(690.00,279.17)(4.000,1.000){2}{\rule{0.964pt}{0.400pt}}
\multiput(698.00,281.61)(1.579,0.447){3}{\rule{1.167pt}{0.108pt}}
\multiput(698.00,280.17)(5.579,3.000){2}{\rule{0.583pt}{0.400pt}}
\put(706,284.17){\rule{1.700pt}{0.400pt}}
\multiput(706.00,283.17)(4.472,2.000){2}{\rule{0.850pt}{0.400pt}}
\put(714,286.17){\rule{1.500pt}{0.400pt}}
\multiput(714.00,285.17)(3.887,2.000){2}{\rule{0.750pt}{0.400pt}}
\multiput(721.00,288.61)(1.579,0.447){3}{\rule{1.167pt}{0.108pt}}
\multiput(721.00,287.17)(5.579,3.000){2}{\rule{0.583pt}{0.400pt}}
\put(729,291.17){\rule{1.700pt}{0.400pt}}
\multiput(729.00,290.17)(4.472,2.000){2}{\rule{0.850pt}{0.400pt}}
\multiput(737.00,293.61)(1.579,0.447){3}{\rule{1.167pt}{0.108pt}}
\multiput(737.00,292.17)(5.579,3.000){2}{\rule{0.583pt}{0.400pt}}
\multiput(745.00,296.61)(1.355,0.447){3}{\rule{1.033pt}{0.108pt}}
\multiput(745.00,295.17)(4.855,3.000){2}{\rule{0.517pt}{0.400pt}}
\multiput(752.00,299.60)(1.066,0.468){5}{\rule{0.900pt}{0.113pt}}
\multiput(752.00,298.17)(6.132,4.000){2}{\rule{0.450pt}{0.400pt}}
\multiput(760.00,303.61)(1.579,0.447){3}{\rule{1.167pt}{0.108pt}}
\multiput(760.00,302.17)(5.579,3.000){2}{\rule{0.583pt}{0.400pt}}
\multiput(768.00,306.60)(1.066,0.468){5}{\rule{0.900pt}{0.113pt}}
\multiput(768.00,305.17)(6.132,4.000){2}{\rule{0.450pt}{0.400pt}}
\multiput(776.00,310.60)(0.920,0.468){5}{\rule{0.800pt}{0.113pt}}
\multiput(776.00,309.17)(5.340,4.000){2}{\rule{0.400pt}{0.400pt}}
\multiput(783.00,314.60)(1.066,0.468){5}{\rule{0.900pt}{0.113pt}}
\multiput(783.00,313.17)(6.132,4.000){2}{\rule{0.450pt}{0.400pt}}
\multiput(791.00,318.59)(0.821,0.477){7}{\rule{0.740pt}{0.115pt}}
\multiput(791.00,317.17)(6.464,5.000){2}{\rule{0.370pt}{0.400pt}}
\multiput(799.00,323.60)(1.066,0.468){5}{\rule{0.900pt}{0.113pt}}
\multiput(799.00,322.17)(6.132,4.000){2}{\rule{0.450pt}{0.400pt}}
\multiput(807.00,327.59)(0.710,0.477){7}{\rule{0.660pt}{0.115pt}}
\multiput(807.00,326.17)(5.630,5.000){2}{\rule{0.330pt}{0.400pt}}
\multiput(814.00,332.59)(0.671,0.482){9}{\rule{0.633pt}{0.116pt}}
\multiput(814.00,331.17)(6.685,6.000){2}{\rule{0.317pt}{0.400pt}}
\multiput(822.00,338.59)(0.821,0.477){7}{\rule{0.740pt}{0.115pt}}
\multiput(822.00,337.17)(6.464,5.000){2}{\rule{0.370pt}{0.400pt}}
\multiput(830.00,343.59)(0.671,0.482){9}{\rule{0.633pt}{0.116pt}}
\multiput(830.00,342.17)(6.685,6.000){2}{\rule{0.317pt}{0.400pt}}
\multiput(838.00,349.59)(0.492,0.485){11}{\rule{0.500pt}{0.117pt}}
\multiput(838.00,348.17)(5.962,7.000){2}{\rule{0.250pt}{0.400pt}}
\multiput(845.00,356.59)(0.671,0.482){9}{\rule{0.633pt}{0.116pt}}
\multiput(845.00,355.17)(6.685,6.000){2}{\rule{0.317pt}{0.400pt}}
\multiput(853.00,362.59)(0.569,0.485){11}{\rule{0.557pt}{0.117pt}}
\multiput(853.00,361.17)(6.844,7.000){2}{\rule{0.279pt}{0.400pt}}
\multiput(861.00,369.59)(0.494,0.488){13}{\rule{0.500pt}{0.117pt}}
\multiput(861.00,368.17)(6.962,8.000){2}{\rule{0.250pt}{0.400pt}}
\multiput(869.00,377.59)(0.492,0.485){11}{\rule{0.500pt}{0.117pt}}
\multiput(869.00,376.17)(5.962,7.000){2}{\rule{0.250pt}{0.400pt}}
\multiput(876.00,384.59)(0.494,0.488){13}{\rule{0.500pt}{0.117pt}}
\multiput(876.00,383.17)(6.962,8.000){2}{\rule{0.250pt}{0.400pt}}
\multiput(884.59,392.00)(0.488,0.560){13}{\rule{0.117pt}{0.550pt}}
\multiput(883.17,392.00)(8.000,7.858){2}{\rule{0.400pt}{0.275pt}}
\multiput(892.59,401.00)(0.488,0.560){13}{\rule{0.117pt}{0.550pt}}
\multiput(891.17,401.00)(8.000,7.858){2}{\rule{0.400pt}{0.275pt}}
\multiput(900.59,410.00)(0.485,0.645){11}{\rule{0.117pt}{0.614pt}}
\multiput(899.17,410.00)(7.000,7.725){2}{\rule{0.400pt}{0.307pt}}
\multiput(907.59,419.00)(0.488,0.626){13}{\rule{0.117pt}{0.600pt}}
\multiput(906.17,419.00)(8.000,8.755){2}{\rule{0.400pt}{0.300pt}}
\multiput(915.59,429.00)(0.488,0.626){13}{\rule{0.117pt}{0.600pt}}
\multiput(914.17,429.00)(8.000,8.755){2}{\rule{0.400pt}{0.300pt}}
\multiput(923.59,439.00)(0.488,0.692){13}{\rule{0.117pt}{0.650pt}}
\multiput(922.17,439.00)(8.000,9.651){2}{\rule{0.400pt}{0.325pt}}
\multiput(931.59,450.00)(0.485,0.798){11}{\rule{0.117pt}{0.729pt}}
\multiput(930.17,450.00)(7.000,9.488){2}{\rule{0.400pt}{0.364pt}}
\multiput(938.59,461.00)(0.488,0.692){13}{\rule{0.117pt}{0.650pt}}
\multiput(937.17,461.00)(8.000,9.651){2}{\rule{0.400pt}{0.325pt}}
\multiput(946.58,472.00)(0.491,0.756){17}{\rule{0.118pt}{0.700pt}}
\multiput(945.17,472.00)(10.000,13.547){2}{\rule{0.400pt}{0.350pt}}
\multiput(956.59,487.00)(0.489,0.961){15}{\rule{0.118pt}{0.856pt}}
\multiput(955.17,487.00)(9.000,15.224){2}{\rule{0.400pt}{0.428pt}}
\multiput(965.58,504.00)(0.491,1.017){17}{\rule{0.118pt}{0.900pt}}
\multiput(964.17,504.00)(10.000,18.132){2}{\rule{0.400pt}{0.450pt}}
\multiput(975.59,524.00)(0.485,1.408){11}{\rule{0.117pt}{1.186pt}}
\multiput(974.17,524.00)(7.000,16.539){2}{\rule{0.400pt}{0.593pt}}
\multiput(982.59,543.00)(0.488,1.352){13}{\rule{0.117pt}{1.150pt}}
\multiput(981.17,543.00)(8.000,18.613){2}{\rule{0.400pt}{0.575pt}}
\multiput(990.59,564.00)(0.488,1.484){13}{\rule{0.117pt}{1.250pt}}
\multiput(989.17,564.00)(8.000,20.406){2}{\rule{0.400pt}{0.625pt}}
\multiput(998.59,587.00)(0.488,1.550){13}{\rule{0.117pt}{1.300pt}}
\multiput(997.17,587.00)(8.000,21.302){2}{\rule{0.400pt}{0.650pt}}
\multiput(1006.59,611.00)(0.488,1.682){13}{\rule{0.117pt}{1.400pt}}
\multiput(1005.17,611.00)(8.000,23.094){2}{\rule{0.400pt}{0.700pt}}
\multiput(1014.59,637.00)(0.488,1.682){13}{\rule{0.117pt}{1.400pt}}
\multiput(1013.17,637.00)(8.000,23.094){2}{\rule{0.400pt}{0.700pt}}
\multiput(1022.59,663.00)(0.485,2.018){11}{\rule{0.117pt}{1.643pt}}
\multiput(1021.17,663.00)(7.000,23.590){2}{\rule{0.400pt}{0.821pt}}
\sbox{\plotpoint}{\rule[-0.500pt]{1.000pt}{1.000pt}}%
\put(518,598){\makebox(0,0)[r]{old P.Th.}}
\multiput(540,598)(20.756,0.000){4}{\usebox{\plotpoint}}
\put(606,598){\usebox{\plotpoint}}
\put(254,182){\usebox{\plotpoint}}
\put(254.00,182.00){\usebox{\plotpoint}}
\multiput(262,183)(20.756,0.000){0}{\usebox{\plotpoint}}
\put(274.66,183.58){\usebox{\plotpoint}}
\multiput(278,184)(20.756,0.000){0}{\usebox{\plotpoint}}
\multiput(285,184)(20.595,2.574){0}{\usebox{\plotpoint}}
\put(295.32,185.00){\usebox{\plotpoint}}
\multiput(301,185)(20.756,0.000){0}{\usebox{\plotpoint}}
\put(316.03,185.88){\usebox{\plotpoint}}
\multiput(317,186)(20.756,0.000){0}{\usebox{\plotpoint}}
\multiput(324,186)(20.595,2.574){0}{\usebox{\plotpoint}}
\put(336.71,187.00){\usebox{\plotpoint}}
\multiput(340,187)(20.595,2.574){0}{\usebox{\plotpoint}}
\multiput(348,188)(20.756,0.000){0}{\usebox{\plotpoint}}
\put(357.39,188.20){\usebox{\plotpoint}}
\multiput(363,189)(20.756,0.000){0}{\usebox{\plotpoint}}
\put(378.09,189.00){\usebox{\plotpoint}}
\multiput(379,189)(20.595,2.574){0}{\usebox{\plotpoint}}
\multiput(387,190)(20.756,0.000){0}{\usebox{\plotpoint}}
\put(398.74,190.53){\usebox{\plotpoint}}
\multiput(402,191)(20.756,0.000){0}{\usebox{\plotpoint}}
\multiput(410,191)(20.595,2.574){0}{\usebox{\plotpoint}}
\put(419.40,192.00){\usebox{\plotpoint}}
\multiput(426,192)(20.595,2.574){0}{\usebox{\plotpoint}}
\put(440.10,193.00){\usebox{\plotpoint}}
\multiput(441,193)(20.595,2.574){0}{\usebox{\plotpoint}}
\multiput(449,194)(20.756,0.000){0}{\usebox{\plotpoint}}
\put(460.76,194.47){\usebox{\plotpoint}}
\multiput(465,195)(20.756,0.000){0}{\usebox{\plotpoint}}
\multiput(472,195)(20.595,2.574){0}{\usebox{\plotpoint}}
\put(481.42,196.00){\usebox{\plotpoint}}
\multiput(488,196)(20.595,2.574){0}{\usebox{\plotpoint}}
\put(502.11,197.00){\usebox{\plotpoint}}
\multiput(504,197)(20.547,2.935){0}{\usebox{\plotpoint}}
\multiput(511,198)(20.595,2.574){0}{\usebox{\plotpoint}}
\put(522.74,199.00){\usebox{\plotpoint}}
\multiput(527,199)(20.595,2.574){0}{\usebox{\plotpoint}}
\multiput(535,200)(20.756,0.000){0}{\usebox{\plotpoint}}
\put(543.43,200.06){\usebox{\plotpoint}}
\multiput(550,201)(20.595,2.574){0}{\usebox{\plotpoint}}
\put(564.05,202.00){\usebox{\plotpoint}}
\multiput(566,202)(20.595,2.574){0}{\usebox{\plotpoint}}
\multiput(574,203)(20.595,2.574){0}{\usebox{\plotpoint}}
\put(584.68,204.00){\usebox{\plotpoint}}
\multiput(589,204)(20.595,2.574){0}{\usebox{\plotpoint}}
\multiput(597,205)(20.595,2.574){0}{\usebox{\plotpoint}}
\put(605.31,206.00){\usebox{\plotpoint}}
\multiput(613,206)(20.547,2.935){0}{\usebox{\plotpoint}}
\put(625.95,207.74){\usebox{\plotpoint}}
\multiput(628,208)(20.756,0.000){0}{\usebox{\plotpoint}}
\multiput(636,208)(20.595,2.574){0}{\usebox{\plotpoint}}
\put(646.60,209.37){\usebox{\plotpoint}}
\multiput(651,210)(20.595,2.574){0}{\usebox{\plotpoint}}
\multiput(659,211)(20.756,0.000){0}{\usebox{\plotpoint}}
\put(667.25,211.03){\usebox{\plotpoint}}
\multiput(675,212)(20.547,2.935){0}{\usebox{\plotpoint}}
\put(687.83,213.73){\usebox{\plotpoint}}
\multiput(690,214)(20.595,2.574){0}{\usebox{\plotpoint}}
\multiput(698,215)(20.756,0.000){0}{\usebox{\plotpoint}}
\put(708.49,215.31){\usebox{\plotpoint}}
\multiput(714,216)(20.547,2.935){0}{\usebox{\plotpoint}}
\multiput(721,217)(20.595,2.574){0}{\usebox{\plotpoint}}
\put(729.06,218.01){\usebox{\plotpoint}}
\multiput(737,219)(20.595,2.574){0}{\usebox{\plotpoint}}
\put(749.65,220.66){\usebox{\plotpoint}}
\multiput(752,221)(20.595,2.574){0}{\usebox{\plotpoint}}
\multiput(760,222)(20.756,0.000){0}{\usebox{\plotpoint}}
\put(770.30,222.29){\usebox{\plotpoint}}
\multiput(776,223)(20.547,2.935){0}{\usebox{\plotpoint}}
\put(790.88,224.98){\usebox{\plotpoint}}
\multiput(791,225)(20.595,2.574){0}{\usebox{\plotpoint}}
\multiput(799,226)(20.595,2.574){0}{\usebox{\plotpoint}}
\put(811.46,227.64){\usebox{\plotpoint}}
\multiput(814,228)(20.595,2.574){0}{\usebox{\plotpoint}}
\multiput(822,229)(20.595,2.574){0}{\usebox{\plotpoint}}
\put(832.05,230.26){\usebox{\plotpoint}}
\multiput(838,231)(20.547,2.935){0}{\usebox{\plotpoint}}
\put(852.63,232.95){\usebox{\plotpoint}}
\multiput(853,233)(20.136,5.034){0}{\usebox{\plotpoint}}
\multiput(861,235)(20.595,2.574){0}{\usebox{\plotpoint}}
\put(873.03,236.58){\usebox{\plotpoint}}
\multiput(876,237)(20.595,2.574){0}{\usebox{\plotpoint}}
\multiput(884,238)(20.595,2.574){0}{\usebox{\plotpoint}}
\put(893.62,239.20){\usebox{\plotpoint}}
\multiput(900,240)(20.547,2.935){0}{\usebox{\plotpoint}}
\put(914.04,242.76){\usebox{\plotpoint}}
\multiput(915,243)(20.595,2.574){0}{\usebox{\plotpoint}}
\multiput(923,244)(20.595,2.574){0}{\usebox{\plotpoint}}
\put(934.61,245.52){\usebox{\plotpoint}}
\multiput(938,246)(20.595,2.574){0}{\usebox{\plotpoint}}
\put(955.08,248.82){\usebox{\plotpoint}}
\multiput(956,249)(20.629,2.292){0}{\usebox{\plotpoint}}
\multiput(965,250)(20.352,4.070){0}{\usebox{\plotpoint}}
\put(975.56,252.08){\usebox{\plotpoint}}
\multiput(982,253)(20.595,2.574){0}{\usebox{\plotpoint}}
\put(996.14,254.77){\usebox{\plotpoint}}
\multiput(998,255)(20.595,2.574){0}{\usebox{\plotpoint}}
\multiput(1006,256)(20.595,2.574){0}{\usebox{\plotpoint}}
\put(1016.74,257.34){\usebox{\plotpoint}}
\multiput(1022,258)(20.547,2.935){0}{\usebox{\plotpoint}}
\put(1029,259){\usebox{\plotpoint}}
\end{picture}
\vspace{8ex}
\setlength{\unitlength}{0.240900pt}
\ifx\plotpoint\undefined\newsavebox{\plotpoint}\fi
\sbox{\plotpoint}{\rule[-0.200pt]{0.400pt}{0.400pt}}%
\begin{picture}(1200,900)(0,0)
\font\gnuplot=cmr10 at 10pt
\gnuplot
\sbox{\plotpoint}{\rule[-0.200pt]{0.400pt}{0.400pt}}%
\put(220.0,113.0){\rule[-0.200pt]{220.664pt}{0.400pt}}
\put(220.0,401.0){\rule[-0.200pt]{4.818pt}{0.400pt}}
\put(198,401){\makebox(0,0)[r]{0.2}}
\put(1116.0,401.0){\rule[-0.200pt]{4.818pt}{0.400pt}}
\put(220.0,688.0){\rule[-0.200pt]{4.818pt}{0.400pt}}
\put(198,688){\makebox(0,0)[r]{0.4}}
\put(1116.0,688.0){\rule[-0.200pt]{4.818pt}{0.400pt}}
\put(449.0,113.0){\rule[-0.200pt]{0.400pt}{4.818pt}}
\put(449,68){\makebox(0,0){0.001}}
\put(449.0,812.0){\rule[-0.200pt]{0.400pt}{4.818pt}}
\put(678.0,113.0){\rule[-0.200pt]{0.400pt}{4.818pt}}
\put(678,68){\makebox(0,0){0.01}}
\put(678.0,812.0){\rule[-0.200pt]{0.400pt}{4.818pt}}
\put(907.0,113.0){\rule[-0.200pt]{0.400pt}{4.818pt}}
\put(907,68){\makebox(0,0){0.1}}
\put(907.0,812.0){\rule[-0.200pt]{0.400pt}{4.818pt}}
\put(220.0,113.0){\rule[-0.200pt]{220.664pt}{0.400pt}}
\put(1136.0,113.0){\rule[-0.200pt]{0.400pt}{173.207pt}}
\put(220.0,832.0){\rule[-0.200pt]{220.664pt}{0.400pt}}
\put(45,472){\makebox(0,0){$\chi_{22}$}}
\put(678,-22){\makebox(0,0){loop size (fm)}}
\put(678,877){\makebox(0,0){$\chi_{22}$ Loop Ratio\,---\,$2^{\rm nd}$ Order}}
\put(220.0,113.0){\rule[-0.200pt]{0.400pt}{173.207pt}}
\put(518,688){\makebox(0,0)[r]{exact}}
\put(562,688){\circle{24}}
\put(1029,650){\circle{24}}
\put(975,495){\circle{24}}
\put(946,448){\circle{24}}
\put(605,259){\circle{24}}
\put(254,208){\circle{24}}
\put(518,643){\makebox(0,0)[r]{new P.Th.}}
\put(540.0,643.0){\rule[-0.200pt]{15.899pt}{0.400pt}}
\put(254,207){\usebox{\plotpoint}}
\put(254,207.17){\rule{1.700pt}{0.400pt}}
\multiput(254.00,206.17)(4.472,2.000){2}{\rule{0.850pt}{0.400pt}}
\put(262,208.67){\rule{1.927pt}{0.400pt}}
\multiput(262.00,208.17)(4.000,1.000){2}{\rule{0.964pt}{0.400pt}}
\put(270,209.67){\rule{1.927pt}{0.400pt}}
\multiput(270.00,209.17)(4.000,1.000){2}{\rule{0.964pt}{0.400pt}}
\put(278,210.67){\rule{1.686pt}{0.400pt}}
\multiput(278.00,210.17)(3.500,1.000){2}{\rule{0.843pt}{0.400pt}}
\put(285,211.67){\rule{1.927pt}{0.400pt}}
\multiput(285.00,211.17)(4.000,1.000){2}{\rule{0.964pt}{0.400pt}}
\put(293,212.67){\rule{1.927pt}{0.400pt}}
\multiput(293.00,212.17)(4.000,1.000){2}{\rule{0.964pt}{0.400pt}}
\put(301,213.67){\rule{1.927pt}{0.400pt}}
\multiput(301.00,213.17)(4.000,1.000){2}{\rule{0.964pt}{0.400pt}}
\put(309,214.67){\rule{1.927pt}{0.400pt}}
\multiput(309.00,214.17)(4.000,1.000){2}{\rule{0.964pt}{0.400pt}}
\put(317,215.67){\rule{1.686pt}{0.400pt}}
\multiput(317.00,215.17)(3.500,1.000){2}{\rule{0.843pt}{0.400pt}}
\put(324,217.17){\rule{1.700pt}{0.400pt}}
\multiput(324.00,216.17)(4.472,2.000){2}{\rule{0.850pt}{0.400pt}}
\put(332,218.67){\rule{1.927pt}{0.400pt}}
\multiput(332.00,218.17)(4.000,1.000){2}{\rule{0.964pt}{0.400pt}}
\put(340,219.67){\rule{1.927pt}{0.400pt}}
\multiput(340.00,219.17)(4.000,1.000){2}{\rule{0.964pt}{0.400pt}}
\put(348,220.67){\rule{1.927pt}{0.400pt}}
\multiput(348.00,220.17)(4.000,1.000){2}{\rule{0.964pt}{0.400pt}}
\put(356,221.67){\rule{1.686pt}{0.400pt}}
\multiput(356.00,221.17)(3.500,1.000){2}{\rule{0.843pt}{0.400pt}}
\put(363,222.67){\rule{1.927pt}{0.400pt}}
\multiput(363.00,222.17)(4.000,1.000){2}{\rule{0.964pt}{0.400pt}}
\put(371,223.67){\rule{1.927pt}{0.400pt}}
\multiput(371.00,223.17)(4.000,1.000){2}{\rule{0.964pt}{0.400pt}}
\put(379,224.67){\rule{1.927pt}{0.400pt}}
\multiput(379.00,224.17)(4.000,1.000){2}{\rule{0.964pt}{0.400pt}}
\put(387,225.67){\rule{1.927pt}{0.400pt}}
\multiput(387.00,225.17)(4.000,1.000){2}{\rule{0.964pt}{0.400pt}}
\put(395,227.17){\rule{1.500pt}{0.400pt}}
\multiput(395.00,226.17)(3.887,2.000){2}{\rule{0.750pt}{0.400pt}}
\put(402,228.67){\rule{1.927pt}{0.400pt}}
\multiput(402.00,228.17)(4.000,1.000){2}{\rule{0.964pt}{0.400pt}}
\put(410,229.67){\rule{1.927pt}{0.400pt}}
\multiput(410.00,229.17)(4.000,1.000){2}{\rule{0.964pt}{0.400pt}}
\put(418,230.67){\rule{1.927pt}{0.400pt}}
\multiput(418.00,230.17)(4.000,1.000){2}{\rule{0.964pt}{0.400pt}}
\put(426,231.67){\rule{1.927pt}{0.400pt}}
\multiput(426.00,231.17)(4.000,1.000){2}{\rule{0.964pt}{0.400pt}}
\put(434,232.67){\rule{1.686pt}{0.400pt}}
\multiput(434.00,232.17)(3.500,1.000){2}{\rule{0.843pt}{0.400pt}}
\put(441,233.67){\rule{1.927pt}{0.400pt}}
\multiput(441.00,233.17)(4.000,1.000){2}{\rule{0.964pt}{0.400pt}}
\put(449,235.17){\rule{1.700pt}{0.400pt}}
\multiput(449.00,234.17)(4.472,2.000){2}{\rule{0.850pt}{0.400pt}}
\put(457,236.67){\rule{1.927pt}{0.400pt}}
\multiput(457.00,236.17)(4.000,1.000){2}{\rule{0.964pt}{0.400pt}}
\put(465,237.67){\rule{1.686pt}{0.400pt}}
\multiput(465.00,237.17)(3.500,1.000){2}{\rule{0.843pt}{0.400pt}}
\put(472,238.67){\rule{1.927pt}{0.400pt}}
\multiput(472.00,238.17)(4.000,1.000){2}{\rule{0.964pt}{0.400pt}}
\put(480,239.67){\rule{1.927pt}{0.400pt}}
\multiput(480.00,239.17)(4.000,1.000){2}{\rule{0.964pt}{0.400pt}}
\put(488,240.67){\rule{1.927pt}{0.400pt}}
\multiput(488.00,240.17)(4.000,1.000){2}{\rule{0.964pt}{0.400pt}}
\put(496,242.17){\rule{1.700pt}{0.400pt}}
\multiput(496.00,241.17)(4.472,2.000){2}{\rule{0.850pt}{0.400pt}}
\put(504,243.67){\rule{1.686pt}{0.400pt}}
\multiput(504.00,243.17)(3.500,1.000){2}{\rule{0.843pt}{0.400pt}}
\put(511,244.67){\rule{1.927pt}{0.400pt}}
\multiput(511.00,244.17)(4.000,1.000){2}{\rule{0.964pt}{0.400pt}}
\put(519,245.67){\rule{1.927pt}{0.400pt}}
\multiput(519.00,245.17)(4.000,1.000){2}{\rule{0.964pt}{0.400pt}}
\put(527,246.67){\rule{1.927pt}{0.400pt}}
\multiput(527.00,246.17)(4.000,1.000){2}{\rule{0.964pt}{0.400pt}}
\put(535,247.67){\rule{1.927pt}{0.400pt}}
\multiput(535.00,247.17)(4.000,1.000){2}{\rule{0.964pt}{0.400pt}}
\put(543,249.17){\rule{1.500pt}{0.400pt}}
\multiput(543.00,248.17)(3.887,2.000){2}{\rule{0.750pt}{0.400pt}}
\put(550,250.67){\rule{1.927pt}{0.400pt}}
\multiput(550.00,250.17)(4.000,1.000){2}{\rule{0.964pt}{0.400pt}}
\put(558,251.67){\rule{1.927pt}{0.400pt}}
\multiput(558.00,251.17)(4.000,1.000){2}{\rule{0.964pt}{0.400pt}}
\put(566,252.67){\rule{1.927pt}{0.400pt}}
\multiput(566.00,252.17)(4.000,1.000){2}{\rule{0.964pt}{0.400pt}}
\put(574,254.17){\rule{1.700pt}{0.400pt}}
\multiput(574.00,253.17)(4.472,2.000){2}{\rule{0.850pt}{0.400pt}}
\put(582,255.67){\rule{1.686pt}{0.400pt}}
\multiput(582.00,255.17)(3.500,1.000){2}{\rule{0.843pt}{0.400pt}}
\put(589,256.67){\rule{1.927pt}{0.400pt}}
\multiput(589.00,256.17)(4.000,1.000){2}{\rule{0.964pt}{0.400pt}}
\put(597,257.67){\rule{1.927pt}{0.400pt}}
\multiput(597.00,257.17)(4.000,1.000){2}{\rule{0.964pt}{0.400pt}}
\put(605,258.67){\rule{1.927pt}{0.400pt}}
\multiput(605.00,258.17)(4.000,1.000){2}{\rule{0.964pt}{0.400pt}}
\put(613,260.17){\rule{1.500pt}{0.400pt}}
\multiput(613.00,259.17)(3.887,2.000){2}{\rule{0.750pt}{0.400pt}}
\put(620,261.67){\rule{1.927pt}{0.400pt}}
\multiput(620.00,261.17)(4.000,1.000){2}{\rule{0.964pt}{0.400pt}}
\put(628,262.67){\rule{1.927pt}{0.400pt}}
\multiput(628.00,262.17)(4.000,1.000){2}{\rule{0.964pt}{0.400pt}}
\put(636,264.17){\rule{1.700pt}{0.400pt}}
\multiput(636.00,263.17)(4.472,2.000){2}{\rule{0.850pt}{0.400pt}}
\put(644,265.67){\rule{1.686pt}{0.400pt}}
\multiput(644.00,265.17)(3.500,1.000){2}{\rule{0.843pt}{0.400pt}}
\put(651,267.17){\rule{1.700pt}{0.400pt}}
\multiput(651.00,266.17)(4.472,2.000){2}{\rule{0.850pt}{0.400pt}}
\put(659,268.67){\rule{1.927pt}{0.400pt}}
\multiput(659.00,268.17)(4.000,1.000){2}{\rule{0.964pt}{0.400pt}}
\put(667,270.17){\rule{1.700pt}{0.400pt}}
\multiput(667.00,269.17)(4.472,2.000){2}{\rule{0.850pt}{0.400pt}}
\put(675,271.67){\rule{1.686pt}{0.400pt}}
\multiput(675.00,271.17)(3.500,1.000){2}{\rule{0.843pt}{0.400pt}}
\put(682,273.17){\rule{1.700pt}{0.400pt}}
\multiput(682.00,272.17)(4.472,2.000){2}{\rule{0.850pt}{0.400pt}}
\put(690,275.17){\rule{1.700pt}{0.400pt}}
\multiput(690.00,274.17)(4.472,2.000){2}{\rule{0.850pt}{0.400pt}}
\put(698,277.17){\rule{1.700pt}{0.400pt}}
\multiput(698.00,276.17)(4.472,2.000){2}{\rule{0.850pt}{0.400pt}}
\put(706,279.17){\rule{1.700pt}{0.400pt}}
\multiput(706.00,278.17)(4.472,2.000){2}{\rule{0.850pt}{0.400pt}}
\put(714,281.17){\rule{1.500pt}{0.400pt}}
\multiput(714.00,280.17)(3.887,2.000){2}{\rule{0.750pt}{0.400pt}}
\multiput(721.00,283.61)(1.579,0.447){3}{\rule{1.167pt}{0.108pt}}
\multiput(721.00,282.17)(5.579,3.000){2}{\rule{0.583pt}{0.400pt}}
\put(729,286.17){\rule{1.700pt}{0.400pt}}
\multiput(729.00,285.17)(4.472,2.000){2}{\rule{0.850pt}{0.400pt}}
\multiput(737.00,288.61)(1.579,0.447){3}{\rule{1.167pt}{0.108pt}}
\multiput(737.00,287.17)(5.579,3.000){2}{\rule{0.583pt}{0.400pt}}
\multiput(745.00,291.61)(1.355,0.447){3}{\rule{1.033pt}{0.108pt}}
\multiput(745.00,290.17)(4.855,3.000){2}{\rule{0.517pt}{0.400pt}}
\multiput(752.00,294.61)(1.579,0.447){3}{\rule{1.167pt}{0.108pt}}
\multiput(752.00,293.17)(5.579,3.000){2}{\rule{0.583pt}{0.400pt}}
\multiput(760.00,297.61)(1.579,0.447){3}{\rule{1.167pt}{0.108pt}}
\multiput(760.00,296.17)(5.579,3.000){2}{\rule{0.583pt}{0.400pt}}
\multiput(768.00,300.60)(1.066,0.468){5}{\rule{0.900pt}{0.113pt}}
\multiput(768.00,299.17)(6.132,4.000){2}{\rule{0.450pt}{0.400pt}}
\multiput(776.00,304.61)(1.355,0.447){3}{\rule{1.033pt}{0.108pt}}
\multiput(776.00,303.17)(4.855,3.000){2}{\rule{0.517pt}{0.400pt}}
\multiput(783.00,307.60)(1.066,0.468){5}{\rule{0.900pt}{0.113pt}}
\multiput(783.00,306.17)(6.132,4.000){2}{\rule{0.450pt}{0.400pt}}
\multiput(791.00,311.60)(1.066,0.468){5}{\rule{0.900pt}{0.113pt}}
\multiput(791.00,310.17)(6.132,4.000){2}{\rule{0.450pt}{0.400pt}}
\multiput(799.00,315.60)(1.066,0.468){5}{\rule{0.900pt}{0.113pt}}
\multiput(799.00,314.17)(6.132,4.000){2}{\rule{0.450pt}{0.400pt}}
\multiput(807.00,319.59)(0.710,0.477){7}{\rule{0.660pt}{0.115pt}}
\multiput(807.00,318.17)(5.630,5.000){2}{\rule{0.330pt}{0.400pt}}
\multiput(814.00,324.59)(0.821,0.477){7}{\rule{0.740pt}{0.115pt}}
\multiput(814.00,323.17)(6.464,5.000){2}{\rule{0.370pt}{0.400pt}}
\multiput(822.00,329.59)(0.821,0.477){7}{\rule{0.740pt}{0.115pt}}
\multiput(822.00,328.17)(6.464,5.000){2}{\rule{0.370pt}{0.400pt}}
\multiput(830.00,334.59)(0.821,0.477){7}{\rule{0.740pt}{0.115pt}}
\multiput(830.00,333.17)(6.464,5.000){2}{\rule{0.370pt}{0.400pt}}
\multiput(838.00,339.59)(0.581,0.482){9}{\rule{0.567pt}{0.116pt}}
\multiput(838.00,338.17)(5.824,6.000){2}{\rule{0.283pt}{0.400pt}}
\multiput(845.00,345.59)(0.821,0.477){7}{\rule{0.740pt}{0.115pt}}
\multiput(845.00,344.17)(6.464,5.000){2}{\rule{0.370pt}{0.400pt}}
\multiput(853.00,350.59)(0.569,0.485){11}{\rule{0.557pt}{0.117pt}}
\multiput(853.00,349.17)(6.844,7.000){2}{\rule{0.279pt}{0.400pt}}
\multiput(861.00,357.59)(0.671,0.482){9}{\rule{0.633pt}{0.116pt}}
\multiput(861.00,356.17)(6.685,6.000){2}{\rule{0.317pt}{0.400pt}}
\multiput(869.00,363.59)(0.492,0.485){11}{\rule{0.500pt}{0.117pt}}
\multiput(869.00,362.17)(5.962,7.000){2}{\rule{0.250pt}{0.400pt}}
\multiput(876.00,370.59)(0.569,0.485){11}{\rule{0.557pt}{0.117pt}}
\multiput(876.00,369.17)(6.844,7.000){2}{\rule{0.279pt}{0.400pt}}
\multiput(884.00,377.59)(0.569,0.485){11}{\rule{0.557pt}{0.117pt}}
\multiput(884.00,376.17)(6.844,7.000){2}{\rule{0.279pt}{0.400pt}}
\multiput(892.00,384.59)(0.494,0.488){13}{\rule{0.500pt}{0.117pt}}
\multiput(892.00,383.17)(6.962,8.000){2}{\rule{0.250pt}{0.400pt}}
\multiput(900.59,392.00)(0.485,0.569){11}{\rule{0.117pt}{0.557pt}}
\multiput(899.17,392.00)(7.000,6.844){2}{\rule{0.400pt}{0.279pt}}
\multiput(907.00,400.59)(0.494,0.488){13}{\rule{0.500pt}{0.117pt}}
\multiput(907.00,399.17)(6.962,8.000){2}{\rule{0.250pt}{0.400pt}}
\multiput(915.59,408.00)(0.488,0.560){13}{\rule{0.117pt}{0.550pt}}
\multiput(914.17,408.00)(8.000,7.858){2}{\rule{0.400pt}{0.275pt}}
\multiput(923.59,417.00)(0.488,0.560){13}{\rule{0.117pt}{0.550pt}}
\multiput(922.17,417.00)(8.000,7.858){2}{\rule{0.400pt}{0.275pt}}
\multiput(931.59,426.00)(0.485,0.721){11}{\rule{0.117pt}{0.671pt}}
\multiput(930.17,426.00)(7.000,8.606){2}{\rule{0.400pt}{0.336pt}}
\multiput(938.59,436.00)(0.488,0.626){13}{\rule{0.117pt}{0.600pt}}
\multiput(937.17,436.00)(8.000,8.755){2}{\rule{0.400pt}{0.300pt}}
\multiput(946.58,446.00)(0.491,0.652){17}{\rule{0.118pt}{0.620pt}}
\multiput(945.17,446.00)(10.000,11.713){2}{\rule{0.400pt}{0.310pt}}
\multiput(956.59,459.00)(0.489,0.786){15}{\rule{0.118pt}{0.722pt}}
\multiput(955.17,459.00)(9.000,12.501){2}{\rule{0.400pt}{0.361pt}}
\multiput(965.58,473.00)(0.491,0.808){17}{\rule{0.118pt}{0.740pt}}
\multiput(964.17,473.00)(10.000,14.464){2}{\rule{0.400pt}{0.370pt}}
\multiput(975.59,489.00)(0.485,1.103){11}{\rule{0.117pt}{0.957pt}}
\multiput(974.17,489.00)(7.000,13.013){2}{\rule{0.400pt}{0.479pt}}
\multiput(982.59,504.00)(0.488,1.088){13}{\rule{0.117pt}{0.950pt}}
\multiput(981.17,504.00)(8.000,15.028){2}{\rule{0.400pt}{0.475pt}}
\multiput(990.59,521.00)(0.488,1.220){13}{\rule{0.117pt}{1.050pt}}
\multiput(989.17,521.00)(8.000,16.821){2}{\rule{0.400pt}{0.525pt}}
\multiput(998.59,540.00)(0.488,1.220){13}{\rule{0.117pt}{1.050pt}}
\multiput(997.17,540.00)(8.000,16.821){2}{\rule{0.400pt}{0.525pt}}
\multiput(1006.59,559.00)(0.488,1.286){13}{\rule{0.117pt}{1.100pt}}
\multiput(1005.17,559.00)(8.000,17.717){2}{\rule{0.400pt}{0.550pt}}
\multiput(1014.59,579.00)(0.488,1.352){13}{\rule{0.117pt}{1.150pt}}
\multiput(1013.17,579.00)(8.000,18.613){2}{\rule{0.400pt}{0.575pt}}
\multiput(1022.59,600.00)(0.485,1.560){11}{\rule{0.117pt}{1.300pt}}
\multiput(1021.17,600.00)(7.000,18.302){2}{\rule{0.400pt}{0.650pt}}
\sbox{\plotpoint}{\rule[-0.500pt]{1.000pt}{1.000pt}}%
\put(518,598){\makebox(0,0)[r]{old P.Th.}}
\multiput(540,598)(20.756,0.000){4}{\usebox{\plotpoint}}
\put(606,598){\usebox{\plotpoint}}
\put(254,199){\usebox{\plotpoint}}
\put(254.00,199.00){\usebox{\plotpoint}}
\multiput(262,200)(20.595,2.574){0}{\usebox{\plotpoint}}
\put(274.63,201.00){\usebox{\plotpoint}}
\multiput(278,201)(20.547,2.935){0}{\usebox{\plotpoint}}
\multiput(285,202)(20.595,2.574){0}{\usebox{\plotpoint}}
\put(295.25,203.00){\usebox{\plotpoint}}
\multiput(301,203)(20.595,2.574){0}{\usebox{\plotpoint}}
\put(315.89,204.86){\usebox{\plotpoint}}
\multiput(317,205)(20.756,0.000){0}{\usebox{\plotpoint}}
\multiput(324,205)(20.595,2.574){0}{\usebox{\plotpoint}}
\put(336.54,206.57){\usebox{\plotpoint}}
\multiput(340,207)(20.756,0.000){0}{\usebox{\plotpoint}}
\multiput(348,207)(20.595,2.574){0}{\usebox{\plotpoint}}
\put(357.20,208.17){\usebox{\plotpoint}}
\multiput(363,209)(20.756,0.000){0}{\usebox{\plotpoint}}
\put(377.84,209.85){\usebox{\plotpoint}}
\multiput(379,210)(20.595,2.574){0}{\usebox{\plotpoint}}
\multiput(387,211)(20.595,2.574){0}{\usebox{\plotpoint}}
\put(398.46,212.00){\usebox{\plotpoint}}
\multiput(402,212)(20.595,2.574){0}{\usebox{\plotpoint}}
\multiput(410,213)(20.595,2.574){0}{\usebox{\plotpoint}}
\put(419.08,214.14){\usebox{\plotpoint}}
\multiput(426,215)(20.756,0.000){0}{\usebox{\plotpoint}}
\put(439.73,215.82){\usebox{\plotpoint}}
\multiput(441,216)(20.595,2.574){0}{\usebox{\plotpoint}}
\multiput(449,217)(20.595,2.574){0}{\usebox{\plotpoint}}
\put(460.32,218.41){\usebox{\plotpoint}}
\multiput(465,219)(20.756,0.000){0}{\usebox{\plotpoint}}
\multiput(472,219)(20.595,2.574){0}{\usebox{\plotpoint}}
\put(480.97,220.12){\usebox{\plotpoint}}
\multiput(488,221)(20.595,2.574){0}{\usebox{\plotpoint}}
\put(501.56,222.70){\usebox{\plotpoint}}
\multiput(504,223)(20.547,2.935){0}{\usebox{\plotpoint}}
\multiput(511,224)(20.595,2.574){0}{\usebox{\plotpoint}}
\put(522.14,225.39){\usebox{\plotpoint}}
\multiput(527,226)(20.595,2.574){0}{\usebox{\plotpoint}}
\put(542.80,227.00){\usebox{\plotpoint}}
\multiput(543,227)(20.547,2.935){0}{\usebox{\plotpoint}}
\multiput(550,228)(20.595,2.574){0}{\usebox{\plotpoint}}
\put(563.26,230.31){\usebox{\plotpoint}}
\multiput(566,231)(20.595,2.574){0}{\usebox{\plotpoint}}
\multiput(574,232)(20.595,2.574){0}{\usebox{\plotpoint}}
\put(583.79,233.26){\usebox{\plotpoint}}
\multiput(589,234)(20.595,2.574){0}{\usebox{\plotpoint}}
\put(604.37,235.92){\usebox{\plotpoint}}
\multiput(605,236)(20.595,2.574){0}{\usebox{\plotpoint}}
\multiput(613,237)(20.547,2.935){0}{\usebox{\plotpoint}}
\put(624.95,238.62){\usebox{\plotpoint}}
\multiput(628,239)(20.136,5.034){0}{\usebox{\plotpoint}}
\multiput(636,241)(20.595,2.574){0}{\usebox{\plotpoint}}
\put(645.36,242.19){\usebox{\plotpoint}}
\multiput(651,243)(20.595,2.574){0}{\usebox{\plotpoint}}
\put(665.79,245.70){\usebox{\plotpoint}}
\multiput(667,246)(20.595,2.574){0}{\usebox{\plotpoint}}
\multiput(675,247)(20.547,2.935){0}{\usebox{\plotpoint}}
\put(686.24,249.06){\usebox{\plotpoint}}
\multiput(690,250)(20.595,2.574){0}{\usebox{\plotpoint}}
\multiput(698,251)(20.136,5.034){0}{\usebox{\plotpoint}}
\put(706.57,253.07){\usebox{\plotpoint}}
\multiput(714,254)(20.547,2.935){0}{\usebox{\plotpoint}}
\put(727.01,256.50){\usebox{\plotpoint}}
\multiput(729,257)(20.136,5.034){0}{\usebox{\plotpoint}}
\multiput(737,259)(20.595,2.574){0}{\usebox{\plotpoint}}
\put(747.30,260.66){\usebox{\plotpoint}}
\multiput(752,262)(20.595,2.574){0}{\usebox{\plotpoint}}
\put(767.57,264.89){\usebox{\plotpoint}}
\multiput(768,265)(20.136,5.034){0}{\usebox{\plotpoint}}
\multiput(776,267)(20.547,2.935){0}{\usebox{\plotpoint}}
\put(787.85,269.21){\usebox{\plotpoint}}
\multiput(791,270)(20.136,5.034){0}{\usebox{\plotpoint}}
\multiput(799,272)(20.136,5.034){0}{\usebox{\plotpoint}}
\put(808.01,274.14){\usebox{\plotpoint}}
\multiput(814,275)(20.136,5.034){0}{\usebox{\plotpoint}}
\put(828.26,278.57){\usebox{\plotpoint}}
\multiput(830,279)(20.136,5.034){0}{\usebox{\plotpoint}}
\multiput(838,281)(19.957,5.702){0}{\usebox{\plotpoint}}
\put(848.33,283.83){\usebox{\plotpoint}}
\multiput(853,285)(20.136,5.034){0}{\usebox{\plotpoint}}
\put(868.47,288.87){\usebox{\plotpoint}}
\multiput(869,289)(19.957,5.702){0}{\usebox{\plotpoint}}
\multiput(876,291)(20.136,5.034){0}{\usebox{\plotpoint}}
\put(888.54,294.14){\usebox{\plotpoint}}
\multiput(892,295)(19.434,7.288){0}{\usebox{\plotpoint}}
\multiput(900,298)(19.957,5.702){0}{\usebox{\plotpoint}}
\put(908.33,300.33){\usebox{\plotpoint}}
\multiput(915,302)(20.136,5.034){0}{\usebox{\plotpoint}}
\put(928.27,305.98){\usebox{\plotpoint}}
\multiput(931,307)(19.957,5.702){0}{\usebox{\plotpoint}}
\multiput(938,309)(20.136,5.034){0}{\usebox{\plotpoint}}
\put(948.22,311.67){\usebox{\plotpoint}}
\multiput(956,314)(19.690,6.563){0}{\usebox{\plotpoint}}
\put(968.01,317.90){\usebox{\plotpoint}}
\multiput(975,320)(19.957,5.702){0}{\usebox{\plotpoint}}
\put(987.79,324.17){\usebox{\plotpoint}}
\multiput(990,325)(20.136,5.034){0}{\usebox{\plotpoint}}
\multiput(998,327)(20.136,5.034){0}{\usebox{\plotpoint}}
\put(1007.84,329.46){\usebox{\plotpoint}}
\multiput(1014,331)(20.136,5.034){0}{\usebox{\plotpoint}}
\put(1027.93,334.69){\usebox{\plotpoint}}
\put(1029,335){\usebox{\plotpoint}}
\end{picture}
\end{center}
\caption{The $\chi_{22}$ Creutz ratio of Wilson loops versus loop
size. Results from Monte Carlo simulations (exact), and from
tadpole-improved (new) and traditional (old) lattice perturbation theory are
shown.}
\label{creutz-fig}
\end{figure}

The problem with traditional lattice-QCD perturbation theory is that the
coupling it uses is much too small. The standard practice was to express
perturbative expansions of short-distance lattice quantities in terms of
the bare coupling~$\alpha_{\rm lat}$ used in the lattice lagrangian. This
practice followed from the notion that the bare coupling in a cutoff theory
is approximately equal to the running coupling evaluated at the cutoff
scale, here~$\alpha_s(\pi/a)$, and therefore that it is the appropriate
coupling for quantities dominated by momenta near the cutoff. In fact the
bare coupling in traditional lattice QCD is much smaller than true
effective coupling at large lattice spacings: for example,
\bearray
\alpha_{\rm lat} &=& \alpha_V(\pi/a) - 4.7\,\alpha_V^2 + \cdots \\
&\le& \half\alpha_V(\pi/a)\qquad\mbox{for $a>.1$\,fm}
\eearray
where $\alpha_V(q)$ is a continuum coupling defined by the static-quark
potential,
\be
V_{\rm Q\overline{Q}}(q) \equiv -4\,\pi\,C_{\rm F} \frac{\alpha_V(q)}{q^2}.
\ee
Consequently $\alpha_{\rm lat}$~expansions, though formally correct, badly
underestimate perturbative effects, and converge poorly.

The anomalously small bare coupling in the traditional lattice theory is a
symptom of the ``tadpole problem''.
As we discuss later, all gluonic operators in lattice QCD
are built from the link operator
\be
U_\mu(x) \equiv \P \e^{-\I \int_x^{x+a\hat\mu}gA\cdot \dd x}
\approx \e^{-\I agA_\mu}
\ee
rather than from the vector potential~$A_\mu$. Thus, for example, the
leading term in the lagrangian that couples quarks and gluons
is~$\psib U_\mu\gamma_\mu\psi/a$.
Such a term contains the usual $\psib gA\cdot\gamma\psi$~vertex, but, in
addition, it contains vertices with any number of additional powers of
$agA_\mu$. These extra vertices are irrelevant for classical fields since
they are suppressed by powers of the lattice spacing. For quantum fields,
however, the situation is quite different since pairs of~$A_\mu$'s, if
contracted with each other, generate ultraviolet divergent factors
of~$1/a^2$ that precisely cancel the extra~$a$'s. Consequently the
contributions generated by the extra vertices are suppressed by powers
of~$g^2$ (not $a$), and turn out to be uncomfortably large. These are the
tadpole contributions.

The tadpoles  result in
large renormalizations\,---\,often as large as a factor of two or
three\,---\,that spoil naive perturbation theory, and with it our
intuition about the connection between lattice operators and the continuum.
However tadpole contributions are generically process independent and so it
is possible to measure their contribution in one quantity and then correct
for them in all other quantities.

The simplest way to do this is to cancel
them out. The mean value~$u_0$ of~$\third\Re\Tr U_\mu$ consists of only
tadpoles and so we can largely cancel the tadpole contributions by dividing
every link operator by~$u_0$. That is, in every lattice operator we replace
\be
U_\mu(x) \to \frac{U_\mu(x)}{u_0}
\ee
where~$u_0$ is computed numerically in a simulation.

The $u_0$'s cancel tadpole
contributions, making lattice operators and perturbation theory far more
continuum-like in their behavior. Thus, for example, the only change
in the standard gluon action when it is tadpole-improved is that the new
bare coupling~$\alpha_{\rm TI}$ is enhanced by a factor of $1/u_0^4$
relative to the coupling~$\alpha_{\rm lat}$ in the unimproved theory:
\be
\alpha_{\rm TI} = \frac{\alpha_{\rm lat}}{u_0^4}.
\ee
Since $u_0^4\!<\!.6$ when $a\!>\!.1$\,fm, the
tadpole-improved coupling is typically more than twice as large for coarse
lattices. Expressing~$\alpha_{\rm TI}$ in terms of the continuum
coupling~$\alpha_V$, we find that now our intuition is satisfied:
\bearray
{\alpha}_{\rm TI} &=& \alpha_V(\pi/a) - .5\,\alpha_V^2 +\cdots \\
&\approx& \alpha_V(\pi/a).
\eearray

Perturbation theory for the Creutz ratio \eq{creutz-ratio}
converges rapidly to the correct answer when it is reexpressed in terms
of~$\alpha_{\rm TI}$. An even better result is obtained if the expansion is
reexpressed as a series in a coupling constant defined in terms of a
physical quantity, like the static-quark potential, where that coupling
constant is measured in a simulation.  By measuring the coupling we
automatically include any large renormalizations of the coupling due to
tadpoles. It is important that the scale~$q^*$ at which the running
coupling constant is evaluated be chosen appropriately for the quantity
being studied\,\cite{gpl93,gpl94}.
When these refinements are added, perturbation theory is
dramatically improved, and, as illustrated in Fig.~\ref{creutz-fig}, is
still quite accurate for loops as large as $1/2$\,fm.

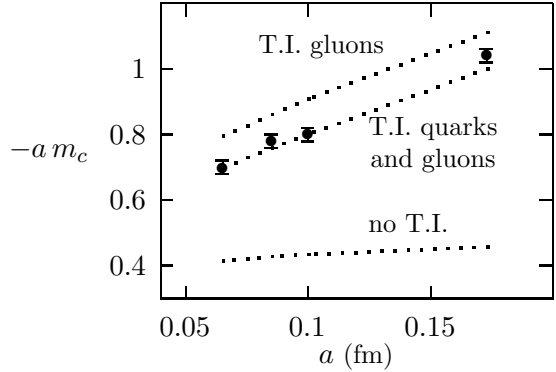
\begin{figure}
\begin{center}
\setlength{\unitlength}{0.240900pt}
\ifx\plotpoint\undefined\newsavebox{\plotpoint}\fi
\sbox{\plotpoint}{\rule[-0.200pt]{0.400pt}{0.400pt}}%
\begin{picture}(900,600)(0,0)
\font\gnuplot=cmr10 at 10pt
\gnuplot
\sbox{\plotpoint}{\rule[-0.200pt]{0.400pt}{0.400pt}}%
\put(220.0,165.0){\rule[-0.200pt]{4.818pt}{0.400pt}}
\put(198,165){\makebox(0,0)[r]{$0.4$}}
\put(816.0,165.0){\rule[-0.200pt]{4.818pt}{0.400pt}}
\put(220.0,268.0){\rule[-0.200pt]{4.818pt}{0.400pt}}
\put(198,268){\makebox(0,0)[r]{$0.6$}}
\put(816.0,268.0){\rule[-0.200pt]{4.818pt}{0.400pt}}
\put(220.0,371.0){\rule[-0.200pt]{4.818pt}{0.400pt}}
\put(198,371){\makebox(0,0)[r]{$0.8$}}
\put(816.0,371.0){\rule[-0.200pt]{4.818pt}{0.400pt}}
\put(220.0,474.0){\rule[-0.200pt]{4.818pt}{0.400pt}}
\put(198,474){\makebox(0,0)[r]{$1$}}
\put(816.0,474.0){\rule[-0.200pt]{4.818pt}{0.400pt}}
\put(259.0,113.0){\rule[-0.200pt]{0.400pt}{4.818pt}}
\put(259,68){\makebox(0,0){$0.05$}}
\put(259.0,557.0){\rule[-0.200pt]{0.400pt}{4.818pt}}
\put(451.0,113.0){\rule[-0.200pt]{0.400pt}{4.818pt}}
\put(451,68){\makebox(0,0){$0.1$}}
\put(451.0,557.0){\rule[-0.200pt]{0.400pt}{4.818pt}}
\put(644.0,113.0){\rule[-0.200pt]{0.400pt}{4.818pt}}
\put(644,68){\makebox(0,0){$0.15$}}
\put(644.0,557.0){\rule[-0.200pt]{0.400pt}{4.818pt}}
\put(220.0,113.0){\rule[-0.200pt]{148.394pt}{0.400pt}}
\put(836.0,113.0){\rule[-0.200pt]{0.400pt}{111.778pt}}
\put(220.0,577.0){\rule[-0.200pt]{148.394pt}{0.400pt}}
\put(45,345){\makebox(0,0){$-a\,m_c$}}
\put(528,23){\makebox(0,0){$a$ (fm)}}
\put(547,232){\makebox(0,0)[l]{no T.I.}}
\put(374,510){\makebox(0,0)[l]{T.I. gluons}}
\put(547,355){\makebox(0,0)[l]{\shortstack{T.I. quarks\\and gluons}}}
\put(220.0,113.0){\rule[-0.200pt]{0.400pt}{111.778pt}}
\put(732,495){\circle*{18}}
\put(451,371){\circle*{18}}
\put(394,360){\circle*{18}}
\put(317,319){\circle*{18}}
\put(732.0,484.0){\rule[-0.200pt]{0.400pt}{5.059pt}}
\put(722.0,484.0){\rule[-0.200pt]{4.818pt}{0.400pt}}
\put(722.0,505.0){\rule[-0.200pt]{4.818pt}{0.400pt}}
\put(451.0,360.0){\rule[-0.200pt]{0.400pt}{5.059pt}}
\put(441.0,360.0){\rule[-0.200pt]{4.818pt}{0.400pt}}
\put(441.0,381.0){\rule[-0.200pt]{4.818pt}{0.400pt}}
\put(394.0,350.0){\rule[-0.200pt]{0.400pt}{5.059pt}}
\put(384.0,350.0){\rule[-0.200pt]{4.818pt}{0.400pt}}
\put(384.0,371.0){\rule[-0.200pt]{4.818pt}{0.400pt}}
\put(317.0,309.0){\rule[-0.200pt]{0.400pt}{5.059pt}}
\put(307.0,309.0){\rule[-0.200pt]{4.818pt}{0.400pt}}
\put(307.0,330.0){\rule[-0.200pt]{4.818pt}{0.400pt}}
\sbox{\plotpoint}{\rule[-0.500pt]{1.000pt}{1.000pt}}%
\put(732,194){\usebox{\plotpoint}}
\multiput(732,194)(-20.737,-0.886){14}{\usebox{\plotpoint}}
\multiput(451,182)(-20.727,-1.091){3}{\usebox{\plotpoint}}
\multiput(394,179)(-20.647,-2.118){2}{\usebox{\plotpoint}}
\multiput(355,175)(-20.691,-1.634){2}{\usebox{\plotpoint}}
\put(317,172){\usebox{\plotpoint}}
\put(732,531){\usebox{\plotpoint}}
\multiput(732,531)(-19.465,-7.204){15}{\usebox{\plotpoint}}
\multiput(451,427)(-19.008,-8.337){3}{\usebox{\plotpoint}}
\multiput(394,402)(-18.845,-8.698){2}{\usebox{\plotpoint}}
\multiput(355,384)(-19.129,-8.054){2}{\usebox{\plotpoint}}
\put(317,368){\usebox{\plotpoint}}
\put(732,474){\usebox{\plotpoint}}
\multiput(732,474)(-19.488,-7.143){15}{\usebox{\plotpoint}}
\multiput(451,371)(-19.476,-7.175){3}{\usebox{\plotpoint}}
\multiput(394,350)(-19.254,-7.751){4}{\usebox{\plotpoint}}
\put(317,319){\usebox{\plotpoint}}
\end{picture}
\end{center}
\caption{The critical bare quark mass for Wilson's lattice quark action
versus lattice spacing. Monte Carlo data points are compared with
perturbation theories in a theory with no tadpole improvement (T.I),
tadpole-improved gluon dynamics, and tadpole-improved quark and gluon
dynamics.}
\label{mc-fig}
\end{figure}

This same conclusion follows from Fig.~\ref{mc-fig} which shows
the value of the bare quark mass needed to obtain zero-mass pions using
Wilson's lattice action for quarks. This quantity diverges linearly
as the lattice spacing vanishes, and so should be quite perturbative. Here
we see dramatic improvements as the tadpoles are removed first from the
gluon action, through use of an improved coupling, and then also from the
quark action.

The Creutz ratio and the critical quark mass are both very
similar to the couplings we need to compute for improved lagrangians.
Tadpole improvement has been very successful in a wide range of
applications.

\begin{exercise} Derive the Feynman rules for lattice $\phi^4$ theory
where
\be
S = \sum_xa^4\,\left\{\frac{-\phi\lder2\phi + m^2\phi^2}{2} +
\frac{\lambda\,\phi^4}{4!}\right\}.
\ee
In particular show that the $\phi$ propagator is
\be
G\propto\frac{a^2}{\sum_\mu\left(2\sin(ap_\mu/2)\right)^2+a^2 m^2}
\ee
\end{exercise}

\subsection{An Aside on ``Perfect Actions''}
Although not  necessary in practice, it is possible to sum all  the
terms in the $a^2$~expansion of a continuum derivative using Fourier
transforms: for example, we replace
\be
\partial^2_x\psi(x_i) \to -\sum_j d^{(2)}_{\rm pf}(x_i\!-\!x_j)\,\psi(x_j)
\ee
where
\be \label{d2slac}
d^{(2)}_{\rm pf} (x_i\!-\!x_j) \equiv
\int\frac{dp}{2\pi}\,p^2\,\e^{\I p(x_i-x_j)}\,\theta(|p|\!<\!\pi/a).
\ee
We do not do this because $d^{(2)}_{\rm pf}(x_i\!-\!x_j)$ falls off only as
$1/|x_i\!-\!x_j|^2$ for large~$|x_i\!-\!x_j|$, resulting in highly nonlocal
actions and operators that are very costly to simulate, particularly when
gauge fields are involved. So  it is generally far
better to truncate the $a^2$~expansion than to sum to all orders. However
this analysis suggests a different strategy for discretization that we now
examine.

The slow fall-off in the all-orders or perfect derivative~$d^{(2)}_{\rm
pf}$ is caused by the abruptness of the lattice
cutoff (the $\theta$~funtion in \eq{d2slac}).
Wilson noticed that
$d^{(2)}$ can be made to vanish faster than any power of
$1/|x_i-x_j|$ if a smoother cutoff is introduced.
In his approach the lattice field at node~$x_i$ is
{\em not\/} equal to the continuum field there, but rather equals the
continuum field averaged over a small volume centered at that node: for
example, the lattice field at site~$x_i$ in one dimension
is\footnote{The smearing discussed here is conceptually
simple, but better smearings, resulting in improved locality, are possible.
In \cite{hasenfratz94} the authors use  stochastic smearing in which $\Psi_i$
equals the blocked continuum field only on average. This introduces a new
parameter that can be tuned to optimize the action.}
\bearray
\Psi_i &=& \frac{1}{a}\,\int_{x_i-a/2}^{x_i+a/2}\psi(x)\,\dd x\\
&=& \int_{-\infty}^{\infty}\frac{dp}{2\pi}\,\e^{\I p x_i}\,\Pi(p)\,\psi(p)
\eearray
where, in the transform, smearing function
\be
\Pi(p) \equiv \frac{2\sin(p a/2)}{p a}
\ee
provides a smooth cutoff at large~$p$.
The lattice action for the smeared fields is
designed so that tree-level Green's
functions built from these fields are {\em exactly\/}
equal to the corresponding Green's functions in the continuum theory, with
the smearing function applied at the ends: for example,
\bearray
\langle \Psi_i\,\Psi_j \rangle
&=& \int_{x_i-a/2}^{x_i+a/2}\!\!\!\!\!\dd x \int_{x_j-a/2}^{x_j+a/2}
\!\!\!\!\!\dd y\,\,\langle\psi(x)\,\psi(y)\rangle \\
&=& \int_{-\infty}^{\infty}\frac{\dd p}{2\pi}\,\e^{\I p
(x_i-x_j)}\,\frac{\Pi^2(p)}{p^2+m^2}.
\eearray
By rewriting the propagator $\langle\Psi_i\,\Psi_j\rangle$ in the form
\be
\int_{-\pi/a}^{\pi/a}\frac{\dd p}{2\pi}\,\e^{\I p(x_i-x_j)}\,\sum_n
\frac{ \Pi^2(p\!+\!2\pi n/a)}{(p\!+\!2\pi n/a)^2+m^2},
\ee
we find that the quadratic terms of this lattice action are
\be
S^{(2)} = \mbox{$\frac{1}{2}$} \sum_{i,j} \Psi_i\,d^{(2)}_{\rm
pa}(x_i\!-\!x_j,m)\,\Psi_j,
\ee
where $d^{(2)}_{\rm pa}(x_i\!-\!x_j,m)$ is
\be
\int_{-\pi/a}^{\pi/a}\!\frac{\dd
p}{2\pi}\,\e^{\I p(x_i-x_j)}\!\!\left(\!\sum_n
\frac{ \Pi^2(p\!+\!2\pi n/a)}{(p\!+\!2\pi
n/a)^2\!+\!m^2}\!\right)^{\!-1}\!\!\!\!\!\!\!.
\ee
It is easily shown that, remarkably, $d^{(2)}_{\rm pa}(x_i\!-\!x_j,m)$
vanishes faster than any power of $1/|x_i\!-\!x_j|$ as the separation
increases. So the derivative terms in such an action are both exact to all
orders in $a$ and quite local\,---\,that is, the classical action is
``perfect.'' Again, this is possible
because the lattice field~$\Psi_i$ is obtained
by smearing the continuum field, which introduces a smooth UV~cutoff in
momentum space rather than the abrupt cutoff of~\eq{d2slac}.

The complication with this approach is that it is difficult to reconstruct
the continuum fields~$\psi(x)$ from the smeared lattice fields~$\Psi_i$.
Consequently nonlinear interactions, currents, and other operators are
generally complicated functions of the~$\Psi_i$, and much more difficult to
design than with the previous approach, where the
lattice field is trivially related to the continuum field. This is an
important issue in QCD where phenomenological studies involve a wide range
of currents and operators. It also means that in practice the ``perfect
action'' used for an interacting theory is not really  perfect
since the arbitrarily complex functions of~$\Psi_i$ that arise in real perfect
actions must be simplified for simulations. Although these difficulties
gradually are being overcome for QCD\,\cite{degrand96,wiese96},
I will not discuss this interesting approach further in these lectures.

\subsection{Summary}
Asymptotic freedom implies that short-distance QCD is simple
(perturbative) while long-distance QCD is difficult (nonperturbative).
The lattice separates short from long distances, allowing us to exploit
this dichotomy to create highly efficient algorithms for solving the
entire theory:
$k\!>\!\pi/a$~QCD is included via corrections~$\delta\Lag$ to the lattice
lagrangian that are computed using perturbation theory; $k\!<\!\pi/a$~QCD
is handled nonperturbatively using Monte Carlo integration. Thus, while we
wish to make the lattice spacing~$a$ as large as possible, we are
constrained by two requirements. First $a$~must be sufficiently small that
our finite-difference approximations for derivatives in the lagrangian and
field equations are sufficiently accurate. Second $a$~must be sufficiently
small that $\pi/a$~is a perturbative momentum. Numerical experiments
indicate that both constraints can be satisfied when $a\approx1/2$\,fm or
smaller, provided all lattice operators are tadpole improved.

It is important to remember that most any perturbative analysis in QCD is
contaminated by nonperturbative effects at some level. This will certainly
be the case for the couplings in~$\delta\Lag$. However we now have
extensive experience showing that such nonperturbative effects are rarely
significant for physical quantities
at the distances relevant to our discussion. Should a
situation arise where this is not the case (and one surely will, some day)
we may have to supplement the perturbative approach
outlined in this section with nonperturbative
techniques. Such situations will not pose a problem if they are relatively
rare, as now seems likely.
Furthermore, as we discuss later, simple nonperturbative techniques
already exist, should they be needed, for tuning the leading correction
terms in both the gluon and quark actions.
\section{Gluon Dynamics on Coarse Lattices}
In this section I  discuss first the construction of accurate
discretizations of the classical theory of gluon dynamics. I then discuss
the changes needed to make a quantum theory, and illustrate the
discussion with Monte Carlo results. Finally I discuss actions for
anisotropic lattices, and the tuning and testing of gluon actions.

\subsection{Classical Gluons}

The continuum action for QCD is
\be
S = \int d^4x\,\half \sum_{\mu,\nu} \Tr \F\mu\nu^2(x)
\ee
where
\be
\F\mu\nu \equiv \partial_\mu A_\nu - \partial_\nu A_\mu + ig [A_\mu,A_\nu]
\ee
is the field tensor, a traceless $3\times3$ hermitian matrix. The defining
characteristic of the theory is its invariance with respect to gauge
transformations where
\be
\F\mu\nu \to \Omega(x)\,\F\mu\nu\,\Omega(x)^\dagger
\ee
and $\Omega(x)$ is an arbitrary $x$-dependent \su\ matrix.

The standard discretization of this theory seems perverse at first sight.
Rather than specifying the gauge field by the values of~$A_\mu(x)$ at the
sites of the lattice, the field is specified by variables on the links
joining the sites. In the classical theory, the ``link variable'' on
the link joining a site at~$x$ to one at $x+a\hat\mu$ is determined by the
line integral of $A_\mu$ along the link:
\be
\U\mu(x) \equiv \P \exp\left(
-i\int_x^{x+a\hat\mu} gA\cdot \dd y \right)
\ee
where the P-operator path-orders the $A_\mu$'s along the integration path.
We use~$U_\mu$'s in place of~$A_\mu$'s on
the lattice, because it is impossible to formulate a lattice version
of QCD directly in terms of~$A_\mu$'s that has exact gauge invariance. The
$U_\mu$'s, on the other hand, transform very simply under a gauge
transformation:
\be\label{gauge-tnfm}
\U\mu(x)\to\Omega(x)\,\U\mu(x)\,\Omega(x+a\hat\mu)^\dagger.
\ee
This makes it easy to build a discrete theory with exact gauge invariance.

A link variable~$\U\mu(x)$ is represented pictorially by a directed line
from~$x$ to~$x+\hat\mu$, where this line is the integration path for the
line integral in the exponent of~$\U\mu(x)$:
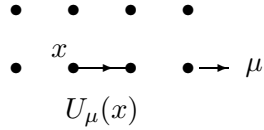
\begin{center}
\setlength{\unitlength}{.03in}
\begin{picture}(75,25)(0,15)
\multiput(25,25)(10,0){4}{\circle*{2}}
\multiput(25,35)(10,0){4}{\circle*{2}}

\put(35,25){\vector(1,0){7}}\put(35,25){\line(1,0){10}}
\put(40,20){\makebox(0,0)[t]{$\U\mu(x)$}}
\put(34,28){\makebox(0,0)[r]{$x$}}
\put(57,25){\vector(1,0){5}}
\put(65,25){\makebox(0,0)[l]{$\mu$}}
\end{picture}
\end{center}
In the conjugate matrix~$\Udag\mu(x)$ the direction of the line integral
is flipped and so we represent~$\Udag\mu(x)$ by a line going backwards
from~$x+\hat\mu$ to~$x$:
\begin{center}
\setlength{\unitlength}{.03in}
\begin{picture}(75,25)(0,15)
\multiput(25,25)(10,0){4}{\circle*{2}}
\multiput(25,35)(10,0){4}{\circle*{2}}
\put(45,25){\vector(-1,0){7}}\put(35,25){\line(1,0){10}}
\put(40,20){\makebox(0,0)[t]{$\Udag\mu(x)$}}
\put(34,28){\makebox(0,0)[r]{$x$}}
\put(57,25){\vector(1,0){5}}
\put(65,25){\makebox(0,0)[l]{$\mu$}}
\end{picture}
\end{center}
A Wilson loop function,
\be
W(\C) \equiv \third\Tr\P\e^{-\I \oint_\C gA\cdot \dd x},
\ee
for any closed path~$\C$ built of links on the lattice can be computed from the
path-ordered product of the~$\U\mu$'s  and~$\Udag\mu$'s associated with each
link. For example, if~$\C$ is the loop
\begin{center}
\setlength{\unitlength}{.03in}
\begin{picture}(40,40)(25,20)
\multiput(25,25)(10,0){4}{\circle*{2}}
\multiput(25,35)(10,0){4}{\circle*{2}}
\multiput(25,45)(10,0){4}{\circle*{2}}
\multiput(25,55)(10,0){4}{\circle*{2}}
\put(35,25){\vector(1,0){7}}\put(35,25){\line(1,0){10}}
\put(45,25){\vector(0,1){7}}\put(45,25){\line(0,1){10}}
\put(45,35){\vector(1,0){7}}\put(45,35){\line(1,0){10}}
\put(55,35){\vector(0,1){7}}\put(55,35){\line(0,1){10}}
\put(55,45){\vector(-1,0){7}}\put(55,45){\line(-1,0){10}}
\put(45,45){\vector(-1,0){7}}\put(45,45){\line(-1,0){10}}
\put(35,45){\vector(0,-1){7}}\put(35,45){\line(0,-1){10}}
\put(35,35){\vector(0,-1){7}}\put(35,35){\line(0,-1){10}}
\put(34,22){\makebox(0,0)[r]{$x$}}
\put(60,25){\vector(1,0){5}}\put(67,25){\makebox(0,0)[l]{$\mu$}}
\put(60,27){\vector(0,1){5}}\put(60,34){\makebox(0,0)[b]{$\nu$}}
\end{picture}
\end{center}
then
\be
W(\C) = \third\Tr\!\left(\U\mu(x)\U\nu(x+a\hat\mu)\ldots\Udag\nu(x)\right).
\ee
Such quantities are obviously invariant under arbitrary
gauge transformations~\eq{gauge-tnfm}.

You might wonder why we go to so much trouble to preserve gauge invariance
when we quite willing give up Lorentz invariance, rotation
invariance, etc. The reason is quite practical. With gauge invariance, the
quark-gluon, three-gluon, and four-gluon couplings in QCD are all equal,
and the bare gluon mass is zero. Without gauge invariance, each of these
couplings must be tuned independently and a gluon mass introduced if one is
to recover QCD. Tuning this many parameters in a numerical simulation is
very expensive. This is not much of a problem in the classical theory,
where approximate gauge invariance keeps the couplings approximately equal;
but it is serious in the quantum theory because quantum fluctuations
(loop-effects) renormalize the various couplings differently in the absence
of exact gauge invariance. So while it is quite possible to formulate
lattice QCD directly in terms of~$\A\mu$'s, the resulting theory would have
only approximate gauge invariance, and thus would be prohibitively expensive
to simulate. Symmetries like Lorentz invariance can be given up with little
cost because the symmetries of the lattice, though far less restrictive, are
still sufficient to prevent the introduction of new interactions with new
couplings (at least to lowest order in~$a$).

We must now build a lattice lagrangian from the link operators. We require
that the lagrangian be gauge invariant, local, and symmetric with respect
to axis interchanges (which is all that is left of Lorentz invariance). The
most local nontrivial gauge invariant object one can build from the link
operators is the ``plaquette operator,'' which involves the product of link
variables around the smallest square at site~$x$ in the $\mu\nu$~plane:
\be
\Pl\mu\nu(x) \equiv \third\Re\Tr\!\left(\U\mu(x)\U\nu(x+a\hat\mu)
\Udag\mu(x+a\hat\mu+a\hat\nu)\Udag\nu(x)\right).
\ee
To see what this object is, consider evaluating the plaquette centered
about a point~$x_0$ for a very smooth weak classical $\A\mu$~field.
In this limit,
\be \label{uplaqa}
\Pl\mu\nu \approx 1
\ee
since
\be
\U\mu \approx \e^{-\I ga\A\mu} \approx 1 .
\ee
Given that $\A\mu$~is slowly varying, its value anywhere on the plaquette
should be accurately specified by its value and derivatives at~$x_0$. Thus
the corrections to \eq{uplaqa} should be a polynomial in~$a$ with
coefficients formed from gauge-invariant combinations of~$\A\mu(x_0)$ and
its derivatives: that is,
\begin{eqnarray}
\Pl\mu\nu \,= \, 1
&-&c_1\,a^4\,\Tr\!\left(g\F\mu\nu(x_0)\right)^2 \nl
&-&c_2\,a^6\,\Tr\!\left(g\F\mu\nu(x_0)(\D_\mu^2+\D_\nu^2)g\F\mu\nu(x_0)
 \right)\nl
&+&\order(a^8)
\label{ope}
\end{eqnarray}
where $c_1$ and $c_2$ are constants, and $\D_\mu$ is the
gauge-covariant derivative.  The leading correction is
order~$a^4$ because $\F\mu\nu^2$ is the lowest-dimension gauge-invariant
combination of derivatives of $\A\mu$, and it has dimension~4.
(There are no $F^3$~terms because
$\Pl\mu\nu$~is invariant under $\U\mu\to\Udag\mu$ or,
equivalently, $F\to -F$.)

It is straightforward to find the coefficients~$c_1$ and~$c_2$. We need
only examine terms in the expansion of~$\Pl\mu\nu$ that are quadratic
in~$A_\mu$; the cubic and quartic parts of $\F\mu\mu^2$ then follow
automatically, by gauge invariance. Because of the trace, the path
ordering is irrelevant to this order. Thus
\bearray
\Pl\mu\nu &=&  \third\Re\Tr\P\e^{-\I \oint_\Box gA\cdot \dd x} \nl
&=&\third\Re\Tr\left[1 - i\oint_\Box gA\cdot \dd x -\half\left(\oint_\Box
gA\cdot \dd x\right)^2 + \order(A^3)\right]
\eearray
where, by Stoke's Theorem,
\bearray
\oint_\Box  A\cdot \dd x &=& \int_{-a/2}^{a/2}\dd x_\mu \dd
x_\nu\left[\partial_\mu
A_\nu(x_0+x) - \partial_\nu A_\mu(x_0+x)\right] \nl
&=& \int_{-a/2}^{a/2}\dd x_\mu \dd x_\nu \left[\F\mu\nu(x_0) +(x_\mu\D_\mu
+x_\nu\D_\mu)\F\mu\nu(x_0) +\cdots\right] \nl
&=& a^2\,\F\mu\nu(x_0)  + \frac{a^4}{24}\,(\D_\mu^2+\D_\nu^2)\F\mu\nu(x_0)
+\order(a^6,A^2).
\eearray
Thus~$c_1 = 1/6$ and~$c_2=1/72$ in~\eq{ope}.

The expansion in \eq{ope} is the classical analogue of an operator product
expansion. Using this expansion, we find that the traditional ``Wilson
action'' for gluons on a lattice,
\be\label{Wil-S}
S_{\rm Wil} = \beta\,\sum_{x,\mu>\nu} \left(1-\Pl\mu\nu(x)\right)
\ee
where $\beta\!=\!6/g^2$, has the correct limit for small lattice spacing
up to corrections of order~$a^2$:
\be
S_{\rm Wil} = \int d^4x\,\sum_{\mu,\nu} \left\{\half\Tr \F\mu\nu^2
+ \frac{a^2}{24}\Tr\F\mu\nu(\D_\mu^2+\D_\nu^2)\F\mu\nu +\cdots\right\}.
\ee

We can cancel the $a^2$~error in the Wilson action by adding other Wilson
loops. For example, the~$2a\times a$``rectangle operator''
\be
\Rt\mu\nu = \third\Re\Tr\!
\setlength{\unitlength}{.015in}
\begin{picture}(75,13)(0,17)
  \put(10,10){\vector(0,1){12.5}}
  \put(10,10){\line(0,1){20}}
  \put(10,30){\vector(1,0){12.5}}
  \put(10,30){\vector(1,0){32.5}}
  \put(10,30){\line(1,0){40}}
  \put(50,30){\vector(0,-1){12.5}}
  \put(50,30){\line(0,-1){20}}
  \put(50,10){\vector(-1,0){12.5}}
  \put(50,10){\vector(-1,0){32.5}}
  \put(50,10){\line(-1,0){40}}
  \put(61,10){\vector(1,0){10}}\put(75,10){\makebox(0,0){$\mu$}}
  \put(60,11){\vector(0,1){10}}\put(60,25){\makebox(0,0){$\nu$}}
\end{picture}
\ee
has expansion
\be
\Rt\mu\nu = 1 -\frac{4}{6}\,a^4\Tr(g\F\mu\nu)^2
-\frac{4}{72}\,a^6\Tr\left(g\F\mu\nu(4\,\D_\mu^2+\D_\nu^2)g\F\mu\nu\right)
-\cdots .
\ee
The mix of $a^4$~terms and $a^6$~terms in the rectangle is different from
that in the plaquette. Therefore we can combine the two operators to
obtain an improved classical lattice action that is accurate up
to~$\order(a^4)$\,\cite{curci83,luscher85a}:
\bearray
S_{\rm classical} &\equiv& -\beta \sum_{x,\mu>\nu}\left\{
\frac{5\Pl\mu\nu}{3} - \frac{\Rt\mu\nu+\Rt\nu\mu}{12}\right\}  + {\rm const}
\\
&=& \int d^4x\,\sum_{\mu,\nu}\half\Tr\F\mu\nu^2 + \order(a^4).
\eearray
This process is the analogue of improving the derivatives in
discretizations of
non-gauge theories.\footnote{An important step that I have not
discussed is to show that the gluon action is positive for any
configuration of link variables. This guarantees that the classical
ground state of the lattice action corresponds to
$\F\mu\nu\!=\!0$. See~\cite{luscher85a} for a detailed discussion.}

\begin{exercise}
The euclidean Green's function or propagator~$G$   for a nonrelativistic
quark in a background gauge field~$A_\mu$ satisfies the equation
\be
\left(\D_t-\frac{\Dv^2}{2M}\right)G(x) = \delta^4(x)
\ee
where $\D_\mu=\partial_\mu - igA_\mu(x)$ is the gauge-covariant derivative.
Show that the static ($M\!=\!\infty$) quark propagator is
\be
G_\infty(\xv,t) = \left[\P\e^{-\I \int^t_0
gA_0(\xv,t)\dd t}\right]^\dagger\,\delta^3(\xv)
\ee
which on the lattice becomes
\be
G_\infty(\xv,t) = \Udag t(\xv,t\!-\!a)\,\Udag t(\xv,t\!-\! 2a)\,\ldots\,
\Udag t(\xv,0).
\ee
Propagation of a static antiquark is described by~$G_\infty^\dagger$ and
therefore the ``static potential''~$V(r)$, which is the energy of a static
quark and antiquark a distance~$r$ apart, is obtained from
\be
W(r,t) \equiv \langle 0| \third\Tr
\setlength{\unitlength}{.02in}
\begin{picture}(50,12)(-5,18)
\multiput(0,10)(10,0){5}{\circle*{2}}
\multiput(0,20)(10,0){5}{\circle*{2}}
\multiput(0,30)(10,0){5}{\circle*{2}}

\multiput(0,10)(10,0){4}{\vector(1,0){7}}
\multiput(0,10)(10,0){4}{\line(1,0){10}}

\multiput(40,10)(0,10){2}{\vector(0,1){7}}
\multiput(40,10)(0,10){2}{\line(0,1){10}}

\multiput(40,30)(-10,0){4}{\vector(-1,0){7}}
\multiput(40,30)(-10,0){4}{\line(-1,0){10}}

\multiput(0,30)(0,-10){2}{\vector(0,-1){7}}
\multiput(0,30)(0,-10){2}{\line(0,-1){10}}

\put(20,0){\makebox(0,0){$t$}}\put(17,0){\vector(-1,0){17}}
\put(23,0){\vector(1,0){17}}

\put(60,20){\makebox(0,0){$r$}}\put(60,23){\vector(0,1){7}}
\put(60,17){\vector(0,-1){7}}

\end{picture}
| 0 \rangle\rule[-.5in]{0pt}{1in}
\ee
where for large~$t$
\be \label{static-V}
W(r,t)\to {\rm const}\,\,\e^{-V(r)\,t} .
\ee
\end{exercise}

\begin{exercise} Defining the ``twisted-rectangle operator''
\be
T_{\mu\nu} = \third\Re\Tr\!
\setlength{\unitlength}{.015in}
\begin{picture}(60,13)(0,17)
  \put(27.5,10){\vector(-1,0){10}}
  \put(27.5,10){\line(-1,0){17.5}}
  \put(10,10){\vector(0,1){12.5}}
  \put(10,10){\line(0,1){20}}
  \put(10,30){\vector(1,0){12.5}}
  \put(10,30){\line(1,0){17.5}}

  \put(32.5,10){\vector(1,0){10}}
  \put(32.5,10){\line(1,0){17.5}}
  \put(50,10){\vector(0,1){12.5}}
  \put(50,10){\line(0,1){20}}
  \put(50,30){\vector(-1,0){12.5}}
  \put(50,30){\line(-1,0){17.5}}

  \put(32.5,10){\line(-1,4){5.0}}
  \put(27.5,10){\line( 1,4){5.0}}
\end{picture}.
\ee
show that
\bearray \label{trt-S}
S_{\rm classical}  &\equiv& -\beta \sum_{x,\mu>\nu}
\left\{\Pl\mu\nu + \frac{T_{\mu\nu}+T_{\nu\mu}}{12}\right\}
+{\rm const}
\\
&=& \int d^4x\,\sum_{\mu,\nu}\half\Tr\F\mu\nu^2 + \order(a^4).
\eearray
This is an alternative to the improved gluon action derived in the previous
exercise.
\end{exercise}

\subsection{Quantum Gluons\protect\footnotemark}
\footnotetext{This section and the next are based on work with M.\,Alford,
W.\,Dimm, G.\,Hockney and P.\,Mackenzie that is described
in~\cite{alford95}.}
In the previous section we derived improved classical actions for gluons
that are accurate through order~$a^4$. We now turn these into quantum
actions. The most important step is to tadpole improve the action by
dividing each link operator~$\U\mu$ by the mean link~$u_0$: for
example, the action built of plaquette and rectangle operators becomes
\be
S = -\beta\sum_{x,\mu>\nu}\left\{\frac{5}{3}\frac{\Pl\mu\nu}{u_0^4}
-\frac{\Rt\mu\nu+\Rt\nu\mu}{12\,u_0^6}\right\}.
\ee
The $u_0$'s cancel lattice tadpole contributions that
otherwise would spoil weak-coupling perturbation theory in the lattice theory
and undermine our procedure for improving the lattice discretization.
Note that $u_0\!\approx\!3/4$ when $a\!=\!.4$\,fm, and therefore the
relative importance of the~$\Rt\mu\nu$'s is larger by a
factor of~$1/u_0^2\!\approx\!2$ than without tadpole improvement.
Without tadpole improvement, we cancel only half of the $a^2$ error.

The mean link~$u_0$ is computed numerically by guessing a value for use in
the action,  measuring the mean link in a simulation, and then readjusting
the value used in the action accordingly. This tuning cycle converges
rapidly to selfconsistent values, and can be done very quickly using small
lattice volumes. The $u_0$'s depend only on lattice spacing, and
become equal to one as the lattice spacing vanishes.

The expectation value of the link operator is gauge dependent. Thus to
minimize gauge artifacts, $u_0$ is commonly defined  as the
Landau-gauge expectation value, $\langle0|\third\Tr\U\mu|0\rangle_{\rm LG}$.
Landau gauge is the axis-symmetric gauge that maximizes $u_0$, thereby
minimizing the tadpole contribution; any tadpole contribution that is left
in Landau gauge cannot be a gauge artifact. An alternative procedure is to
define
$u_0$ as the fourth root of the plaquette expectation value,
\be
u_0 = \langle 0|\Pl\mu\nu |0\rangle^{1/4}.
\ee
This definition gives almost identical results and is more convenient for
numerical work since gauge fixing is unnecessary.

Tadpole improvement is the first step in a systematic procedure for
improving the action. The next step is to add in renormalizations due to
contributions from $k\!>\!\pi/a$~physics not already included in the
tadpole improvement. These renormalizations induce
$a^2\,\alpha_s(\pi/a)$~corrections,
\bearray
\delta\Lag &=&
\alpha_s\,r_1\,a^2\sum_{\mu,\nu}\Tr\!(\F\mu\nu\D_\mu^2\F\mu\nu)
\nl
&+&\alpha_s\,r_2\,a^2\sum_{\mu,\nu}\Tr\!(\D_\mu\F\nu\sigma
\D_\mu\F\nu\sigma)
\nl
&+&\alpha_s\,r_3\,a^2\sum_{\mu,\nu}\Tr\!(\D_\mu\F\mu\sigma
\D_\nu\F\nu\sigma)
\nl
&+&\cdots,
\eearray
that must be removed. The last term is harmless; its coefficient can be set
to zero by a change of field variable (in the path integral) of the form
\be \label{Atransform}
A_\mu \to A_\mu + a^2\,\alpha_s\,f(\alpha_s)\,\sum_\nu \D_\nu F_{\nu\mu}.
\ee
Since changing integration variables does not change the value of an integral,
such field transformations must leave the physics unchanged.\footnote{One
must, of course, include the jacobian for the transformation in the transformed
path integral. This contributes only in one-loop order and higher; it has no
effect on tree-level calculations.} Operators that can be removed by a field
transformation are called ``redundant.'' The other corrections are removed by
renormalizing the coefficient of the rectangle operator~$\Rt\mu\nu$ in the
action, and by adding an additional operator. One choice for the extra operator
is
\be
C_{\mu\nu\sigma}  \equiv \third\Re\Tr\!\!
\setlength{\unitlength}{.015in}
\begin{picture}(60,20)(0,17)
  \put(10,10){\vector(0,1){12.5}}
  \put(10,10){\line(0,1){20}}
  \put(10,30){\vector(2,1){10}}
  \put(10,30){\line(2,1){15}}
  \put(25.2,37.6){\vector(1,0){12.5}}
  \put(25.2,37.6){\line(1,0){20}}
  \put(45.2,37.6){\vector(0,-1){12.5}}
  \put(45.2,37.6){\line(0,-1){20}}
  \put(45.2,17.6){\vector(-2,-1){10}}
  \put(45.2,17.6){\line(-2,-1){15}}
  \put(30,10){\vector(-1,0){12.5}}
  \put(30,10){\line(-1,0){20}}
\end{picture}.
\ee
Then the action, correct up to~$\order(a^2\alpha_s^2,a^4)$,
is\,\cite{luscher85b}
\be \label{LW-S}
S = -\beta\sum_{x,\mu>\nu}\left\{\frac{5}{3}\frac{\Pl\mu\nu}{u_0^4}
-r_{\rm g}\,\frac{\Rt\mu\nu+\Rt\nu\mu}{12\,u_0^6}\right\}
+c_{\rm g}\,\beta\sum_{x,\mu>\nu>\sigma} \frac{C_{\mu\nu\sigma}}{u_0^6},
\ee
where
\bearray
r_{\rm g} &=& 1 + .48\,\alpha_s(\pi/a) \\
c_{\rm g} &=& .055\,\alpha_s(\pi/a).
\eearray

The coefficients~$r_{\rm g}$ and~$c_{\rm g}$ are computed by ``matching''
physical
quantities, like low-energy scattering amplitudes, computed using
perturbation theory in the lattice theory with the analogous quantity in
the continuum theory. The lattice result depends upon~$r_{\rm g}$ and~$c_{\rm
g}$;
these parameters are tuned until the lattice amplitude agrees with the
continuum amplitude to the order in~$a$ and~$\alpha_s$ required:
\be
T_{\rm lat}(r_{\rm g},c_{\rm g}) = T_{\rm contin}.
\ee
Note that tadpole improvement has a big effect on these coefficients.
Without tadpole improvement, $r_{\rm g}=1+2\alpha_s$; that is, the coefficient
of
the radiative correction is four times larger.
Tadpole improvement automatically supplies 75\% of the one-loop
contribution needed without improvement.
Since $\alpha_s\!\approx\!0.3$, the unimproved expansion for~$r_{\rm g}$ is not
particularly convergent. However, with tadpole improvement, the one-loop
correction is only about 10--20\% of~$r_{\rm g}$. Indeed, for most
current applications, one-loop corrections to tadpole-improved actions
are negligible.

Finally note that the twisted-rectangle action~\eq{trt-S} must also be
tadpole improved:
\be \label{twisted-S}
S_{\rm trt}  \equiv -\beta \sum_{x,\mu>\nu}
\left\{\frac{\Pl\mu\nu}{u_0^4} +
\frac{T_{\mu\nu}+T_{\nu\mu}}{12\,u_0^8} \right\}.
\ee
The correction term in this action has two more~$u_0$'s than that
in our other action. This makes~$S_{\rm trt}$ much more dependent upon
tadpole improvement. Comparing results generated by our two actions
provides a sensitive test of tadpole improvement. The $\alpha_s$
corrections to this action have not yet been computed, but are probably
unnecessary.

\begin{exercise} Show that the gauge that maximizes
$\langle0|\third\Tr\U\mu|0\rangle$ becomes Landau gauge ($\partial\cdot
A\!=\!0$) in the $a\!\to\!0$~limit.
\end{exercise}

\begin{exercise} Tadpole diagrams are only important because they are
ultraviolet divergent; the loop integrations generate factors
of~$1/a^2$ that cancel the explicit~$a$'s from the link
operator~$U_\mu$. Consequently such contributions can be significantly
reduced if the effective ultraviolet cutoff for the gluons is
lowered. Lowering the cutoff by a factor of two, for example, would reduce
tadpole contributions by a factor of four, rendering them harmless. This
reduction is easily accomplished by smearing, where, for example,
\be
A_\mu(x) \to  \left(1-\frac{\epsilon\,a^4 {\rm
D}^4}{n}\right)^n\,A_\mu(x).
\ee
The low-momentum components of the smeared $A_\mu$~field are identical, up
to $\order(a^4)$~corrections, with the unsmeared field; but at high
momentum the smearing factor acts as an ultraviolet cutoff
\be
\left(1-\frac{\epsilon\,a^4\,{\rm
D}^4}{n}\right)^n \to
\left(1-\frac{\epsilon\,a^4\,
p^4}{n}\right)^n
\approx
\e^{-\epsilon p^4 a^4}.
\ee
Tadpole improvement should be unnecessary when the smeared field is used
in quark actions, etc., in place of the original field.
Parameter~$\epsilon$ is chosen to give the appropriate cutoff,
while~$n$ is chosen sufficiently large that smearing factor approximately
exponentiates as indicated.  Design and test an implementation of this
smearing scheme for the link variables~$U_\mu$.
\end{exercise}

\begin{exercise} Lattice $\phi^4$~theory has $a^2$~corrections of the form
\be
\delta\Lag = c_1\,a^2\,\sum_\mu\left(\partial_\mu^2\phi\right)^2
+ c_2\,a^2\,\phi^6
+ c_3\,a^2\,\left(\partial^2\phi\right)^2
+ c_4\,a^2\,\phi^3\partial^2\phi.
\ee
Find a field transformation
\be
\phi\to\tilde\phi = \phi + a^2\,f(\phi,\partial_\mu^2\phi\ldots)
\ee
that removes the $c_3$~and $c_4$~terms (that is, the redundant operators).

Note that~$\phi\to\phi+\delta\phi$ implies
\be
S[\phi]\to S + \delta\phi\,\frac{\delta S}{\delta\phi}.
\ee
Thus the leading redundant operators can be rearranged so that they are
proportional to $\delta S/\delta \phi$. Since the classical field equation
is $\delta S/\delta\phi = 0$, redundant operators are often said to have no
effect ``because of the equations of motion.'' More precisely it is because
they can be removed by a change of integration variable in the path integral.

Show that the jacobian~$|\delta\phi/\delta\tilde\phi|$ has no effect at
tree-level, but must be included when quantum loop corrections are
important. (Hint: Try computing a simple scattering amplitude using
perturbation theory in the original and in the transformed theory.)
\end{exercise}

\subsection{Tests of the Gluon Action}
Having a procedure for systematically improving the gluon action, we need
to establish whether the improvements really do improve simulations.

\begin{figure}
\begin{center}
\setlength{\unitlength}{0.240900pt}
\ifx\plotpoint\undefined\newsavebox{\plotpoint}\fi
\sbox{\plotpoint}{\rule[-0.175pt]{0.350pt}{0.350pt}}%
\begin{picture}(1200,900)(0,0)
\tenrm
\sbox{\plotpoint}{\rule[-0.175pt]{0.350pt}{0.350pt}}%
\put(264,158){\rule[-0.175pt]{210.065pt}{0.350pt}}
\put(264,158){\rule[-0.175pt]{0.350pt}{151.526pt}}
\put(264,284){\rule[-0.175pt]{4.818pt}{0.350pt}}
\put(242,284){\makebox(0,0)[r]{$1$}}
\put(1116,284){\rule[-0.175pt]{4.818pt}{0.350pt}}
\put(264,410){\rule[-0.175pt]{4.818pt}{0.350pt}}
\put(242,410){\makebox(0,0)[r]{$2$}}
\put(1116,410){\rule[-0.175pt]{4.818pt}{0.350pt}}
\put(264,535){\rule[-0.175pt]{4.818pt}{0.350pt}}
\put(242,535){\makebox(0,0)[r]{$3$}}
\put(1116,535){\rule[-0.175pt]{4.818pt}{0.350pt}}
\put(264,661){\rule[-0.175pt]{4.818pt}{0.350pt}}
\put(242,661){\makebox(0,0)[r]{$4$}}
\put(1116,661){\rule[-0.175pt]{4.818pt}{0.350pt}}
\put(438,158){\rule[-0.175pt]{0.350pt}{4.818pt}}
\put(438,113){\makebox(0,0){$1$}}
\put(438,767){\rule[-0.175pt]{0.350pt}{4.818pt}}
\put(613,158){\rule[-0.175pt]{0.350pt}{4.818pt}}
\put(613,113){\makebox(0,0){$2$}}
\put(613,767){\rule[-0.175pt]{0.350pt}{4.818pt}}
\put(787,158){\rule[-0.175pt]{0.350pt}{4.818pt}}
\put(787,113){\makebox(0,0){$3$}}
\put(787,767){\rule[-0.175pt]{0.350pt}{4.818pt}}
\put(962,158){\rule[-0.175pt]{0.350pt}{4.818pt}}
\put(962,113){\makebox(0,0){$4$}}
\put(962,767){\rule[-0.175pt]{0.350pt}{4.818pt}}
\put(264,158){\rule[-0.175pt]{210.065pt}{0.350pt}}
\put(1136,158){\rule[-0.175pt]{0.350pt}{151.526pt}}
\put(264,787){\rule[-0.175pt]{210.065pt}{0.350pt}}
\put(45,472){\makebox(0,0)[l]{\shortstack{$a\,V(r)$}}}
\put(700,68){\makebox(0,0){$r/a$}}
\put(395,693){\makebox(0,0)[l]{a) Wilson Action}}
\put(264,158){\rule[-0.175pt]{0.350pt}{151.526pt}}
\put(438,293){\circle{12}}
\put(511,392){\circle{12}}
\put(566,478){\circle{12}}
\put(613,427){\circle{12}}
\put(654,510){\circle{12}}
\put(787,553){\circle{12}}
\put(962,636){\circle{12}}
\put(438,293){\usebox{\plotpoint}}
\put(428,293){\rule[-0.175pt]{4.818pt}{0.350pt}}
\put(428,293){\rule[-0.175pt]{4.818pt}{0.350pt}}
\put(511,388){\rule[-0.175pt]{0.350pt}{1.927pt}}
\put(501,388){\rule[-0.175pt]{4.818pt}{0.350pt}}
\put(501,396){\rule[-0.175pt]{4.818pt}{0.350pt}}
\put(566,473){\rule[-0.175pt]{0.350pt}{2.409pt}}
\put(556,473){\rule[-0.175pt]{4.818pt}{0.350pt}}
\put(556,483){\rule[-0.175pt]{4.818pt}{0.350pt}}
\put(613,426){\rule[-0.175pt]{0.350pt}{0.482pt}}
\put(603,426){\rule[-0.175pt]{4.818pt}{0.350pt}}
\put(603,428){\rule[-0.175pt]{4.818pt}{0.350pt}}
\put(654,508){\rule[-0.175pt]{0.350pt}{1.204pt}}
\put(644,508){\rule[-0.175pt]{4.818pt}{0.350pt}}
\put(644,513){\rule[-0.175pt]{4.818pt}{0.350pt}}
\put(787,545){\rule[-0.175pt]{0.350pt}{3.854pt}}
\put(777,545){\rule[-0.175pt]{4.818pt}{0.350pt}}
\put(777,561){\rule[-0.175pt]{4.818pt}{0.350pt}}
\put(962,586){\rule[-0.175pt]{0.350pt}{24.090pt}}
\put(952,586){\rule[-0.175pt]{4.818pt}{0.350pt}}
\put(952,686){\rule[-0.175pt]{4.818pt}{0.350pt}}
\sbox{\plotpoint}{\rule[-0.250pt]{0.500pt}{0.500pt}}%
\put(326,158){\usebox{\plotpoint}}
\put(335,176){\usebox{\plotpoint}}
\put(346,194){\usebox{\plotpoint}}
\put(358,210){\usebox{\plotpoint}}
\put(371,226){\usebox{\plotpoint}}
\put(385,242){\usebox{\plotpoint}}
\put(399,257){\usebox{\plotpoint}}
\put(414,271){\usebox{\plotpoint}}
\put(429,285){\usebox{\plotpoint}}
\put(445,299){\usebox{\plotpoint}}
\put(461,312){\usebox{\plotpoint}}
\put(477,325){\usebox{\plotpoint}}
\put(493,338){\usebox{\plotpoint}}
\put(510,351){\usebox{\plotpoint}}
\put(526,364){\usebox{\plotpoint}}
\put(543,376){\usebox{\plotpoint}}
\put(560,388){\usebox{\plotpoint}}
\put(576,401){\usebox{\plotpoint}}
\put(593,413){\usebox{\plotpoint}}
\put(610,425){\usebox{\plotpoint}}
\put(627,437){\usebox{\plotpoint}}
\put(644,449){\usebox{\plotpoint}}
\put(660,461){\usebox{\plotpoint}}
\put(677,473){\usebox{\plotpoint}}
\put(695,485){\usebox{\plotpoint}}
\put(711,497){\usebox{\plotpoint}}
\put(728,509){\usebox{\plotpoint}}
\put(745,521){\usebox{\plotpoint}}
\put(762,533){\usebox{\plotpoint}}
\put(779,545){\usebox{\plotpoint}}
\put(796,557){\usebox{\plotpoint}}
\put(813,568){\usebox{\plotpoint}}
\put(830,580){\usebox{\plotpoint}}
\put(847,592){\usebox{\plotpoint}}
\put(864,604){\usebox{\plotpoint}}
\put(882,615){\usebox{\plotpoint}}
\put(899,627){\usebox{\plotpoint}}
\put(916,639){\usebox{\plotpoint}}
\put(932,651){\usebox{\plotpoint}}
\put(950,663){\usebox{\plotpoint}}
\put(967,674){\usebox{\plotpoint}}
\put(984,686){\usebox{\plotpoint}}
\put(1001,698){\usebox{\plotpoint}}
\put(1018,710){\usebox{\plotpoint}}
\put(1035,721){\usebox{\plotpoint}}
\put(1053,733){\usebox{\plotpoint}}
\put(1070,745){\usebox{\plotpoint}}
\put(1087,756){\usebox{\plotpoint}}
\put(1104,768){\usebox{\plotpoint}}
\put(1121,780){\usebox{\plotpoint}}
\put(1131,787){\usebox{\plotpoint}}
\end{picture}
\setlength{\unitlength}{0.240900pt}
\ifx\plotpoint\undefined\newsavebox{\plotpoint}\fi
\sbox{\plotpoint}{\rule[-0.175pt]{0.350pt}{0.350pt}}%
\begin{picture}(1200,900)(0,0)
\tenrm
\sbox{\plotpoint}{\rule[-0.175pt]{0.350pt}{0.350pt}}%
\put(264,158){\rule[-0.175pt]{210.065pt}{0.350pt}}
\put(264,158){\rule[-0.175pt]{0.350pt}{151.526pt}}
\put(264,284){\rule[-0.175pt]{4.818pt}{0.350pt}}
\put(242,284){\makebox(0,0)[r]{$1$}}
\put(1116,284){\rule[-0.175pt]{4.818pt}{0.350pt}}
\put(264,410){\rule[-0.175pt]{4.818pt}{0.350pt}}
\put(242,410){\makebox(0,0)[r]{$2$}}
\put(1116,410){\rule[-0.175pt]{4.818pt}{0.350pt}}
\put(264,535){\rule[-0.175pt]{4.818pt}{0.350pt}}
\put(242,535){\makebox(0,0)[r]{$3$}}
\put(1116,535){\rule[-0.175pt]{4.818pt}{0.350pt}}
\put(264,661){\rule[-0.175pt]{4.818pt}{0.350pt}}
\put(242,661){\makebox(0,0)[r]{$4$}}
\put(1116,661){\rule[-0.175pt]{4.818pt}{0.350pt}}
\put(438,158){\rule[-0.175pt]{0.350pt}{4.818pt}}
\put(438,113){\makebox(0,0){$1$}}
\put(438,767){\rule[-0.175pt]{0.350pt}{4.818pt}}
\put(613,158){\rule[-0.175pt]{0.350pt}{4.818pt}}
\put(613,113){\makebox(0,0){$2$}}
\put(613,767){\rule[-0.175pt]{0.350pt}{4.818pt}}
\put(787,158){\rule[-0.175pt]{0.350pt}{4.818pt}}
\put(787,113){\makebox(0,0){$3$}}
\put(787,767){\rule[-0.175pt]{0.350pt}{4.818pt}}
\put(962,158){\rule[-0.175pt]{0.350pt}{4.818pt}}
\put(962,113){\makebox(0,0){$4$}}
\put(962,767){\rule[-0.175pt]{0.350pt}{4.818pt}}
\put(264,158){\rule[-0.175pt]{210.065pt}{0.350pt}}
\put(1136,158){\rule[-0.175pt]{0.350pt}{151.526pt}}
\put(264,787){\rule[-0.175pt]{210.065pt}{0.350pt}}
\put(45,472){\makebox(0,0)[l]{\shortstack{$a\,V(r)$}}}
\put(700,68){\makebox(0,0){$r/a$}}
\put(395,693){\makebox(0,0)[l]{b) Improved Action}}
\put(264,158){\rule[-0.175pt]{0.350pt}{151.526pt}}
\put(438,270){\circle{12}}
\put(511,332){\circle{12}}
\put(566,377){\circle{12}}
\put(613,402){\circle{12}}
\put(654,429){\circle{12}}
\put(757,496){\circle{12}}
\put(787,530){\circle{12}}
\put(962,669){\circle{12}}
\put(438,269){\usebox{\plotpoint}}
\put(428,269){\rule[-0.175pt]{4.818pt}{0.350pt}}
\put(428,270){\rule[-0.175pt]{4.818pt}{0.350pt}}
\put(511,332){\usebox{\plotpoint}}
\put(501,332){\rule[-0.175pt]{4.818pt}{0.350pt}}
\put(501,332){\rule[-0.175pt]{4.818pt}{0.350pt}}
\put(566,376){\rule[-0.175pt]{0.350pt}{0.482pt}}
\put(556,376){\rule[-0.175pt]{4.818pt}{0.350pt}}
\put(556,378){\rule[-0.175pt]{4.818pt}{0.350pt}}
\put(613,401){\rule[-0.175pt]{0.350pt}{0.482pt}}
\put(603,401){\rule[-0.175pt]{4.818pt}{0.350pt}}
\put(603,403){\rule[-0.175pt]{4.818pt}{0.350pt}}
\put(654,427){\rule[-0.175pt]{0.350pt}{0.964pt}}
\put(644,427){\rule[-0.175pt]{4.818pt}{0.350pt}}
\put(644,431){\rule[-0.175pt]{4.818pt}{0.350pt}}
\put(757,495){\rule[-0.175pt]{0.350pt}{0.482pt}}
\put(747,495){\rule[-0.175pt]{4.818pt}{0.350pt}}
\put(747,497){\rule[-0.175pt]{4.818pt}{0.350pt}}
\put(787,524){\rule[-0.175pt]{0.350pt}{3.132pt}}
\put(777,524){\rule[-0.175pt]{4.818pt}{0.350pt}}
\put(777,537){\rule[-0.175pt]{4.818pt}{0.350pt}}
\put(962,656){\rule[-0.175pt]{0.350pt}{6.022pt}}
\put(952,656){\rule[-0.175pt]{4.818pt}{0.350pt}}
\put(952,681){\rule[-0.175pt]{4.818pt}{0.350pt}}
\sbox{\plotpoint}{\rule[-0.250pt]{0.500pt}{0.500pt}}%
\put(339,158){\usebox{\plotpoint}}
\put(349,175){\usebox{\plotpoint}}
\put(362,192){\usebox{\plotpoint}}
\put(375,208){\usebox{\plotpoint}}
\put(388,223){\usebox{\plotpoint}}
\put(404,238){\usebox{\plotpoint}}
\put(419,252){\usebox{\plotpoint}}
\put(434,266){\usebox{\plotpoint}}
\put(450,279){\usebox{\plotpoint}}
\put(466,292){\usebox{\plotpoint}}
\put(482,305){\usebox{\plotpoint}}
\put(498,318){\usebox{\plotpoint}}
\put(515,331){\usebox{\plotpoint}}
\put(531,344){\usebox{\plotpoint}}
\put(548,356){\usebox{\plotpoint}}
\put(565,368){\usebox{\plotpoint}}
\put(582,380){\usebox{\plotpoint}}
\put(598,392){\usebox{\plotpoint}}
\put(615,405){\usebox{\plotpoint}}
\put(632,417){\usebox{\plotpoint}}
\put(649,429){\usebox{\plotpoint}}
\put(666,441){\usebox{\plotpoint}}
\put(683,453){\usebox{\plotpoint}}
\put(700,465){\usebox{\plotpoint}}
\put(717,477){\usebox{\plotpoint}}
\put(734,489){\usebox{\plotpoint}}
\put(751,501){\usebox{\plotpoint}}
\put(768,512){\usebox{\plotpoint}}
\put(785,524){\usebox{\plotpoint}}
\put(802,536){\usebox{\plotpoint}}
\put(819,548){\usebox{\plotpoint}}
\put(836,559){\usebox{\plotpoint}}
\put(853,571){\usebox{\plotpoint}}
\put(870,583){\usebox{\plotpoint}}
\put(887,595){\usebox{\plotpoint}}
\put(904,607){\usebox{\plotpoint}}
\put(921,618){\usebox{\plotpoint}}
\put(938,630){\usebox{\plotpoint}}
\put(956,642){\usebox{\plotpoint}}
\put(973,654){\usebox{\plotpoint}}
\put(990,665){\usebox{\plotpoint}}
\put(1007,677){\usebox{\plotpoint}}
\put(1024,689){\usebox{\plotpoint}}
\put(1041,700){\usebox{\plotpoint}}
\put(1059,712){\usebox{\plotpoint}}
\put(1075,724){\usebox{\plotpoint}}
\put(1093,735){\usebox{\plotpoint}}
\put(1110,747){\usebox{\plotpoint}}
\put(1127,759){\usebox{\plotpoint}}
\put(1136,765){\usebox{\plotpoint}}
\end{picture}
\end{center}
\caption{Static-quark potential computed on $6^4$ lattices with $a\approx
0.4$\,fm using the Wilson action and the
improved action. The dotted line
is the standard infrared parameterization for the continuum potential,
$V(r)=Kr-\pi/12r + c$, adjusted to fit the on-axis values of the
potential. }
\label{V-fig}
\end{figure}
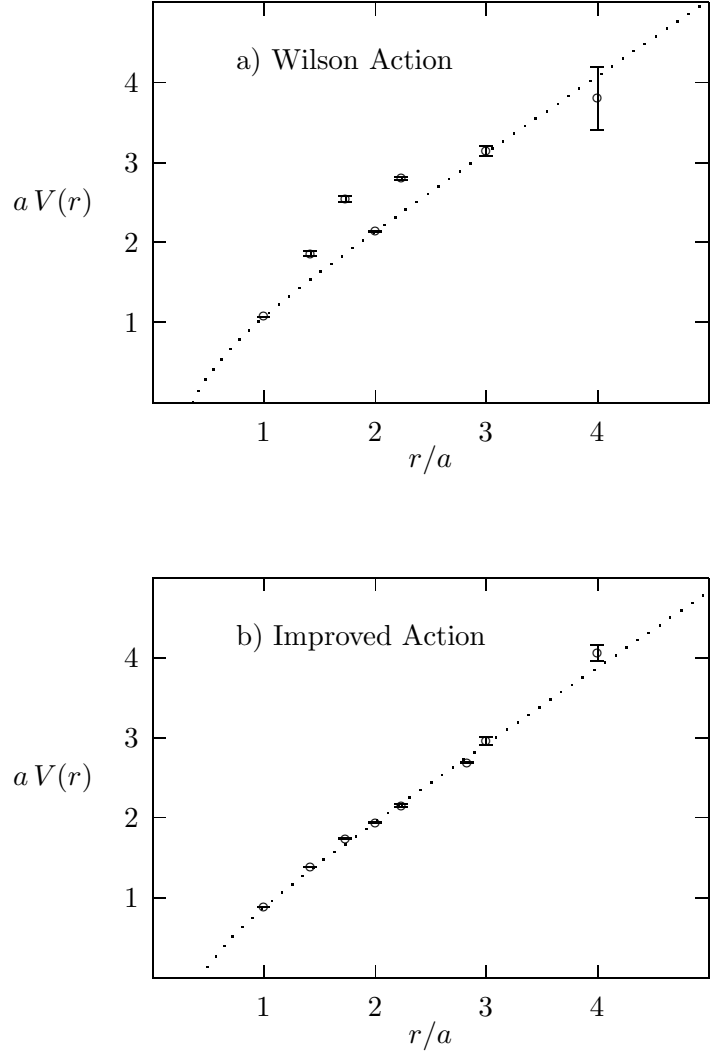

The effects of the $\order(a^2)$~improvements in the gluon action are
immediately apparent if one compares the static-quark
potential computed, using~\eq{static-V}, with the Wilson action~\eq{Wil-S}
and with the improved action~\eq{LW-S}\,\cite{alford95}.
Monte Carlo simulation results for
these potentials are plotted in Fig.~\ref{V-fig}, together with the
continuum potential. The Wilson action has errors as large as~40\%. These
are reduced to only 4--5\% when the improved action is used instead.

The dominant errors in the uncorrected simulation of~$V(r)$ reflect a
failure of rotational invariance, which is expected since the $a^2$~error in
the Wilson action~\eq{Wil-S} is neither Lorentz nor rotationally invariant.
The points at $r=a,2a,3a\ldots$ are for separations that are parallel to an
axis of the lattice, while the points at
$r=\sqrt2a,\sqrt3a\ldots$ are for diagonal separations between the static
quark and antiquark. In Table~\ref{dV-table}, I tabulate the error~$\Delta
V$ in the potential at $r=\sqrt3a$ for a variety gluon actions, with and
without tadpole improvement and one-loop radiative corrections.
As expected, the correction
term in the action is significantly underestimated without tadpole
improvement. The tadpole-improved action is very accurate both with
and without one-loop corrections, suggesting that $\order(a^2\alpha_s)$
corrections are comparable to those of $\order(a^4)$ and therefore
unimportant for most current applications.

\begin{table}
\caption{Error in the static quark potential at $V(\protect\sqrt 3a)$ for a
variety of gluon actions. The lattice spacing in each case is $a\approx
.4$\,fm; $K$ is the slope of the linear part of the static potential.}
\label{dV-table}
\begin{center}
\begin{tabular}{lc} \hline
\\
Action &  $\Delta V(\sqrt{3} a)/ K\sqrt{3} a$ \nl \hline
unimproved (Wilson) & .41 (2) \nl
\mbox{} \nl
tree-level improved, no tadpole improvement & .15 (1) \nl
one-loop improved, no tadpole improvement & .12 (2) \nl
\mbox{} \nl
tree-level improved, with tadpole improvement & .05 (1) \nl
one-loop improved, with tadpole improvement & .04 (1) \nl
\mbox{} \nl
twisted-rectangle correction, with tadpole improvement &
 .04 (2)\nl \hline
\end{tabular}
\end{center}
\end{table}

I also include in Table~\ref{dV-table} results obtained using the
twisted-rectangle action $S_{\rm trt}$ in~\eq{twisted-S}.  This action
gives a potential that is essentially identical to that obtained with the
other improved action. The $a^4$~errors are very different in the two
actions; their agreement suggests that $a^4$~errors are small. It also is
striking confirmation of the importance of tadpole improvement since the
$u_0$'s more than double the size of the correction term in the
twisted-rectangle action when $a= 0.4$\,fm.

Another test of our improved action is to compute physical quantities on
lattices with several different lattice spacings, looking for evidence of
finite-$a$ errors. These are consistently small. For example, the
temperature at which gluonic QCD has a deconfining phase transition can be
computed using the improved the action. One finds\,\cite{bliss96}:
\be
T_c = \cases{298\,(6)\,{\rm MeV} & \mbox{for $a\!=\!.32$\,fm}\cr
309(6)\,(6)\,{\rm MeV} & \mbox{for $a\!=\!.21$\,fm.}\cr
303\,(8)\,{\rm MeV} & \mbox{for $a\!=\!.16$\,fm.}\cr}
\ee
The results all agree to within statistical errors of a few percent.

\subsection{Anisotropic Lattices and Tuned Gluon Actions\protect\footnotemark}
\footnotetext{This section is based on work with M.\,Alford,
T.\,Klassen, C.\,Morningstar, M.\,Peardon and H.\,Trottier.}
As a final illustration of improved gluon actions with
tadpole-improved operators, I now show results of simulations employing
anisotropic lattices with  temporal lattice spacings~$a_t$ that are smaller
than the spatial lattice spacings~$a_s$. Such anisotropic lattices lead to
greatly improved signal-to-noise in Monte Carlo simulations of
hadrons; they are also useful for designing improved quark actions, as I
discuss later. On an anisotropic lattice, our rectangle-improved action
becomes
\begin{eqnarray} \label{Sd}
{ S} &=& - \beta \sum_{x,\,s>s^\prime}
\frac{a_t}{a_s} \,\left\{
\frac{5}{3}  \frac{P_{ss^\prime}}{u_s^4}
- \frac{1}{12} \frac{R_{ss^\prime}}{u_s^6}
- \frac{1}{12} \frac{R_{s^\prime s}}{u_s^6}
\right\} \\ \nonumber
&& - \beta \sum_{x,\,s}
\frac{a_s}{a_t} \,\left\{
\frac{5}{3}  \frac{P_{st}}{u_s^2 u_t^2}
- \frac{1}{12} \frac{R_{st}}{u_s^4 u_t^2}
- \frac{1}{12} \frac{R_{ts}}{u_t^4 u_s^2}
\right\},
\end{eqnarray}
where $s$ and $s^\prime$ are spatial directions.
The mean links~$u_t$ and~$u_s$ are different on anisotropic lattices. When
$a_t\le a_s/2$, $u_t$~is very close to its continuum value of one. Thus
we take
\begin{eqnarray}
  u_t &=& 1 \\
  u_s &=& \langle P_{ss^\prime} \rangle^{1/4}.
\end{eqnarray}
The same~$u_s$, to within a few percent, can be obtained
from the link expectation value in Coulomb gauge, which is the natural gauge
choice for anisotropic lattices with small~$a_t$'s.

The anisotropic simulation can be tested by comparing the static quark
potential computed from Wilson loops in different orientations: loops in
the $x$--$t$~plane, loops in the $x$--$y$~plane, and loops in the
$t$--$x$~plane (with $x$~playing the role of time). If the anisotropic
theory is tadpole-improved all three potentials agree (Fig.~\ref{aniso-V}).
Without tadpole-improvement the slopes disagree by 20\%
(Fig.~\ref{aniso-V-noti}).

\begin{figure}
\begin{center}
\setlength{\unitlength}{0.240900pt}
\ifx\plotpoint\undefined\newsavebox{\plotpoint}\fi
\sbox{\plotpoint}{\rule[-0.200pt]{0.400pt}{0.400pt}}%
\begin{picture}(1200,900)(0,0)
\font\gnuplot=cmr10 at 10pt
\gnuplot
\sbox{\plotpoint}{\rule[-0.200pt]{0.400pt}{0.400pt}}%
\put(220.0,113.0){\rule[-0.200pt]{220.664pt}{0.400pt}}
\put(220.0,113.0){\rule[-0.200pt]{0.400pt}{184.048pt}}
\put(220.0,266.0){\rule[-0.200pt]{4.818pt}{0.400pt}}
\put(198,266){\makebox(0,0)[r]{$1$}}
\put(1116.0,266.0){\rule[-0.200pt]{4.818pt}{0.400pt}}
\put(220.0,419.0){\rule[-0.200pt]{4.818pt}{0.400pt}}
\put(198,419){\makebox(0,0)[r]{$2$}}
\put(1116.0,419.0){\rule[-0.200pt]{4.818pt}{0.400pt}}
\put(220.0,571.0){\rule[-0.200pt]{4.818pt}{0.400pt}}
\put(198,571){\makebox(0,0)[r]{$3$}}
\put(1116.0,571.0){\rule[-0.200pt]{4.818pt}{0.400pt}}
\put(220.0,724.0){\rule[-0.200pt]{4.818pt}{0.400pt}}
\put(198,724){\makebox(0,0)[r]{$4$}}
\put(1116.0,724.0){\rule[-0.200pt]{4.818pt}{0.400pt}}
\put(403.0,113.0){\rule[-0.200pt]{0.400pt}{4.818pt}}
\put(403,68){\makebox(0,0){$1$}}
\put(403.0,857.0){\rule[-0.200pt]{0.400pt}{4.818pt}}
\put(586.0,113.0){\rule[-0.200pt]{0.400pt}{4.818pt}}
\put(586,68){\makebox(0,0){$2$}}
\put(586.0,857.0){\rule[-0.200pt]{0.400pt}{4.818pt}}
\put(770.0,113.0){\rule[-0.200pt]{0.400pt}{4.818pt}}
\put(770,68){\makebox(0,0){$3$}}
\put(770.0,857.0){\rule[-0.200pt]{0.400pt}{4.818pt}}
\put(953.0,113.0){\rule[-0.200pt]{0.400pt}{4.818pt}}
\put(953,68){\makebox(0,0){$4$}}
\put(953.0,857.0){\rule[-0.200pt]{0.400pt}{4.818pt}}
\put(220.0,113.0){\rule[-0.200pt]{220.664pt}{0.400pt}}
\put(1136.0,113.0){\rule[-0.200pt]{0.400pt}{184.048pt}}
\put(220.0,877.0){\rule[-0.200pt]{220.664pt}{0.400pt}}
\put(45,495){\makebox(0,0){$a_s\,V$}}
\put(678,23){\makebox(0,0){$r/a_s$}}
\put(220.0,113.0){\rule[-0.200pt]{0.400pt}{184.048pt}}
\put(449,801){\makebox(0,0)[r]{from $W_{xt}$}}
\put(493,801){\circle*{18}}
\put(403,245){\circle*{18}}
\put(479,311){\circle*{18}}
\put(537,355){\circle*{18}}
\put(586,376){\circle*{18}}
\put(630,410){\circle*{18}}
\put(738,478){\circle*{18}}
\put(770,511){\circle*{18}}
\put(953,645){\circle*{18}}
\put(471.0,801.0){\rule[-0.200pt]{15.899pt}{0.400pt}}
\put(471.0,791.0){\rule[-0.200pt]{0.400pt}{4.818pt}}
\put(537.0,791.0){\rule[-0.200pt]{0.400pt}{4.818pt}}
\put(403,245){\usebox{\plotpoint}}
\put(393.0,245.0){\rule[-0.200pt]{4.818pt}{0.400pt}}
\put(393.0,245.0){\rule[-0.200pt]{4.818pt}{0.400pt}}
\put(479.0,311.0){\usebox{\plotpoint}}
\put(469.0,311.0){\rule[-0.200pt]{4.818pt}{0.400pt}}
\put(469.0,312.0){\rule[-0.200pt]{4.818pt}{0.400pt}}
\put(537.0,354.0){\rule[-0.200pt]{0.400pt}{0.482pt}}
\put(527.0,354.0){\rule[-0.200pt]{4.818pt}{0.400pt}}
\put(527.0,356.0){\rule[-0.200pt]{4.818pt}{0.400pt}}
\put(586.0,375.0){\rule[-0.200pt]{0.400pt}{0.482pt}}
\put(576.0,375.0){\rule[-0.200pt]{4.818pt}{0.400pt}}
\put(576.0,377.0){\rule[-0.200pt]{4.818pt}{0.400pt}}
\put(630.0,409.0){\rule[-0.200pt]{0.400pt}{0.723pt}}
\put(620.0,409.0){\rule[-0.200pt]{4.818pt}{0.400pt}}
\put(620.0,412.0){\rule[-0.200pt]{4.818pt}{0.400pt}}
\put(738.0,474.0){\rule[-0.200pt]{0.400pt}{1.927pt}}
\put(728.0,474.0){\rule[-0.200pt]{4.818pt}{0.400pt}}
\put(728.0,482.0){\rule[-0.200pt]{4.818pt}{0.400pt}}
\put(770.0,508.0){\rule[-0.200pt]{0.400pt}{1.204pt}}
\put(760.0,508.0){\rule[-0.200pt]{4.818pt}{0.400pt}}
\put(760.0,513.0){\rule[-0.200pt]{4.818pt}{0.400pt}}
\put(953.0,636.0){\rule[-0.200pt]{0.400pt}{4.336pt}}
\put(943.0,636.0){\rule[-0.200pt]{4.818pt}{0.400pt}}
\put(943.0,654.0){\rule[-0.200pt]{4.818pt}{0.400pt}}
\put(449,756){\makebox(0,0)[r]{from $W_{tx}$}}
\put(493,756){\circle{24}}
\put(312,170){\circle{24}}
\put(403,245){\circle{24}}
\put(495,311){\circle{24}}
\put(586,373){\circle{24}}
\put(678,442){\circle{24}}
\put(770,506){\circle{24}}
\put(449,711){\makebox(0,0)[r]{from $W_{xy}$}}
\put(493,711){\raisebox{-.8pt}{\makebox(0,0){$\Box$}}}
\put(403,245){\raisebox{-.8pt}{\makebox(0,0){$\Box$}}}
\put(479,311){\raisebox{-.8pt}{\makebox(0,0){$\Box$}}}
\put(586,375){\raisebox{-.8pt}{\makebox(0,0){$\Box$}}}
\put(630,410){\raisebox{-.8pt}{\makebox(0,0){$\Box$}}}
\put(738,480){\raisebox{-.8pt}{\makebox(0,0){$\Box$}}}
\put(770,518){\raisebox{-.8pt}{\makebox(0,0){$\Box$}}}
\put(953,633){\raisebox{-.8pt}{\makebox(0,0){$\Box$}}}
\sbox{\plotpoint}{\rule[-0.500pt]{1.000pt}{1.000pt}}%
\put(449,666){\makebox(0,0)[r]{fit}}
\multiput(471,666)(20.756,0.000){4}{\usebox{\plotpoint}}
\put(537,666){\usebox{\plotpoint}}
\put(264.00,113.00){\usebox{\plotpoint}}
\put(275.46,130.25){\usebox{\plotpoint}}
\multiput(276,131)(12.453,16.604){0}{\usebox{\plotpoint}}
\put(288.23,146.59){\usebox{\plotpoint}}
\put(302.58,161.58){\usebox{\plotpoint}}
\multiput(303,162)(15.427,13.885){0}{\usebox{\plotpoint}}
\put(317.74,175.74){\usebox{\plotpoint}}
\multiput(322,180)(15.513,13.789){0}{\usebox{\plotpoint}}
\put(333.13,189.66){\usebox{\plotpoint}}
\put(349.41,202.53){\usebox{\plotpoint}}
\multiput(350,203)(16.383,12.743){0}{\usebox{\plotpoint}}
\put(365.43,215.71){\usebox{\plotpoint}}
\multiput(368,218)(16.383,12.743){0}{\usebox{\plotpoint}}
\put(381.84,228.39){\usebox{\plotpoint}}
\multiput(387,232)(16.383,12.743){0}{\usebox{\plotpoint}}
\put(398.41,240.88){\usebox{\plotpoint}}
\multiput(405,246)(16.383,12.743){0}{\usebox{\plotpoint}}
\put(414.83,253.58){\usebox{\plotpoint}}
\put(431.54,265.87){\usebox{\plotpoint}}
\multiput(433,267)(17.270,11.513){0}{\usebox{\plotpoint}}
\put(448.39,277.97){\usebox{\plotpoint}}
\multiput(451,280)(17.004,11.902){0}{\usebox{\plotpoint}}
\put(465.14,290.22){\usebox{\plotpoint}}
\multiput(470,294)(17.270,11.513){0}{\usebox{\plotpoint}}
\put(481.98,302.32){\usebox{\plotpoint}}
\multiput(488,307)(17.798,10.679){0}{\usebox{\plotpoint}}
\put(499.16,313.90){\usebox{\plotpoint}}
\put(515.54,326.65){\usebox{\plotpoint}}
\multiput(516,327)(17.270,11.513){0}{\usebox{\plotpoint}}
\put(532.67,338.37){\usebox{\plotpoint}}
\multiput(535,340)(17.270,11.513){0}{\usebox{\plotpoint}}
\put(549.60,350.36){\usebox{\plotpoint}}
\multiput(553,353)(16.383,12.743){0}{\usebox{\plotpoint}}
\put(566.33,362.60){\usebox{\plotpoint}}
\multiput(572,366)(16.383,12.743){0}{\usebox{\plotpoint}}
\put(583.28,374.52){\usebox{\plotpoint}}
\multiput(590,379)(16.383,12.743){0}{\usebox{\plotpoint}}
\put(600.09,386.66){\usebox{\plotpoint}}
\put(617.18,398.37){\usebox{\plotpoint}}
\multiput(618,399)(17.270,11.513){0}{\usebox{\plotpoint}}
\put(634.03,410.47){\usebox{\plotpoint}}
\multiput(636,412)(17.798,10.679){0}{\usebox{\plotpoint}}
\put(651.21,422.05){\usebox{\plotpoint}}
\multiput(655,425)(17.270,11.513){0}{\usebox{\plotpoint}}
\put(668.05,434.15){\usebox{\plotpoint}}
\multiput(673,438)(17.798,10.679){0}{\usebox{\plotpoint}}
\put(685.23,445.74){\usebox{\plotpoint}}
\multiput(692,451)(17.270,11.513){0}{\usebox{\plotpoint}}
\put(702.08,457.84){\usebox{\plotpoint}}
\put(719.19,469.51){\usebox{\plotpoint}}
\multiput(720,470)(16.383,12.743){0}{\usebox{\plotpoint}}
\put(736.00,481.66){\usebox{\plotpoint}}
\multiput(738,483)(17.270,11.513){0}{\usebox{\plotpoint}}
\put(753.17,493.32){\usebox{\plotpoint}}
\multiput(757,496)(17.270,11.513){0}{\usebox{\plotpoint}}
\put(770.15,505.23){\usebox{\plotpoint}}
\multiput(775,509)(17.270,11.513){0}{\usebox{\plotpoint}}
\put(787.11,517.18){\usebox{\plotpoint}}
\multiput(794,522)(17.270,11.513){0}{\usebox{\plotpoint}}
\put(804.21,528.94){\usebox{\plotpoint}}
\multiput(812,535)(17.270,11.513){0}{\usebox{\plotpoint}}
\put(821.06,541.04){\usebox{\plotpoint}}
\put(838.23,552.63){\usebox{\plotpoint}}
\multiput(840,554)(17.270,11.513){0}{\usebox{\plotpoint}}
\put(855.08,564.73){\usebox{\plotpoint}}
\multiput(858,567)(17.798,10.679){0}{\usebox{\plotpoint}}
\put(872.26,576.31){\usebox{\plotpoint}}
\multiput(877,580)(17.270,11.513){0}{\usebox{\plotpoint}}
\put(889.10,588.41){\usebox{\plotpoint}}
\multiput(895,593)(17.798,10.679){0}{\usebox{\plotpoint}}
\put(906.35,599.90){\usebox{\plotpoint}}
\multiput(914,605)(16.383,12.743){0}{\usebox{\plotpoint}}
\put(923.13,612.09){\usebox{\plotpoint}}
\put(940.27,623.79){\usebox{\plotpoint}}
\multiput(942,625)(17.270,11.513){0}{\usebox{\plotpoint}}
\put(957.18,635.81){\usebox{\plotpoint}}
\multiput(960,638)(17.270,11.513){0}{\usebox{\plotpoint}}
\put(974.46,647.28){\usebox{\plotpoint}}
\multiput(979,650)(16.383,12.743){0}{\usebox{\plotpoint}}
\put(991.38,659.25){\usebox{\plotpoint}}
\multiput(997,663)(16.383,12.743){0}{\usebox{\plotpoint}}
\put(1008.23,671.34){\usebox{\plotpoint}}
\multiput(1016,676)(17.270,11.513){0}{\usebox{\plotpoint}}
\put(1025.69,682.54){\usebox{\plotpoint}}
\put(1042.51,694.67){\usebox{\plotpoint}}
\multiput(1043,695)(17.004,11.902){0}{\usebox{\plotpoint}}
\put(1059.62,706.41){\usebox{\plotpoint}}
\multiput(1062,708)(16.383,12.743){0}{\usebox{\plotpoint}}
\put(1076.41,718.60){\usebox{\plotpoint}}
\multiput(1080,721)(17.798,10.679){0}{\usebox{\plotpoint}}
\put(1093.77,729.93){\usebox{\plotpoint}}
\multiput(1099,734)(17.270,11.513){0}{\usebox{\plotpoint}}
\put(1110.61,742.03){\usebox{\plotpoint}}
\multiput(1117,747)(17.798,10.679){0}{\usebox{\plotpoint}}
\put(1127.83,753.56){\usebox{\plotpoint}}
\put(1136,759){\usebox{\plotpoint}}
\end{picture}
\end{center}
\caption{The static-quark potential computed on an anisotropic lattice
in different orientations. Results are shown for $a_t/a_s$ equal $1/2$.
Except as shown, analysis
errors are smaller than the plot symbols. The fit
in each plot is to the open circles; the fitting function is
$V(r)=Kr-b/r+c$.}\label{aniso-V}
\end{figure}
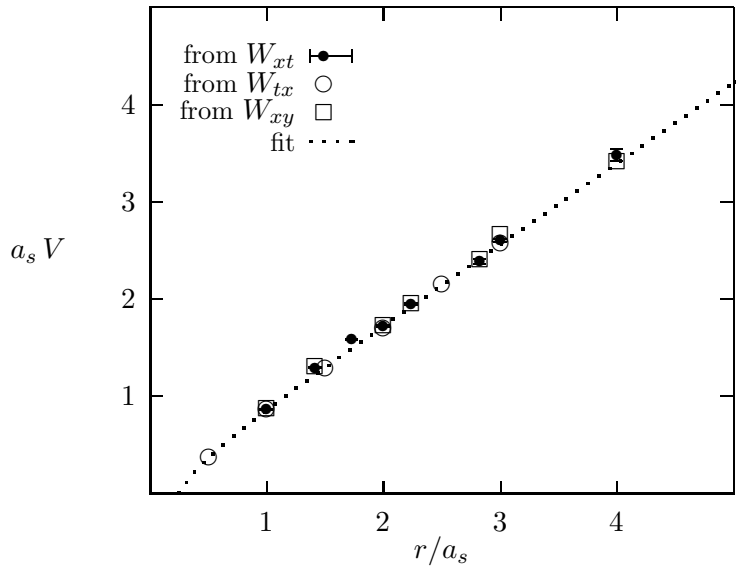
\begin{figure}
\begin{center}
\setlength{\unitlength}{0.240900pt}
\ifx\plotpoint\undefined\newsavebox{\plotpoint}\fi
\sbox{\plotpoint}{\rule[-0.200pt]{0.400pt}{0.400pt}}%
\begin{picture}(1200,900)(0,0)
\font\gnuplot=cmr10 at 10pt
\gnuplot
\sbox{\plotpoint}{\rule[-0.200pt]{0.400pt}{0.400pt}}%
\put(220.0,113.0){\rule[-0.200pt]{220.664pt}{0.400pt}}
\put(220.0,113.0){\rule[-0.200pt]{0.400pt}{184.048pt}}
\put(220.0,266.0){\rule[-0.200pt]{4.818pt}{0.400pt}}
\put(198,266){\makebox(0,0)[r]{$1$}}
\put(1116.0,266.0){\rule[-0.200pt]{4.818pt}{0.400pt}}
\put(220.0,419.0){\rule[-0.200pt]{4.818pt}{0.400pt}}
\put(198,419){\makebox(0,0)[r]{$2$}}
\put(1116.0,419.0){\rule[-0.200pt]{4.818pt}{0.400pt}}
\put(220.0,571.0){\rule[-0.200pt]{4.818pt}{0.400pt}}
\put(198,571){\makebox(0,0)[r]{$3$}}
\put(1116.0,571.0){\rule[-0.200pt]{4.818pt}{0.400pt}}
\put(220.0,724.0){\rule[-0.200pt]{4.818pt}{0.400pt}}
\put(198,724){\makebox(0,0)[r]{$4$}}
\put(1116.0,724.0){\rule[-0.200pt]{4.818pt}{0.400pt}}
\put(403.0,113.0){\rule[-0.200pt]{0.400pt}{4.818pt}}
\put(403,68){\makebox(0,0){$1$}}
\put(403.0,857.0){\rule[-0.200pt]{0.400pt}{4.818pt}}
\put(586.0,113.0){\rule[-0.200pt]{0.400pt}{4.818pt}}
\put(586,68){\makebox(0,0){$2$}}
\put(586.0,857.0){\rule[-0.200pt]{0.400pt}{4.818pt}}
\put(770.0,113.0){\rule[-0.200pt]{0.400pt}{4.818pt}}
\put(770,68){\makebox(0,0){$3$}}
\put(770.0,857.0){\rule[-0.200pt]{0.400pt}{4.818pt}}
\put(953.0,113.0){\rule[-0.200pt]{0.400pt}{4.818pt}}
\put(953,68){\makebox(0,0){$4$}}
\put(953.0,857.0){\rule[-0.200pt]{0.400pt}{4.818pt}}
\put(220.0,113.0){\rule[-0.200pt]{220.664pt}{0.400pt}}
\put(1136.0,113.0){\rule[-0.200pt]{0.400pt}{184.048pt}}
\put(220.0,877.0){\rule[-0.200pt]{220.664pt}{0.400pt}}
\put(45,495){\makebox(0,0){$a_s\,V$}}
\put(678,23){\makebox(0,0){$r/a_s$}}
\put(220.0,113.0){\rule[-0.200pt]{0.400pt}{184.048pt}}
\put(449,801){\makebox(0,0)[r]{from $W_{xt}$}}
\put(493,801){\circle*{18}}
\put(403,230){\circle*{18}}
\put(479,295){\circle*{18}}
\put(537,337){\circle*{18}}
\put(586,343){\circle*{18}}
\put(630,383){\circle*{18}}
\put(738,452){\circle*{18}}
\put(770,455){\circle*{18}}
\put(953,590){\circle*{18}}
\put(471.0,801.0){\rule[-0.200pt]{15.899pt}{0.400pt}}
\put(471.0,791.0){\rule[-0.200pt]{0.400pt}{4.818pt}}
\put(537.0,791.0){\rule[-0.200pt]{0.400pt}{4.818pt}}
\put(403.0,230.0){\usebox{\plotpoint}}
\put(393.0,230.0){\rule[-0.200pt]{4.818pt}{0.400pt}}
\put(393.0,231.0){\rule[-0.200pt]{4.818pt}{0.400pt}}
\put(479.0,294.0){\usebox{\plotpoint}}
\put(469.0,294.0){\rule[-0.200pt]{4.818pt}{0.400pt}}
\put(469.0,295.0){\rule[-0.200pt]{4.818pt}{0.400pt}}
\put(537.0,335.0){\rule[-0.200pt]{0.400pt}{0.964pt}}
\put(527.0,335.0){\rule[-0.200pt]{4.818pt}{0.400pt}}
\put(527.0,339.0){\rule[-0.200pt]{4.818pt}{0.400pt}}
\put(586.0,340.0){\rule[-0.200pt]{0.400pt}{1.204pt}}
\put(576.0,340.0){\rule[-0.200pt]{4.818pt}{0.400pt}}
\put(576.0,345.0){\rule[-0.200pt]{4.818pt}{0.400pt}}
\put(630.0,380.0){\rule[-0.200pt]{0.400pt}{1.686pt}}
\put(620.0,380.0){\rule[-0.200pt]{4.818pt}{0.400pt}}
\put(620.0,387.0){\rule[-0.200pt]{4.818pt}{0.400pt}}
\put(738.0,446.0){\rule[-0.200pt]{0.400pt}{2.891pt}}
\put(728.0,446.0){\rule[-0.200pt]{4.818pt}{0.400pt}}
\put(728.0,458.0){\rule[-0.200pt]{4.818pt}{0.400pt}}
\put(770.0,449.0){\rule[-0.200pt]{0.400pt}{2.891pt}}
\put(760.0,449.0){\rule[-0.200pt]{4.818pt}{0.400pt}}
\put(760.0,461.0){\rule[-0.200pt]{4.818pt}{0.400pt}}
\put(953.0,581.0){\rule[-0.200pt]{0.400pt}{4.336pt}}
\put(943.0,581.0){\rule[-0.200pt]{4.818pt}{0.400pt}}
\put(943.0,599.0){\rule[-0.200pt]{4.818pt}{0.400pt}}
\put(449,756){\makebox(0,0)[r]{from $W_{tx}$}}
\put(493,756){\circle{24}}
\put(312,165){\circle{24}}
\put(403,230){\circle{24}}
\put(495,291){\circle{24}}
\put(586,351){\circle{24}}
\put(678,403){\circle{24}}
\put(770,461){\circle{24}}
\put(449,711){\makebox(0,0)[r]{from $W_{xy}$}}
\put(493,711){\raisebox{-.8pt}{\makebox(0,0){$\Box$}}}
\put(403,230){\raisebox{-.8pt}{\makebox(0,0){$\Box$}}}
\put(479,308){\raisebox{-.8pt}{\makebox(0,0){$\Box$}}}
\put(586,369){\raisebox{-.8pt}{\makebox(0,0){$\Box$}}}
\put(630,427){\raisebox{-.8pt}{\makebox(0,0){$\Box$}}}
\put(738,510){\raisebox{-.8pt}{\makebox(0,0){$\Box$}}}
\put(770,525){\raisebox{-.8pt}{\makebox(0,0){$\Box$}}}
\sbox{\plotpoint}{\rule[-0.500pt]{1.000pt}{1.000pt}}%
\put(449,666){\makebox(0,0)[r]{fit}}
\multiput(471,666)(20.756,0.000){4}{\usebox{\plotpoint}}
\put(537,666){\usebox{\plotpoint}}
\put(261.00,113.00){\usebox{\plotpoint}}
\put(273.62,129.38){\usebox{\plotpoint}}
\multiput(276,132)(14.676,14.676){0}{\usebox{\plotpoint}}
\put(288.17,144.17){\usebox{\plotpoint}}
\multiput(294,150)(15.513,13.789){0}{\usebox{\plotpoint}}
\put(303.37,158.29){\usebox{\plotpoint}}
\put(319.64,171.17){\usebox{\plotpoint}}
\multiput(322,173)(16.383,12.743){0}{\usebox{\plotpoint}}
\put(336.03,183.91){\usebox{\plotpoint}}
\multiput(340,187)(17.798,10.679){0}{\usebox{\plotpoint}}
\put(353.21,195.49){\usebox{\plotpoint}}
\multiput(359,200)(17.270,11.513){0}{\usebox{\plotpoint}}
\put(370.05,207.60){\usebox{\plotpoint}}
\multiput(377,213)(17.798,10.679){0}{\usebox{\plotpoint}}
\put(387.23,219.18){\usebox{\plotpoint}}
\put(404.02,231.35){\usebox{\plotpoint}}
\multiput(405,232)(17.270,11.513){0}{\usebox{\plotpoint}}
\put(421.52,242.51){\usebox{\plotpoint}}
\multiput(424,244)(17.270,11.513){0}{\usebox{\plotpoint}}
\put(438.86,253.91){\usebox{\plotpoint}}
\multiput(442,256)(17.270,11.513){0}{\usebox{\plotpoint}}
\put(456.05,265.54){\usebox{\plotpoint}}
\multiput(461,269)(17.270,11.513){0}{\usebox{\plotpoint}}
\put(473.24,277.16){\usebox{\plotpoint}}
\multiput(479,281)(17.270,11.513){0}{\usebox{\plotpoint}}
\put(490.59,288.55){\usebox{\plotpoint}}
\multiput(498,293)(17.270,11.513){0}{\usebox{\plotpoint}}
\put(508.08,299.72){\usebox{\plotpoint}}
\multiput(516,305)(17.270,11.513){0}{\usebox{\plotpoint}}
\put(525.36,311.22){\usebox{\plotpoint}}
\put(542.92,322.28){\usebox{\plotpoint}}
\multiput(544,323)(18.144,10.080){0}{\usebox{\plotpoint}}
\put(560.62,333.08){\usebox{\plotpoint}}
\multiput(562,334)(17.798,10.679){0}{\usebox{\plotpoint}}
\put(578.18,344.12){\usebox{\plotpoint}}
\multiput(581,346)(17.270,11.513){0}{\usebox{\plotpoint}}
\put(595.45,355.64){\usebox{\plotpoint}}
\multiput(599,358)(17.798,10.679){0}{\usebox{\plotpoint}}
\put(613.02,366.68){\usebox{\plotpoint}}
\multiput(618,370)(17.270,11.513){0}{\usebox{\plotpoint}}
\put(630.29,378.19){\usebox{\plotpoint}}
\multiput(636,382)(17.798,10.679){0}{\usebox{\plotpoint}}
\put(647.86,389.24){\usebox{\plotpoint}}
\multiput(655,394)(18.144,10.080){0}{\usebox{\plotpoint}}
\put(665.56,400.04){\usebox{\plotpoint}}
\multiput(673,405)(17.798,10.679){0}{\usebox{\plotpoint}}
\put(683.13,411.08){\usebox{\plotpoint}}
\put(700.40,422.60){\usebox{\plotpoint}}
\multiput(701,423)(17.270,11.513){0}{\usebox{\plotpoint}}
\put(717.90,433.74){\usebox{\plotpoint}}
\multiput(720,435)(17.270,11.513){0}{\usebox{\plotpoint}}
\put(735.23,445.15){\usebox{\plotpoint}}
\multiput(738,447)(18.144,10.080){0}{\usebox{\plotpoint}}
\put(753.12,455.67){\usebox{\plotpoint}}
\multiput(757,458)(17.270,11.513){0}{\usebox{\plotpoint}}
\put(770.50,467.00){\usebox{\plotpoint}}
\multiput(775,470)(17.270,11.513){0}{\usebox{\plotpoint}}
\put(787.89,478.33){\usebox{\plotpoint}}
\multiput(794,482)(17.270,11.513){0}{\usebox{\plotpoint}}
\put(805.34,489.56){\usebox{\plotpoint}}
\multiput(812,494)(18.144,10.080){0}{\usebox{\plotpoint}}
\put(823.10,500.26){\usebox{\plotpoint}}
\multiput(831,505)(17.270,11.513){0}{\usebox{\plotpoint}}
\put(840.61,511.40){\usebox{\plotpoint}}
\put(857.88,522.92){\usebox{\plotpoint}}
\multiput(858,523)(17.798,10.679){0}{\usebox{\plotpoint}}
\put(875.44,533.96){\usebox{\plotpoint}}
\multiput(877,535)(18.144,10.080){0}{\usebox{\plotpoint}}
\put(893.15,544.76){\usebox{\plotpoint}}
\multiput(895,546)(17.798,10.679){0}{\usebox{\plotpoint}}
\put(910.71,555.81){\usebox{\plotpoint}}
\multiput(914,558)(17.270,11.513){0}{\usebox{\plotpoint}}
\put(927.98,567.32){\usebox{\plotpoint}}
\multiput(932,570)(17.798,10.679){0}{\usebox{\plotpoint}}
\put(945.73,578.07){\usebox{\plotpoint}}
\multiput(951,581)(17.270,11.513){0}{\usebox{\plotpoint}}
\put(963.25,589.17){\usebox{\plotpoint}}
\multiput(969,593)(17.798,10.679){0}{\usebox{\plotpoint}}
\put(980.82,600.21){\usebox{\plotpoint}}
\multiput(988,605)(17.270,11.513){0}{\usebox{\plotpoint}}
\put(998.14,611.63){\usebox{\plotpoint}}
\multiput(1006,616)(17.798,10.679){0}{\usebox{\plotpoint}}
\put(1016.09,622.06){\usebox{\plotpoint}}
\put(1033.36,633.57){\usebox{\plotpoint}}
\multiput(1034,634)(17.270,11.513){0}{\usebox{\plotpoint}}
\put(1050.86,644.72){\usebox{\plotpoint}}
\multiput(1053,646)(17.270,11.513){0}{\usebox{\plotpoint}}
\put(1068.51,655.61){\usebox{\plotpoint}}
\multiput(1071,657)(17.270,11.513){0}{\usebox{\plotpoint}}
\put(1086.08,666.65){\usebox{\plotpoint}}
\multiput(1090,669)(17.270,11.513){0}{\usebox{\plotpoint}}
\put(1103.46,677.97){\usebox{\plotpoint}}
\multiput(1108,681)(17.270,11.513){0}{\usebox{\plotpoint}}
\put(1121.01,689.01){\usebox{\plotpoint}}
\multiput(1127,692)(17.270,11.513){0}{\usebox{\plotpoint}}
\put(1136,698){\usebox{\plotpoint}}
\end{picture}
\end{center}
\caption{The static-quark potential computed on an anisotropic lattice
in different orientations, as in the previous figure.  The action used
here was not tadpole
improved. The input value for $a_t/a_s$ is $1/2$.}
\label{aniso-V-noti}
\end{figure}
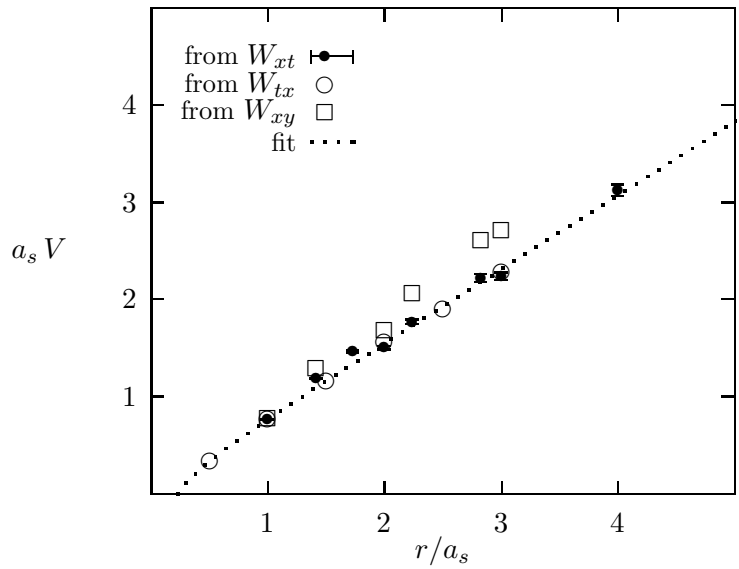

The coefficients of the correction terms in the gluon action for
anisotropic lattices are known only to leading order
in~$\alpha_s$. While it is clear from the simulations
that tadpole improvement captures the
bulk of the renormalizations, these same simulations provide a handle
on further radiative corrections. This is because the leading
corrections in the gluon action affect the extent to which continuum
symmetries, like rotation invariance, are restored. We can use
symmetry restoration as a criterion for tuning the couplings beyond
their tadpole-improved tree-level values. Thus, for example, the
static potential at $\rv=(a_s,2a_s,2a_s)$ should agree exactly with
that at $\rv=(3a_s,0,0)$. One might tune
the value of the spatial mean link~$u_s$ used in the action
until this is the case. Similarly one would like
the anistropy~$a_t/a_s$ built into the action to agree with the
renormalized anisotropy coming out of the simulations. The
ratio~$u_t/u_s$ can easily be tuned to make the two agree exactly.
Such tuning does not capture all of the $\order(\alpha_s)$ corrections
except in the simplest of situations. However it may be a useful
intermediate step between tree-level tadpole improvement and a full
perturbative analysis of the radiative corrections.

\begin{exercise}
One defect of our improved gluon action is that the gluon spectrum has
ghosts. These are caused by the
$R_{ts}$~terms in the action, which extend two steps in the time
direction. (Numerical ghosts, like these, are discussed in greater detail
in the section on light quarks.)
These gluon ghosts, being at high energies (of order $2.6/a_t$), should have
little effect on simulation results, particularly for smaller~$a_t$'s; this
is confirmed by simulations. Nevertheless one might wish to remove them
completely. This is easily done,  when $a_t$~is  small compared with~$a_s$,
by dropping the
$R_{ts}$ terms in the action. This introduces new errors of order $a_t^2$ but
these will be negligible relative to the $a_s^4$~errors provided $a_t$~is
small enough. Modify the $st$~terms
in the action  so that all
$a_s^2$ errors are still canceled in the absence of the $R_{ts}$
contribution.
\end{exercise}

\subsection{Summary}
In this section I have outlined how one designs  actions for
simulating gluon dynamics on coarse lattices. The simulation results show
that accurate simulations are possible even on lattices as coarse as
$a\!=\!.4$\,fm using very simple actions. Tadpole improvement is essential,
but one-loop radiative corrections to the action are not too important.
Rotational invariance of the potential and, for anisotropic lattices,
space-time interchange symmetry provide sensitive tests of
improved gluon actions, and can be used to tune the leading correction
terms.

The coarse lattices make simulations very fast.
Most of the simulation results shown here were obtained  using an IBM
RS6000/250 desktop workstation, which is powered by a personal-computer CPU
(66MHz PowerPC). Any of these results could be generated easily on a
high-end personal computer.
\section{Light Quarks on Coarse Lattices\protect\footnotemark}
\footnotetext{This section is based on work with M.\,Alford and
T.\,Klassen.}
In this section I develop techniques for designing very accurate
discretizations of the Dirac equation for quarks in a gluonic field. I begin
by reviewing various resolutions of the ``doubling problem'' which plagues
even the simplest discretization. I then discuss two currently popular
discretizations, and finally new discretizations that are significantly more
accurate.

\subsection{Naive Discretization and Doubling}
The euclidean Dirac lagrangian in the continuum is
\be
\Lag = \psib\left(\D\cdot\gamma+m\right)\psi
\ee
where
\be
\left\{\gamma_\mu,\gamma_\nu\right\} = 2\,\delta_{\mu\nu}
\ee
and I take
\be
\gamma_t = \left(\begin{array}{rr}1&0\\0& -1\end{array}\right)
\qquad
\gamma_s = \left(\begin{array}{cc}0&\sigma_s\\\sigma_s&0\end{array}\right)
\qquad
\gamma_5 \equiv \gamma_t\gamma_x\gamma_y\gamma_z = i
\left(\begin{array}{rr}0&1\\  -1&0 \end{array}\right)
\ee
where, again, $s$ and $t$ as subscripts signify spatial and temporal
indices respectively.
The obvious discretization is
\be \label{naive-dirac}
\Lag_\lat = \psib\left(\Delta\cdot\gamma + m +\order(a^2)\right)\psi
\ee
where now $\Delta_\mu$ is a (tadpole-improved) gauge-covariant derivative:
\be
\Delta_\mu\psi(x) \equiv
\frac{\U\mu(x)\,\psi(x+a\hat\mu) - \Udag\mu(x-a\hat\mu)\,\psi(x-a\hat\mu)}%
     {2\,a\,u_0}.
\ee
The other finite-difference operators, $\lder2_\mu$, $\lder3_\mu$\,\ldots, can
all be made gauge-covariant by adding~$\U\mu$'s as in~$\Delta_\mu$; in what
follows I will always use these symbols to refer to the gauge-covariant
differences.

Unfortunately this simple lattice Dirac action has a very serious pathology.
To see this consider the $\pv\!=\!0$ spin-up solution
of the corresponding Dirac equation without gluon fields:
\be
\psi_\uparrow (t)=
\e^{-Et} \left[\begin{array}{c}1\\0\\0\\0\end{array}\right]
\ee
(There is no~$\I$ in the exponent because~$t$ is euclidean; $E$~is the
minkowski energy.) Substituting into the Dirac equation,
\be
\left(\Delta\cdot\gamma+m\right)\,\psi_\uparrow(t) = 0,
\ee
we obtain
\be
\left(\Delta_t+m\right)\e^{-Et} = 0 \qquad \Longrightarrow \qquad
\frac{\e^{-aE} - \e^{aE}}{2\,a} + m = 0.
\ee
This equation is quadratic in $z\!\equiv\!\exp(aE)$ and therefore it has two
solutions for~$E$. We only want one! The two solutions are
\be
E = \cases{ m +\order(a^2) & \mbox{(normal)}\cr
            -m +\I\pi/a + \order(a^2) & \mbox{(doubler)}}.
\ee
The first is the solution we want; the second is known as a ``doubler.''

The possibility of a second solution for the energy is obvious from
inspection of the lattice Dirac equation. The problem lies in the
$\Delta_t\psi$ term. This links
$\psi$'s two lattice spacings apart which means that the dispersion
relation is second order in~$z\!\equiv\!\exp(aE)$. In general one expects
$n$~roots if the $\psi$'s are spread over $n$~lattice spacings\,---\,a
potential problem when we go to improve the discretization.
The doubler cannot be
ignored. It is related to the normal solution by a symmetry of the gluon-free
lattice theory: if
$\psi(\xv,t)$~is a solution of the lattice Dirac equation then so is
\be
\tilde\psi(\xv,t) \equiv i\gamma_5\gamma_t\,(-1)^{t/a}\,\psi(\xv,t).
\ee
This symmetry operation turns the normal solution into the doubler solution.
Consequently the doubler acts as an extra flavor of quark, which is
degenerate with the normal quark (when there are no gauge fields).
Furthermore, in euclidean space there is nothing unique about the time
direction. Thus there is an analogous doubling symmetry for each direction,
and successive transformations involving two or more directions also are
symmetries. This implies that there are fifteen doublers in all.

The root of the doubler problem lies in the finite-difference approximation
for  first-order derivatives. Fourier transforming $\partial$ and $\Delta$
we obtain
\bearray
\partial &\to& \I\, p \\
\Delta &\to& \I\, \sin(pa)/a.
\eearray
These agree well for low momenta, but, as $p$~approaches the ultraviolet
cutoff~$\pi/a$, the transform of~$\Delta$ vanishes rather than becoming
large. As a consequence, for example,  high-momentum states can have very
low or even vanishing energies. This is disastrous. Note that the same
problem does not arise for second-order derivatives:
\bearray
\partial^2 &\to& -p^2 \\
\lder2 &\to&  -\left(\frac{2}{a}\,\sin(pa/2)\right)^2.
\eearray
Although the transform of~$\lder2$ becomes quite different from~$p^2$ at
large~$p$, it nevertheless remains large, going to its maximum of~$4/a^2$ at
the ultraviolet cutoff. Second-order (and higher even-order) derivatives do
not cause doubling problems.

I will discuss two approaches to the doubling problem. One involves adding
operators to the quark action that are almost redundant but
remove the doublers or drive
them to high energies. The other involves ``staggering'' the quark degrees
of freedom on the lattice.

\begin{exercise} Show that gauge transformation
\be
\psi(x) \to \Omega(x)\,\psi(x)\qquad
\U\mu(x)\to\Omega(x)\,\U\mu(x)\,\Omega(x+a\hat\mu)^\dagger.
\ee
implies that
\be
\Delta_\mu\psi(x) \to \Omega(x)\,\Delta_\mu\psi(x).
\ee
Also show that (setting $u_0\!=\!1$)
\be
\Delta_\mu = \frac{\e^{a\D_\mu} - \e^{-a\D_\mu}}{2\,a} .
\ee
(Hint: Why can't there be $\F\mu\nu$'s on the right-hand side?) This
relation is crucial to whot follows; find a different definition of
the the covariant difference operator~$\Delta$ for which this relation
is incorrect in~$\order(a)$.
\end{exercise}

\subsection{Field Transformations to Remove Doublers}
The first approach to removing the doublers is to introduce
into the action an operator that is redundant up to~$\order(a^2)$ and
has higher order non-redundant parts that drive the
doubler to high energies. A redundant operator remember, is
one that can be eliminated from the action by a change
of variables on the quark fields in the path integral,
and hence has no effect on any physical observables.
For example, consider the field transformation from
the continuum field~$\psi_\ct$ to a new field~$\psi$ where
\bearray
\psi_\ct &=&\left[1
-\frac{r\,a}{4}\,(\Delta\cdot\gamma-m_\ct)\right]\,\psi\nl
&\equiv&\Omega\psi,
\eearray
and
\be
\psib_\ct = \psib\,\Omega
\ee
With this transformation our naive lattice action becomes
\bearray
\lefteqn{\psib_\ct\left(\Delta\cdot\gamma+m_c\right)\psi_\ct =
\psib\,\Omega\,\left(\Delta\cdot\gamma+m_c\right)\,\Omega\,\psi}
\qquad\qquad&&\\
&=& \psib \left[ \Delta\cdot\gamma+m
-\frac{r\,a}{2}\left(\Delta\cdot\Delta + \frac{\sigma\cdot gF}{2}\right)
\right] \psi + \order(a^2)
\eearray
where $m\equiv m_\ct+ram_\ct^2/2$. In deriving the final formula we used the
lattice version of the identity
\be
\left(\D\cdot\gamma\right)^2 = \D\cdot\D + \frac{\sigma\cdot gF}{2}
\ee
where
\be
\sigma_{\mu\nu} \equiv -\frac{i}{2}\left[\gamma_\mu,\gamma_\nu\right].
\ee
The new $\order(a)$~term in the transformed lagrangian does not
introduce new $\order(a)$~errors because it results from a field
transformation; it is redundant in the sense discussed above for gluon
operators. In its present form this term
also has no effect on the doubling problem. The trick is to replace the
operator~$\Delta\cdot\Delta$, which vanishes at the ultraviolet cutoff, with
$\lder2$, which is the same up to $\order(a^2)$~errors in the infrared but
large at the ultraviolet cutoff. The $\lder2$ pushes the doublers off to
high energies, of order~$1/a$, where they are harmless. Thus an
$\order(a)$-accurate quark action that is doubler free is
\bearray
\Lag_{\rm SW} &=& \psib \left[ \Delta\cdot\gamma+m
-\frac{r\,a}{2}\left(\lder2 + \frac{\sigma\cdot gF}{2}\right)
\right] \psi \\
&=& \psib_\ct\left(\D\cdot\gamma+m_c\right)\psi_\ct + \order(a^2),
\eearray
where, again,
\be
\psi_\ct = \Omega\,\psi \qquad m = m_\ct+ram_\ct^2/2.
\ee
This is the Sheikholeslami-Wohlert (SW) quark action\,\cite{sw85}.
It becomes
Wilson's original action if the $\sigma\cdot gF$~term is dropped
(resulting in $\order(a^1)$~errors).

We can check that the doublers are removed by again substituting the
$\pv\!=\!0$~solution into the corresponding Dirac equation. We obtain
\bearray
\lefteqn{\left(\Delta_t+m-\frac{ra}{2}\,\lder{2}\right)\e^{-Et} = 0}
\qquad \quad &&  \nl
\qquad \qquad &&
\Longrightarrow \quad
(1-r)\,\e^{-aE} - (1+r)\,\e^{aE} +2\,r+2\,a\,m = 0.
\eearray
which has roots
\be
E = \cases{ m_c +\order(a^2) & \mbox{(normal)}\cr
            \left(1/a\right)\,\ln\left[(r-1)/(r+1)\right] + \cdots &
\mbox{(ghost)}}.
\ee
The doubler has been replaced by a high-energy ``ghost'' particle, as
desired. The ghost goes away completely for~$r\!=\!1$.

\begin{exercise} Create a lattice version of $\F\mu\nu$ for use in the
SW~quark action. Build your operator out of tadpole-improved link operators.
Try designing a discretization that is accurate up to corrections
of~$\order(a^4)$.
\end{exercise}

\subsection{$a^3$-Accurate Lattice Dirac Theory}
Our numerical experiments in the first Section suggest that the $a^2$~errors
in the SW~quark action might still be quite large when the lattice spacing
is of order .4\,fm. Thus it is worthwhile developing a quark action that has
errors only of order~$a^4$ and higher.

A  attempt at such an improved Dirac theory is the action
\be \label{naive-a3}
\psib \left[ \Delta\cdot\gamma -
\frac{a^2}{6}\,\lder{3}\cdot\gamma + m_c
\right] \psi
\ee
where the $a^2$~correction cancels the leading errors induced by the
$\Delta\cdot\gamma$~term. Not surprisingly this action is plagued with
doublers and ghosts. Since the temporal derivatives extend two steps in each
direction, the energy dispersion relation has four branches instead of one.

There are several approaches one can take to solving this doubler disaster.
The ``D234'' quark actions rely upon three ingredients\,\cite{alford96}:
\begin{enumerate}
\item anisotropic lattices, with $a_t\!\le\!a_s/2$, which push
ghost states to higher energies, where they decouple. This works because
ghost energies are governed by the temporal lattice spacing and grow like
$1/a_t$ when $a_t$~is reduced. Reducing the temporal lattice spacing also
makes Monte Carlo measurements of hadron propagators much more efficient.
\item redundant operators like
\be
-\frac{r\,a_t}{2}\,\left(\lder{2}+\frac{\sigma\cdot gF}{2} + \cdots\right)
\ee
in the quark action, which can be tuned to remove some of the doublers and
ghosts.
\item ``irrelevant'' operators like
\be
s\,a_t^3\,\lder{4}_t
\ee
in the quark action, which have negligible effect on low-momentum physics (if
$a_t\!\le\!a_2/2$)  but are useful at high energies for removing doublers and
ghosts. Terms involving higher-order spatial derivatives, like
$a_s^4\lder{6}_s$, can also be added.
\end{enumerate}
A variety of actions can be created by combining differing amounts of
redundant and irrelevant operators. In general one tries to tune these
additions to minimize the number and importance of ghost states, to give the
best low-momentum behavior, and to minimize pathologies at high momentum.

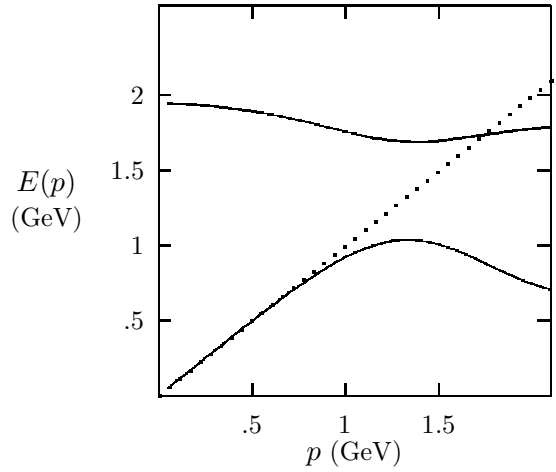
\begin{figure}
\begin{center}
\setlength{\unitlength}{0.240900pt}
\ifx\plotpoint\undefined\newsavebox{\plotpoint}\fi
\sbox{\plotpoint}{\rule[-0.200pt]{0.400pt}{0.400pt}}%
\begin{picture}(900,749)(0,0)
\font\gnuplot=cmr10 at 10pt
\gnuplot
\sbox{\plotpoint}{\rule[-0.200pt]{0.400pt}{0.400pt}}%
\put(220.0,113.0){\rule[-0.200pt]{148.394pt}{0.400pt}}
\put(220.0,113.0){\rule[-0.200pt]{0.400pt}{147.672pt}}
\put(220.0,231.0){\rule[-0.200pt]{4.818pt}{0.400pt}}
\put(198,231){\makebox(0,0)[r]{.5}}
\put(816.0,231.0){\rule[-0.200pt]{4.818pt}{0.400pt}}
\put(220.0,349.0){\rule[-0.200pt]{4.818pt}{0.400pt}}
\put(198,349){\makebox(0,0)[r]{1}}
\put(816.0,349.0){\rule[-0.200pt]{4.818pt}{0.400pt}}
\put(220.0,467.0){\rule[-0.200pt]{4.818pt}{0.400pt}}
\put(198,467){\makebox(0,0)[r]{1.5}}
\put(816.0,467.0){\rule[-0.200pt]{4.818pt}{0.400pt}}
\put(220.0,585.0){\rule[-0.200pt]{4.818pt}{0.400pt}}
\put(198,585){\makebox(0,0)[r]{2}}
\put(816.0,585.0){\rule[-0.200pt]{4.818pt}{0.400pt}}
\put(367.0,113.0){\rule[-0.200pt]{0.400pt}{4.818pt}}
\put(367,68){\makebox(0,0){.5}}
\put(367.0,706.0){\rule[-0.200pt]{0.400pt}{4.818pt}}
\put(513.0,113.0){\rule[-0.200pt]{0.400pt}{4.818pt}}
\put(513,68){\makebox(0,0){1}}
\put(513.0,706.0){\rule[-0.200pt]{0.400pt}{4.818pt}}
\put(660.0,113.0){\rule[-0.200pt]{0.400pt}{4.818pt}}
\put(660,68){\makebox(0,0){1.5}}
\put(660.0,706.0){\rule[-0.200pt]{0.400pt}{4.818pt}}
\put(220.0,113.0){\rule[-0.200pt]{148.394pt}{0.400pt}}
\put(836.0,113.0){\rule[-0.200pt]{0.400pt}{147.672pt}}
\put(220.0,726.0){\rule[-0.200pt]{148.394pt}{0.400pt}}
\put(45,419){\makebox(0,0){\shortstack{$E(p)$\\(GeV)}}}
\put(528,23){\makebox(0,0){$p$ (GeV)}}
\put(220.0,113.0){\rule[-0.200pt]{0.400pt}{147.672pt}}
\put(235,125){\usebox{\plotpoint}}
\multiput(235.00,125.58)(0.582,0.492){21}{\rule{0.567pt}{0.119pt}}
\multiput(235.00,124.17)(12.824,12.000){2}{\rule{0.283pt}{0.400pt}}
\multiput(249.00,137.58)(0.684,0.492){19}{\rule{0.645pt}{0.118pt}}
\multiput(249.00,136.17)(13.660,11.000){2}{\rule{0.323pt}{0.400pt}}
\multiput(264.00,148.58)(0.625,0.492){21}{\rule{0.600pt}{0.119pt}}
\multiput(264.00,147.17)(13.755,12.000){2}{\rule{0.300pt}{0.400pt}}
\multiput(279.00,160.58)(0.582,0.492){21}{\rule{0.567pt}{0.119pt}}
\multiput(279.00,159.17)(12.824,12.000){2}{\rule{0.283pt}{0.400pt}}
\multiput(293.00,172.58)(0.625,0.492){21}{\rule{0.600pt}{0.119pt}}
\multiput(293.00,171.17)(13.755,12.000){2}{\rule{0.300pt}{0.400pt}}
\multiput(308.00,184.58)(0.684,0.492){19}{\rule{0.645pt}{0.118pt}}
\multiput(308.00,183.17)(13.660,11.000){2}{\rule{0.323pt}{0.400pt}}
\multiput(323.00,195.58)(0.582,0.492){21}{\rule{0.567pt}{0.119pt}}
\multiput(323.00,194.17)(12.824,12.000){2}{\rule{0.283pt}{0.400pt}}
\multiput(337.00,207.58)(0.625,0.492){21}{\rule{0.600pt}{0.119pt}}
\multiput(337.00,206.17)(13.755,12.000){2}{\rule{0.300pt}{0.400pt}}
\multiput(352.00,219.58)(0.684,0.492){19}{\rule{0.645pt}{0.118pt}}
\multiput(352.00,218.17)(13.660,11.000){2}{\rule{0.323pt}{0.400pt}}
\multiput(367.00,230.58)(0.582,0.492){21}{\rule{0.567pt}{0.119pt}}
\multiput(367.00,229.17)(12.824,12.000){2}{\rule{0.283pt}{0.400pt}}
\multiput(381.00,242.58)(0.684,0.492){19}{\rule{0.645pt}{0.118pt}}
\multiput(381.00,241.17)(13.660,11.000){2}{\rule{0.323pt}{0.400pt}}
\multiput(396.00,253.58)(0.684,0.492){19}{\rule{0.645pt}{0.118pt}}
\multiput(396.00,252.17)(13.660,11.000){2}{\rule{0.323pt}{0.400pt}}
\multiput(411.00,264.58)(0.637,0.492){19}{\rule{0.609pt}{0.118pt}}
\multiput(411.00,263.17)(12.736,11.000){2}{\rule{0.305pt}{0.400pt}}
\multiput(425.00,275.58)(0.756,0.491){17}{\rule{0.700pt}{0.118pt}}
\multiput(425.00,274.17)(13.547,10.000){2}{\rule{0.350pt}{0.400pt}}
\multiput(440.00,285.58)(0.756,0.491){17}{\rule{0.700pt}{0.118pt}}
\multiput(440.00,284.17)(13.547,10.000){2}{\rule{0.350pt}{0.400pt}}
\multiput(455.00,295.58)(0.704,0.491){17}{\rule{0.660pt}{0.118pt}}
\multiput(455.00,294.17)(12.630,10.000){2}{\rule{0.330pt}{0.400pt}}
\multiput(469.00,305.59)(0.844,0.489){15}{\rule{0.767pt}{0.118pt}}
\multiput(469.00,304.17)(13.409,9.000){2}{\rule{0.383pt}{0.400pt}}
\multiput(484.00,314.59)(0.844,0.489){15}{\rule{0.767pt}{0.118pt}}
\multiput(484.00,313.17)(13.409,9.000){2}{\rule{0.383pt}{0.400pt}}
\multiput(499.00,323.59)(0.890,0.488){13}{\rule{0.800pt}{0.117pt}}
\multiput(499.00,322.17)(12.340,8.000){2}{\rule{0.400pt}{0.400pt}}
\multiput(513.00,331.59)(1.103,0.485){11}{\rule{0.957pt}{0.117pt}}
\multiput(513.00,330.17)(13.013,7.000){2}{\rule{0.479pt}{0.400pt}}
\multiput(528.00,338.59)(1.304,0.482){9}{\rule{1.100pt}{0.116pt}}
\multiput(528.00,337.17)(12.717,6.000){2}{\rule{0.550pt}{0.400pt}}
\multiput(543.00,344.59)(1.489,0.477){7}{\rule{1.220pt}{0.115pt}}
\multiput(543.00,343.17)(11.468,5.000){2}{\rule{0.610pt}{0.400pt}}
\multiput(557.00,349.60)(2.090,0.468){5}{\rule{1.600pt}{0.113pt}}
\multiput(557.00,348.17)(11.679,4.000){2}{\rule{0.800pt}{0.400pt}}
\multiput(572.00,353.61)(3.141,0.447){3}{\rule{2.100pt}{0.108pt}}
\multiput(572.00,352.17)(10.641,3.000){2}{\rule{1.050pt}{0.400pt}}
\put(587,356.17){\rule{2.900pt}{0.400pt}}
\multiput(587.00,355.17)(7.981,2.000){2}{\rule{1.450pt}{0.400pt}}
\put(616,356.67){\rule{3.614pt}{0.400pt}}
\multiput(616.00,357.17)(7.500,-1.000){2}{\rule{1.807pt}{0.400pt}}
\multiput(631.00,355.95)(2.918,-0.447){3}{\rule{1.967pt}{0.108pt}}
\multiput(631.00,356.17)(9.918,-3.000){2}{\rule{0.983pt}{0.400pt}}
\multiput(645.00,352.95)(3.141,-0.447){3}{\rule{2.100pt}{0.108pt}}
\multiput(645.00,353.17)(10.641,-3.000){2}{\rule{1.050pt}{0.400pt}}
\multiput(660.00,349.93)(1.601,-0.477){7}{\rule{1.300pt}{0.115pt}}
\multiput(660.00,350.17)(12.302,-5.000){2}{\rule{0.650pt}{0.400pt}}
\multiput(675.00,344.93)(1.489,-0.477){7}{\rule{1.220pt}{0.115pt}}
\multiput(675.00,345.17)(11.468,-5.000){2}{\rule{0.610pt}{0.400pt}}
\multiput(689.00,339.93)(1.304,-0.482){9}{\rule{1.100pt}{0.116pt}}
\multiput(689.00,340.17)(12.717,-6.000){2}{\rule{0.550pt}{0.400pt}}
\multiput(704.00,333.93)(1.103,-0.485){11}{\rule{0.957pt}{0.117pt}}
\multiput(704.00,334.17)(13.013,-7.000){2}{\rule{0.479pt}{0.400pt}}
\multiput(719.00,326.93)(1.026,-0.485){11}{\rule{0.900pt}{0.117pt}}
\multiput(719.00,327.17)(12.132,-7.000){2}{\rule{0.450pt}{0.400pt}}
\multiput(733.00,319.93)(1.103,-0.485){11}{\rule{0.957pt}{0.117pt}}
\multiput(733.00,320.17)(13.013,-7.000){2}{\rule{0.479pt}{0.400pt}}
\multiput(748.00,312.93)(1.103,-0.485){11}{\rule{0.957pt}{0.117pt}}
\multiput(748.00,313.17)(13.013,-7.000){2}{\rule{0.479pt}{0.400pt}}
\multiput(763.00,305.93)(1.214,-0.482){9}{\rule{1.033pt}{0.116pt}}
\multiput(763.00,306.17)(11.855,-6.000){2}{\rule{0.517pt}{0.400pt}}
\multiput(777.00,299.93)(1.103,-0.485){11}{\rule{0.957pt}{0.117pt}}
\multiput(777.00,300.17)(13.013,-7.000){2}{\rule{0.479pt}{0.400pt}}
\multiput(792.00,292.93)(1.601,-0.477){7}{\rule{1.300pt}{0.115pt}}
\multiput(792.00,293.17)(12.302,-5.000){2}{\rule{0.650pt}{0.400pt}}
\multiput(807.00,287.93)(1.489,-0.477){7}{\rule{1.220pt}{0.115pt}}
\multiput(807.00,288.17)(11.468,-5.000){2}{\rule{0.610pt}{0.400pt}}
\multiput(821.00,282.94)(2.090,-0.468){5}{\rule{1.600pt}{0.113pt}}
\multiput(821.00,283.17)(11.679,-4.000){2}{\rule{0.800pt}{0.400pt}}
\put(601.0,358.0){\rule[-0.200pt]{3.613pt}{0.400pt}}
\put(235,572){\usebox{\plotpoint}}
\put(235,570.67){\rule{3.373pt}{0.400pt}}
\multiput(235.00,571.17)(7.000,-1.000){2}{\rule{1.686pt}{0.400pt}}
\put(264,569.67){\rule{3.614pt}{0.400pt}}
\multiput(264.00,570.17)(7.500,-1.000){2}{\rule{1.807pt}{0.400pt}}
\put(279,568.67){\rule{3.373pt}{0.400pt}}
\multiput(279.00,569.17)(7.000,-1.000){2}{\rule{1.686pt}{0.400pt}}
\put(293,567.67){\rule{3.614pt}{0.400pt}}
\multiput(293.00,568.17)(7.500,-1.000){2}{\rule{1.807pt}{0.400pt}}
\put(308,566.17){\rule{3.100pt}{0.400pt}}
\multiput(308.00,567.17)(8.566,-2.000){2}{\rule{1.550pt}{0.400pt}}
\put(323,564.17){\rule{2.900pt}{0.400pt}}
\multiput(323.00,565.17)(7.981,-2.000){2}{\rule{1.450pt}{0.400pt}}
\put(337,562.17){\rule{3.100pt}{0.400pt}}
\multiput(337.00,563.17)(8.566,-2.000){2}{\rule{1.550pt}{0.400pt}}
\put(352,560.17){\rule{3.100pt}{0.400pt}}
\multiput(352.00,561.17)(8.566,-2.000){2}{\rule{1.550pt}{0.400pt}}
\put(367,558.17){\rule{2.900pt}{0.400pt}}
\multiput(367.00,559.17)(7.981,-2.000){2}{\rule{1.450pt}{0.400pt}}
\multiput(381.00,556.95)(3.141,-0.447){3}{\rule{2.100pt}{0.108pt}}
\multiput(381.00,557.17)(10.641,-3.000){2}{\rule{1.050pt}{0.400pt}}
\multiput(396.00,553.95)(3.141,-0.447){3}{\rule{2.100pt}{0.108pt}}
\multiput(396.00,554.17)(10.641,-3.000){2}{\rule{1.050pt}{0.400pt}}
\multiput(411.00,550.95)(2.918,-0.447){3}{\rule{1.967pt}{0.108pt}}
\multiput(411.00,551.17)(9.918,-3.000){2}{\rule{0.983pt}{0.400pt}}
\multiput(425.00,547.95)(3.141,-0.447){3}{\rule{2.100pt}{0.108pt}}
\multiput(425.00,548.17)(10.641,-3.000){2}{\rule{1.050pt}{0.400pt}}
\multiput(440.00,544.95)(3.141,-0.447){3}{\rule{2.100pt}{0.108pt}}
\multiput(440.00,545.17)(10.641,-3.000){2}{\rule{1.050pt}{0.400pt}}
\multiput(455.00,541.94)(1.943,-0.468){5}{\rule{1.500pt}{0.113pt}}
\multiput(455.00,542.17)(10.887,-4.000){2}{\rule{0.750pt}{0.400pt}}
\multiput(469.00,537.94)(2.090,-0.468){5}{\rule{1.600pt}{0.113pt}}
\multiput(469.00,538.17)(11.679,-4.000){2}{\rule{0.800pt}{0.400pt}}
\multiput(484.00,533.95)(3.141,-0.447){3}{\rule{2.100pt}{0.108pt}}
\multiput(484.00,534.17)(10.641,-3.000){2}{\rule{1.050pt}{0.400pt}}
\multiput(499.00,530.94)(1.943,-0.468){5}{\rule{1.500pt}{0.113pt}}
\multiput(499.00,531.17)(10.887,-4.000){2}{\rule{0.750pt}{0.400pt}}
\multiput(513.00,526.95)(3.141,-0.447){3}{\rule{2.100pt}{0.108pt}}
\multiput(513.00,527.17)(10.641,-3.000){2}{\rule{1.050pt}{0.400pt}}
\multiput(528.00,523.94)(2.090,-0.468){5}{\rule{1.600pt}{0.113pt}}
\multiput(528.00,524.17)(11.679,-4.000){2}{\rule{0.800pt}{0.400pt}}
\multiput(543.00,519.95)(2.918,-0.447){3}{\rule{1.967pt}{0.108pt}}
\multiput(543.00,520.17)(9.918,-3.000){2}{\rule{0.983pt}{0.400pt}}
\put(557,516.17){\rule{3.100pt}{0.400pt}}
\multiput(557.00,517.17)(8.566,-2.000){2}{\rule{1.550pt}{0.400pt}}
\put(572,514.17){\rule{3.100pt}{0.400pt}}
\multiput(572.00,515.17)(8.566,-2.000){2}{\rule{1.550pt}{0.400pt}}
\put(587,512.67){\rule{3.373pt}{0.400pt}}
\multiput(587.00,513.17)(7.000,-1.000){2}{\rule{1.686pt}{0.400pt}}
\put(601,511.67){\rule{3.614pt}{0.400pt}}
\multiput(601.00,512.17)(7.500,-1.000){2}{\rule{1.807pt}{0.400pt}}
\put(249.0,571.0){\rule[-0.200pt]{3.613pt}{0.400pt}}
\put(645,511.67){\rule{3.614pt}{0.400pt}}
\multiput(645.00,511.17)(7.500,1.000){2}{\rule{1.807pt}{0.400pt}}
\put(660,513.17){\rule{3.100pt}{0.400pt}}
\multiput(660.00,512.17)(8.566,2.000){2}{\rule{1.550pt}{0.400pt}}
\put(675,515.17){\rule{2.900pt}{0.400pt}}
\multiput(675.00,514.17)(7.981,2.000){2}{\rule{1.450pt}{0.400pt}}
\put(689,517.17){\rule{3.100pt}{0.400pt}}
\multiput(689.00,516.17)(8.566,2.000){2}{\rule{1.550pt}{0.400pt}}
\put(704,519.17){\rule{3.100pt}{0.400pt}}
\multiput(704.00,518.17)(8.566,2.000){2}{\rule{1.550pt}{0.400pt}}
\put(719,521.17){\rule{2.900pt}{0.400pt}}
\multiput(719.00,520.17)(7.981,2.000){2}{\rule{1.450pt}{0.400pt}}
\put(733,523.17){\rule{3.100pt}{0.400pt}}
\multiput(733.00,522.17)(8.566,2.000){2}{\rule{1.550pt}{0.400pt}}
\put(748,525.17){\rule{3.100pt}{0.400pt}}
\multiput(748.00,524.17)(8.566,2.000){2}{\rule{1.550pt}{0.400pt}}
\put(763,527.17){\rule{2.900pt}{0.400pt}}
\multiput(763.00,526.17)(7.981,2.000){2}{\rule{1.450pt}{0.400pt}}
\put(777,529.17){\rule{3.100pt}{0.400pt}}
\multiput(777.00,528.17)(8.566,2.000){2}{\rule{1.550pt}{0.400pt}}
\put(792,530.67){\rule{3.614pt}{0.400pt}}
\multiput(792.00,530.17)(7.500,1.000){2}{\rule{1.807pt}{0.400pt}}
\put(807,532.17){\rule{2.900pt}{0.400pt}}
\multiput(807.00,531.17)(7.981,2.000){2}{\rule{1.450pt}{0.400pt}}
\put(821,533.67){\rule{3.614pt}{0.400pt}}
\multiput(821.00,533.17)(7.500,1.000){2}{\rule{1.807pt}{0.400pt}}
\put(616.0,512.0){\rule[-0.200pt]{6.986pt}{0.400pt}}
\sbox{\plotpoint}{\rule[-0.500pt]{1.000pt}{1.000pt}}%
\put(220,113){\usebox{\plotpoint}}
\put(220.00,113.00){\usebox{\plotpoint}}
\multiput(226,118)(15.945,13.287){0}{\usebox{\plotpoint}}
\put(236.18,125.98){\usebox{\plotpoint}}
\multiput(239,128)(15.945,13.287){0}{\usebox{\plotpoint}}
\multiput(245,133)(15.945,13.287){0}{\usebox{\plotpoint}}
\put(252.28,139.07){\usebox{\plotpoint}}
\multiput(257,143)(16.889,12.064){0}{\usebox{\plotpoint}}
\put(268.62,151.85){\usebox{\plotpoint}}
\multiput(270,153)(15.945,13.287){0}{\usebox{\plotpoint}}
\multiput(276,158)(15.945,13.287){0}{\usebox{\plotpoint}}
\put(284.56,165.14){\usebox{\plotpoint}}
\multiput(288,168)(16.889,12.064){0}{\usebox{\plotpoint}}
\put(300.90,177.92){\usebox{\plotpoint}}
\multiput(301,178)(15.945,13.287){0}{\usebox{\plotpoint}}
\multiput(307,183)(15.945,13.287){0}{\usebox{\plotpoint}}
\put(317.07,190.91){\usebox{\plotpoint}}
\multiput(320,193)(15.945,13.287){0}{\usebox{\plotpoint}}
\multiput(326,198)(15.945,13.287){0}{\usebox{\plotpoint}}
\put(333.18,203.98){\usebox{\plotpoint}}
\multiput(338,208)(15.945,13.287){0}{\usebox{\plotpoint}}
\put(349.43,216.88){\usebox{\plotpoint}}
\multiput(351,218)(15.945,13.287){0}{\usebox{\plotpoint}}
\multiput(357,223)(15.945,13.287){0}{\usebox{\plotpoint}}
\put(365.46,230.05){\usebox{\plotpoint}}
\multiput(369,233)(16.889,12.064){0}{\usebox{\plotpoint}}
\put(381.80,242.83){\usebox{\plotpoint}}
\multiput(382,243)(15.945,13.287){0}{\usebox{\plotpoint}}
\multiput(388,248)(15.945,13.287){0}{\usebox{\plotpoint}}
\put(397.74,256.12){\usebox{\plotpoint}}
\multiput(400,258)(16.889,12.064){0}{\usebox{\plotpoint}}
\multiput(407,263)(15.945,13.287){0}{\usebox{\plotpoint}}
\put(414.08,268.90){\usebox{\plotpoint}}
\multiput(419,273)(15.945,13.287){0}{\usebox{\plotpoint}}
\put(430.32,281.80){\usebox{\plotpoint}}
\multiput(432,283)(15.945,13.287){0}{\usebox{\plotpoint}}
\multiput(438,288)(15.945,13.287){0}{\usebox{\plotpoint}}
\put(446.36,294.97){\usebox{\plotpoint}}
\multiput(450,298)(15.945,13.287){0}{\usebox{\plotpoint}}
\put(462.68,307.77){\usebox{\plotpoint}}
\multiput(463,308)(15.945,13.287){0}{\usebox{\plotpoint}}
\multiput(469,313)(15.945,13.287){0}{\usebox{\plotpoint}}
\put(478.64,321.03){\usebox{\plotpoint}}
\multiput(481,323)(16.889,12.064){0}{\usebox{\plotpoint}}
\multiput(488,328)(15.945,13.287){0}{\usebox{\plotpoint}}
\put(494.98,333.81){\usebox{\plotpoint}}
\multiput(500,338)(15.945,13.287){0}{\usebox{\plotpoint}}
\put(510.92,347.10){\usebox{\plotpoint}}
\multiput(512,348)(16.889,12.064){0}{\usebox{\plotpoint}}
\multiput(519,353)(15.945,13.287){0}{\usebox{\plotpoint}}
\put(527.26,359.88){\usebox{\plotpoint}}
\multiput(531,363)(15.945,13.287){0}{\usebox{\plotpoint}}
\put(543.57,372.69){\usebox{\plotpoint}}
\multiput(544,373)(15.945,13.287){0}{\usebox{\plotpoint}}
\multiput(550,378)(15.945,13.287){0}{\usebox{\plotpoint}}
\put(559.54,385.95){\usebox{\plotpoint}}
\multiput(562,388)(15.945,13.287){0}{\usebox{\plotpoint}}
\multiput(568,393)(16.889,12.064){0}{\usebox{\plotpoint}}
\put(575.88,398.73){\usebox{\plotpoint}}
\multiput(581,403)(15.945,13.287){0}{\usebox{\plotpoint}}
\put(591.82,412.02){\usebox{\plotpoint}}
\multiput(593,413)(16.889,12.064){0}{\usebox{\plotpoint}}
\multiput(600,418)(15.945,13.287){0}{\usebox{\plotpoint}}
\put(608.16,424.80){\usebox{\plotpoint}}
\multiput(612,428)(15.945,13.287){0}{\usebox{\plotpoint}}
\multiput(618,433)(15.945,13.287){0}{\usebox{\plotpoint}}
\put(624.11,438.08){\usebox{\plotpoint}}
\multiput(631,443)(15.945,13.287){0}{\usebox{\plotpoint}}
\put(640.44,450.86){\usebox{\plotpoint}}
\multiput(643,453)(15.945,13.287){0}{\usebox{\plotpoint}}
\multiput(649,458)(16.889,12.064){0}{\usebox{\plotpoint}}
\put(656.77,463.65){\usebox{\plotpoint}}
\multiput(662,468)(15.945,13.287){0}{\usebox{\plotpoint}}
\put(672.72,476.93){\usebox{\plotpoint}}
\multiput(674,478)(15.945,13.287){0}{\usebox{\plotpoint}}
\multiput(680,483)(16.889,12.064){0}{\usebox{\plotpoint}}
\put(689.06,489.71){\usebox{\plotpoint}}
\multiput(693,493)(15.945,13.287){0}{\usebox{\plotpoint}}
\multiput(699,498)(15.945,13.287){0}{\usebox{\plotpoint}}
\put(705.00,503.00){\usebox{\plotpoint}}
\multiput(712,508)(15.945,13.287){0}{\usebox{\plotpoint}}
\put(721.34,515.78){\usebox{\plotpoint}}
\multiput(724,518)(15.945,13.287){0}{\usebox{\plotpoint}}
\multiput(730,523)(15.945,13.287){0}{\usebox{\plotpoint}}
\put(737.36,528.97){\usebox{\plotpoint}}
\multiput(743,533)(15.945,13.287){0}{\usebox{\plotpoint}}
\put(753.62,541.85){\usebox{\plotpoint}}
\multiput(755,543)(15.945,13.287){0}{\usebox{\plotpoint}}
\multiput(761,548)(16.889,12.064){0}{\usebox{\plotpoint}}
\put(769.95,554.63){\usebox{\plotpoint}}
\multiput(774,558)(15.945,13.287){0}{\usebox{\plotpoint}}
\put(785.90,567.92){\usebox{\plotpoint}}
\multiput(786,568)(15.945,13.287){0}{\usebox{\plotpoint}}
\multiput(792,573)(16.889,12.064){0}{\usebox{\plotpoint}}
\put(802.24,580.70){\usebox{\plotpoint}}
\multiput(805,583)(15.945,13.287){0}{\usebox{\plotpoint}}
\multiput(811,588)(15.945,13.287){0}{\usebox{\plotpoint}}
\put(818.25,593.89){\usebox{\plotpoint}}
\multiput(824,598)(15.945,13.287){0}{\usebox{\plotpoint}}
\put(834.52,606.76){\usebox{\plotpoint}}
\put(836,608){\usebox{\plotpoint}}
\end{picture}
\end{center}
\caption{Dispersion relation for a free massless
lattice quarks described by the D234x
action. A low-energy, physical branch and a high-energy ghost branch
are shown, together with the continuum dispersion relation (dotted
line). Energies are for momenta proportional to $(1,1,0)$.}
\label{quark-E(p)-fig}
\end{figure}

A simple example of a D234-type action is obtained by generalizing our
derivation of the SW~action. We begin with a field
transformation~$\psi_\ct\to\psi$ where
\be
\psi_\ct = \Omega\,\psi
\ee
with
\bearray
\Omega &\equiv& 1 - X - \frac{X^2}{2} -\frac{X^3}{2}\\
&=& \sqrt{1-2X} + \order(X^4)
\eearray
and
\bearray
X &\equiv& \frac{r\,a_t}{4}\,\left(\Delta\cdot\gamma -
\sum_\mu\frac{a_\mu^2}{6}\,\lder{3}_\mu\,\gamma_\mu - m_\ct \right) \\
&=& \frac{r\,a_t}{4}\,\left(D\cdot\gamma - m_\ct + \order(a^4)\right).
\eearray
Applying this transformation to the naive improved action~\eq{naive-a3}
and adding an irrelevant $a_t^3$~operator, we  obtain\footnote{The
designers of this action, having examined a large number of
similar actions, all very effective, are currently grappling with a
nomenclature crisis. I refer to the particular action shown here as
the ``D234x'' action, but this is certainly not its final name. Most of
what I say about this action is true of almost all other D234 actions,
and so ``D234x'' can usually be taken to refer to a generic D234 action.}
\bearray\label{D234x-action}
\Lag_{\rm D234x} &=& \psib\left[
\Delta\cdot\gamma - \sum_\mu\frac{a_\mu^2}{6}\,\lder{3}_\mu\,\gamma_\mu
+ m \right. \nl
&& - \frac{r\,a_t}{2}\left(\lder{2} -
\sum_\mu\frac{a_\mu^2}{12}\,\lder{4}_\mu +\frac{\sigma\cdot gF}{2}\right)
\nl
&& \left. + s \, a_t^3\,\lder{4}_t
\rule[-3ex]{0em}{7ex}\right] \,\psi .
\eearray
This action is the same as the continuum action up to corrections of
order~$a_t^3$ and~$a_s^4$. It is easy to show that the choice
\be
s = \frac{1}{12}-\frac{r}{24}
\ee
removes one of the ghost/doubler states, leaving two
ghosts that have high energies for any value of~$r$ near~1. If in
addition $r=2/3$ is chosen,
a second ghost disappears and  only one remains (with an
energy of order $2/a_t$.)
The free-quark dispersion relation for this set of parameters is
shown in Fig.~\ref{quark-E(p)-fig} for a lattice with
$a_s\!=\!2\,a_t\!=\!.4$\,fm; the ghost is roughly 2\,GeV above the real quark
and so harmless in most applications.

\subsection{Tests of Quark Action}
Highly improved actions like the D234x~action~\eq{D234x-action}
have a large number of bare coupling constants that must be computed in
perturbation theory\,---\,one for each term in the action. In the previous
section we worked only to lowest order (tree-level) in perturbation theory.
However our discussion of the gluon action suggests that this is
sufficiently accurate provided all operators are tadpole improved; the
renormalizations due to $k\!>\!\pi/a$~physics affect phyical
quantities by only a few percent.
How can we ascertain whether the same is true of our quark actions?

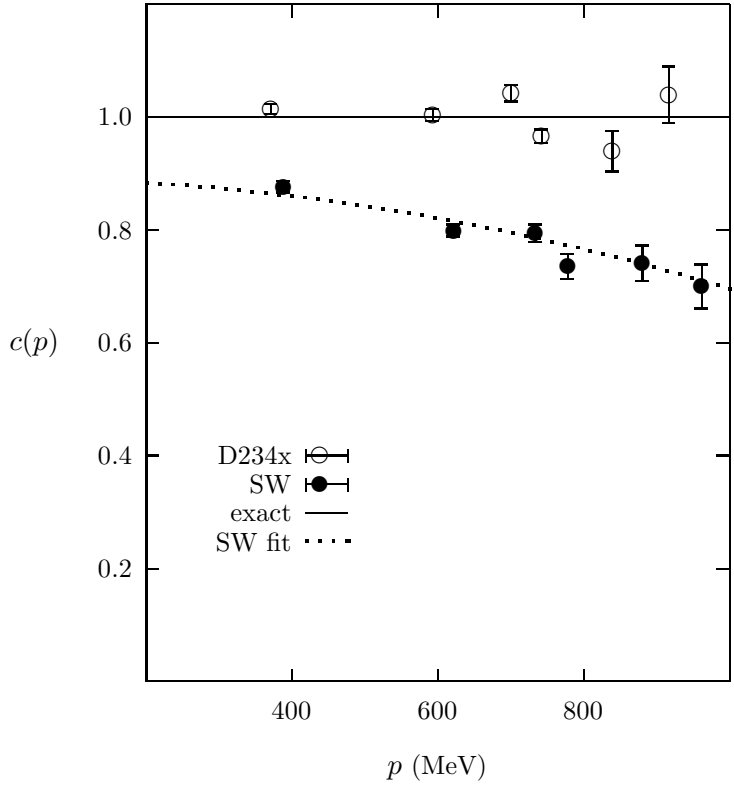
\begin{figure}
\begin{center}
\setlength{\unitlength}{0.240900pt}
\ifx\plotpoint\undefined\newsavebox{\plotpoint}\fi
\sbox{\plotpoint}{\rule[-0.200pt]{0.400pt}{0.400pt}}%
\begin{picture}(1200,1200)(0,0)
\font\gnuplot=cmr10 at 10pt
\gnuplot
\sbox{\plotpoint}{\rule[-0.200pt]{0.400pt}{0.400pt}}%
\put(220.0,113.0){\rule[-0.200pt]{220.664pt}{0.400pt}}
\put(220.0,290.0){\rule[-0.200pt]{4.818pt}{0.400pt}}
\put(198,290){\makebox(0,0)[r]{0.2}}
\put(1116.0,290.0){\rule[-0.200pt]{4.818pt}{0.400pt}}
\put(220.0,468.0){\rule[-0.200pt]{4.818pt}{0.400pt}}
\put(198,468){\makebox(0,0)[r]{0.4}}
\put(1116.0,468.0){\rule[-0.200pt]{4.818pt}{0.400pt}}
\put(220.0,645.0){\rule[-0.200pt]{4.818pt}{0.400pt}}
\put(198,645){\makebox(0,0)[r]{0.6}}
\put(1116.0,645.0){\rule[-0.200pt]{4.818pt}{0.400pt}}
\put(220.0,822.0){\rule[-0.200pt]{4.818pt}{0.400pt}}
\put(198,822){\makebox(0,0)[r]{0.8}}
\put(1116.0,822.0){\rule[-0.200pt]{4.818pt}{0.400pt}}
\put(220.0,1000.0){\rule[-0.200pt]{4.818pt}{0.400pt}}
\put(198,1000){\makebox(0,0)[r]{1.0}}
\put(1116.0,1000.0){\rule[-0.200pt]{4.818pt}{0.400pt}}
\put(449.0,113.0){\rule[-0.200pt]{0.400pt}{4.818pt}}
\put(449,68){\makebox(0,0){400}}
\put(449.0,1157.0){\rule[-0.200pt]{0.400pt}{4.818pt}}
\put(678.0,113.0){\rule[-0.200pt]{0.400pt}{4.818pt}}
\put(678,68){\makebox(0,0){600}}
\put(678.0,1157.0){\rule[-0.200pt]{0.400pt}{4.818pt}}
\put(907.0,113.0){\rule[-0.200pt]{0.400pt}{4.818pt}}
\put(907,68){\makebox(0,0){800}}
\put(907.0,1157.0){\rule[-0.200pt]{0.400pt}{4.818pt}}
\put(220.0,113.0){\rule[-0.200pt]{220.664pt}{0.400pt}}
\put(1136.0,113.0){\rule[-0.200pt]{0.400pt}{256.318pt}}
\put(220.0,1177.0){\rule[-0.200pt]{220.664pt}{0.400pt}}
\put(45,645){\makebox(0,0){$c(p)$}}
\put(678,-22){\makebox(0,0){$p$ (MeV)}}
\put(220.0,113.0){\rule[-0.200pt]{0.400pt}{256.318pt}}
\put(449,468){\makebox(0,0)[r]{D234x}}
\put(493,468){\circle{24}}
\put(416,1012){\circle{24}}
\put(670,1003){\circle{24}}
\put(793,1037){\circle{24}}
\put(841,970){\circle{24}}
\put(952,946){\circle{24}}
\put(1041,1034){\circle{24}}
\put(471.0,468.0){\rule[-0.200pt]{15.899pt}{0.400pt}}
\put(471.0,458.0){\rule[-0.200pt]{0.400pt}{4.818pt}}
\put(537.0,458.0){\rule[-0.200pt]{0.400pt}{4.818pt}}
\put(416.0,1004.0){\rule[-0.200pt]{0.400pt}{3.854pt}}
\put(406.0,1004.0){\rule[-0.200pt]{4.818pt}{0.400pt}}
\put(406.0,1020.0){\rule[-0.200pt]{4.818pt}{0.400pt}}
\put(670.0,994.0){\rule[-0.200pt]{0.400pt}{4.336pt}}
\put(660.0,994.0){\rule[-0.200pt]{4.818pt}{0.400pt}}
\put(660.0,1012.0){\rule[-0.200pt]{4.818pt}{0.400pt}}
\put(793.0,1024.0){\rule[-0.200pt]{0.400pt}{6.263pt}}
\put(783.0,1024.0){\rule[-0.200pt]{4.818pt}{0.400pt}}
\put(783.0,1050.0){\rule[-0.200pt]{4.818pt}{0.400pt}}
\put(841.0,959.0){\rule[-0.200pt]{0.400pt}{5.059pt}}
\put(831.0,959.0){\rule[-0.200pt]{4.818pt}{0.400pt}}
\put(831.0,980.0){\rule[-0.200pt]{4.818pt}{0.400pt}}
\put(952.0,914.0){\rule[-0.200pt]{0.400pt}{15.418pt}}
\put(942.0,914.0){\rule[-0.200pt]{4.818pt}{0.400pt}}
\put(942.0,978.0){\rule[-0.200pt]{4.818pt}{0.400pt}}
\put(1041.0,990.0){\rule[-0.200pt]{0.400pt}{21.440pt}}
\put(1031.0,990.0){\rule[-0.200pt]{4.818pt}{0.400pt}}
\put(1031.0,1079.0){\rule[-0.200pt]{4.818pt}{0.400pt}}
\put(449,423){\makebox(0,0)[r]{SW}}
\put(493,423){\circle*{24}}
\put(436,890){\circle*{24}}
\put(703,821){\circle*{24}}
\put(831,817){\circle*{24}}
\put(882,765){\circle*{24}}
\put(999,770){\circle*{24}}
\put(1092,734){\circle*{24}}
\put(471.0,423.0){\rule[-0.200pt]{15.899pt}{0.400pt}}
\put(471.0,413.0){\rule[-0.200pt]{0.400pt}{4.818pt}}
\put(537.0,413.0){\rule[-0.200pt]{0.400pt}{4.818pt}}
\put(436.0,881.0){\rule[-0.200pt]{0.400pt}{4.336pt}}
\put(426.0,881.0){\rule[-0.200pt]{4.818pt}{0.400pt}}
\put(426.0,899.0){\rule[-0.200pt]{4.818pt}{0.400pt}}
\put(703.0,812.0){\rule[-0.200pt]{0.400pt}{4.577pt}}
\put(693.0,812.0){\rule[-0.200pt]{4.818pt}{0.400pt}}
\put(693.0,831.0){\rule[-0.200pt]{4.818pt}{0.400pt}}
\put(831.0,803.0){\rule[-0.200pt]{0.400pt}{6.745pt}}
\put(821.0,803.0){\rule[-0.200pt]{4.818pt}{0.400pt}}
\put(821.0,831.0){\rule[-0.200pt]{4.818pt}{0.400pt}}
\put(882.0,745.0){\rule[-0.200pt]{0.400pt}{9.395pt}}
\put(872.0,745.0){\rule[-0.200pt]{4.818pt}{0.400pt}}
\put(872.0,784.0){\rule[-0.200pt]{4.818pt}{0.400pt}}
\put(999.0,742.0){\rule[-0.200pt]{0.400pt}{13.490pt}}
\put(989.0,742.0){\rule[-0.200pt]{4.818pt}{0.400pt}}
\put(989.0,798.0){\rule[-0.200pt]{4.818pt}{0.400pt}}
\put(1092.0,699.0){\rule[-0.200pt]{0.400pt}{16.622pt}}
\put(1082.0,699.0){\rule[-0.200pt]{4.818pt}{0.400pt}}
\put(1082.0,768.0){\rule[-0.200pt]{4.818pt}{0.400pt}}
\put(449,378){\makebox(0,0)[r]{exact}}
\put(471.0,378.0){\rule[-0.200pt]{15.899pt}{0.400pt}}
\put(220,1000){\usebox{\plotpoint}}
\put(220.0,1000.0){\rule[-0.200pt]{220.664pt}{0.400pt}}
\sbox{\plotpoint}{\rule[-0.500pt]{1.000pt}{1.000pt}}%
\put(449,333){\makebox(0,0)[r]{SW fit}}
\multiput(471,333)(20.756,0.000){4}{\usebox{\plotpoint}}
\put(537,333){\usebox{\plotpoint}}
\put(220,896){\usebox{\plotpoint}}
\put(220.00,896.00){\usebox{\plotpoint}}
\multiput(229,895)(20.756,0.000){0}{\usebox{\plotpoint}}
\put(240.69,894.81){\usebox{\plotpoint}}
\multiput(248,894)(20.756,0.000){0}{\usebox{\plotpoint}}
\put(261.37,893.51){\usebox{\plotpoint}}
\multiput(266,893)(20.652,-2.065){0}{\usebox{\plotpoint}}
\put(282.05,892.00){\usebox{\plotpoint}}
\multiput(285,892)(20.629,-2.292){0}{\usebox{\plotpoint}}
\put(302.70,890.03){\usebox{\plotpoint}}
\multiput(303,890)(20.652,-2.065){0}{\usebox{\plotpoint}}
\multiput(313,889)(20.756,0.000){0}{\usebox{\plotpoint}}
\put(323.39,888.85){\usebox{\plotpoint}}
\multiput(331,888)(20.629,-2.292){0}{\usebox{\plotpoint}}
\put(344.03,886.60){\usebox{\plotpoint}}
\multiput(350,886)(20.629,-2.292){0}{\usebox{\plotpoint}}
\put(364.66,884.37){\usebox{\plotpoint}}
\multiput(368,884)(20.629,-2.292){0}{\usebox{\plotpoint}}
\put(385.30,882.17){\usebox{\plotpoint}}
\multiput(387,882)(20.629,-2.292){0}{\usebox{\plotpoint}}
\multiput(396,881)(20.629,-2.292){0}{\usebox{\plotpoint}}
\put(405.93,879.90){\usebox{\plotpoint}}
\multiput(414,879)(20.652,-2.065){0}{\usebox{\plotpoint}}
\put(426.57,877.71){\usebox{\plotpoint}}
\multiput(433,877)(20.629,-2.292){0}{\usebox{\plotpoint}}
\put(447.20,875.42){\usebox{\plotpoint}}
\multiput(451,875)(20.652,-2.065){0}{\usebox{\plotpoint}}
\put(467.84,873.24){\usebox{\plotpoint}}
\multiput(470,873)(20.261,-4.503){0}{\usebox{\plotpoint}}
\multiput(479,871)(20.629,-2.292){0}{\usebox{\plotpoint}}
\put(488.30,869.97){\usebox{\plotpoint}}
\multiput(498,869)(20.629,-2.292){0}{\usebox{\plotpoint}}
\put(508.91,867.58){\usebox{\plotpoint}}
\multiput(516,866)(20.629,-2.292){0}{\usebox{\plotpoint}}
\put(529.41,864.56){\usebox{\plotpoint}}
\multiput(535,864)(20.629,-2.292){0}{\usebox{\plotpoint}}
\put(549.94,861.68){\usebox{\plotpoint}}
\multiput(553,861)(20.629,-2.292){0}{\usebox{\plotpoint}}
\put(570.40,858.32){\usebox{\plotpoint}}
\multiput(572,858)(20.629,-2.292){0}{\usebox{\plotpoint}}
\multiput(581,857)(20.261,-4.503){0}{\usebox{\plotpoint}}
\put(590.84,854.91){\usebox{\plotpoint}}
\multiput(599,854)(20.352,-4.070){0}{\usebox{\plotpoint}}
\put(611.34,851.74){\usebox{\plotpoint}}
\multiput(618,851)(20.261,-4.503){0}{\usebox{\plotpoint}}
\put(631.80,848.47){\usebox{\plotpoint}}
\multiput(636,848)(20.352,-4.070){0}{\usebox{\plotpoint}}
\put(652.30,845.30){\usebox{\plotpoint}}
\multiput(655,845)(20.261,-4.503){0}{\usebox{\plotpoint}}
\put(672.61,841.09){\usebox{\plotpoint}}
\multiput(673,841)(20.652,-2.065){0}{\usebox{\plotpoint}}
\multiput(683,840)(20.261,-4.503){0}{\usebox{\plotpoint}}
\put(693.06,837.77){\usebox{\plotpoint}}
\multiput(701,836)(20.261,-4.503){0}{\usebox{\plotpoint}}
\put(713.38,833.66){\usebox{\plotpoint}}
\multiput(720,833)(20.261,-4.503){0}{\usebox{\plotpoint}}
\put(733.77,829.94){\usebox{\plotpoint}}
\multiput(738,829)(20.261,-4.503){0}{\usebox{\plotpoint}}
\put(754.06,825.59){\usebox{\plotpoint}}
\multiput(757,825)(20.261,-4.503){0}{\usebox{\plotpoint}}
\put(774.49,822.06){\usebox{\plotpoint}}
\multiput(775,822)(20.261,-4.503){0}{\usebox{\plotpoint}}
\multiput(784,820)(20.352,-4.070){0}{\usebox{\plotpoint}}
\put(794.80,817.82){\usebox{\plotpoint}}
\multiput(803,816)(20.261,-4.503){0}{\usebox{\plotpoint}}
\put(815.06,813.32){\usebox{\plotpoint}}
\multiput(821,812)(20.352,-4.070){0}{\usebox{\plotpoint}}
\put(835.25,808.58){\usebox{\plotpoint}}
\multiput(840,807)(20.261,-4.503){0}{\usebox{\plotpoint}}
\put(855.37,803.58){\usebox{\plotpoint}}
\multiput(858,803)(20.352,-4.070){0}{\usebox{\plotpoint}}
\put(875.68,799.29){\usebox{\plotpoint}}
\multiput(877,799)(20.261,-4.503){0}{\usebox{\plotpoint}}
\multiput(886,797)(20.261,-4.503){0}{\usebox{\plotpoint}}
\put(895.92,794.72){\usebox{\plotpoint}}
\multiput(905,792)(20.261,-4.503){0}{\usebox{\plotpoint}}
\put(916.01,789.55){\usebox{\plotpoint}}
\multiput(923,788)(20.261,-4.503){0}{\usebox{\plotpoint}}
\put(936.19,784.74){\usebox{\plotpoint}}
\multiput(942,783)(20.261,-4.503){0}{\usebox{\plotpoint}}
\put(956.34,779.81){\usebox{\plotpoint}}
\multiput(960,779)(19.690,-6.563){0}{\usebox{\plotpoint}}
\put(976.37,774.53){\usebox{\plotpoint}}
\multiput(979,774)(19.690,-6.563){0}{\usebox{\plotpoint}}
\put(996.38,769.14){\usebox{\plotpoint}}
\multiput(997,769)(19.690,-6.563){0}{\usebox{\plotpoint}}
\multiput(1006,766)(20.352,-4.070){0}{\usebox{\plotpoint}}
\put(1016.42,763.86){\usebox{\plotpoint}}
\multiput(1025,761)(20.261,-4.503){0}{\usebox{\plotpoint}}
\put(1036.36,758.21){\usebox{\plotpoint}}
\multiput(1043,756)(20.352,-4.070){0}{\usebox{\plotpoint}}
\put(1056.38,752.88){\usebox{\plotpoint}}
\multiput(1062,751)(19.690,-6.563){0}{\usebox{\plotpoint}}
\put(1076.21,746.84){\usebox{\plotpoint}}
\multiput(1080,746)(19.880,-5.964){0}{\usebox{\plotpoint}}
\put(1096.10,740.97){\usebox{\plotpoint}}
\multiput(1099,740)(20.261,-4.503){0}{\usebox{\plotpoint}}
\put(1116.05,735.32){\usebox{\plotpoint}}
\multiput(1117,735)(19.880,-5.964){0}{\usebox{\plotpoint}}
\put(1135.83,729.06){\usebox{\plotpoint}}
\put(1136,729){\usebox{\plotpoint}}
\end{picture}
\end{center}
\caption{The speed of light computed using the D234x and SW quark
actions from the energy and three momenta of a meson. The fit to the
SW data is $\tilde{c}_0+k\, (ap)^2$ where $\tilde{c}_0$ and $k$ are
constants.}
\label{c(p)-fig}
\end{figure}

One powerful approach is to verify that continuum symmetries are restored by
the corrections. For example, the leading  error in the discretized Dirac
operator, $\Delta\cdot\gamma+m$, breaks Lorentz invariance. The
$a_\mu^2\lder{3}_\mu\,\gamma_\mu$ term in the D234x~action is meant to cancel
this error. We can check that this term has done its job by comparing the
dispersion relation of a meson (or baryon) computed in a simulation with
the Lorentz-invariant dispersion relation of the continuum. In a
Lorentz-invariant theory, a hadron's energy~$E_{\rm h}(\pv)$
and its three momentum~$\pv$ are such that the combination
\be
c^2(\pv) \equiv \frac{E_{\rm h}^2(\pv) - m_{\rm h}^2}{\pv^2}
\ee
is independent of~$\pv$ and equal to the square of the speed of light
($=\!1$). In Fig.~\ref{c(p)-fig} I show D234x~simulation results for
meson's~$c_(\pv)$. The lattice for this simulation is very coarse, with
$a_s\!=\!2\,a_t\!=\!.4$\,fm, and yet $c(\pv)$ is within a few percent of
one all the way out to momenta of order 1\,GeV ($\approx 2/a_s$). For
comparison I include the same
quantity calculated using the SW~action (on an isotropic lattice), which has
no correction for the $a^2$~error; it has 25\% errors by
$p\!=\!1$\,GeV, and the momentum dependence of the discrepancy is consistent
with the quadratic errors expected. This
comparison shows that the coefficient of the of the
$a^2$ correction in the D234x~action is not renormalized by more than
10--20\%, and that this renormalization will have no more than a few percent
effect on low-momentum physics. The momenta used for the Monte Carlo
points shown in this plot include momenta that are parallel to a lattice
axis, and momenta that point along lattice diagonals; the fact that
different momenta agree to within a few percent shows that rotation
invariance is also restored by the correction term.

The $c(\pv)$~plot provides a second important check on the D234x~action. On
anisotropic lattices, the symmetry between space and time is not
exact. Consequently the spatial and temporal parts of the leading Dirac
operator~$\Delta\cdot\gamma$ are renormalized differently:
\be
D\cdot\gamma \to \Delta_t\,\gamma_t + c_0\,\Deltav\cdot\gammav,
\ee
where $c_0$ is a bare speed-of-light,
\be
c_0 = 1 + {\rm const}\,\alpha_s +\cdots.
\ee
Such a renormalization would shift the low-momentum values of $c(\pv)$ away
from one. In fact, with only our tree-level action, $c(\pv)=1.025(20)$ at
low momenta; parameter~$c_0$ must be within a percent of one. Note
that this success depends crucially on tadpole improvement; without tadpole
improvement $c(\pv)=1.2$ at low momenta on our $a_s\!=\!.4$\,fm lattice.

The only other potentially large renormalization is in the $\order(r\,a_t)$
terms induced by the field transformation. Again there is a symmetry
associated with this term: field-redefinition or ``$r$~symmetry''.
The operator multiplied by~$r$ is redundant and so should have no effect
on the low-momentum physics. Consequently the low-momentum physics should
be invariant under changes in the value of~$r$. If the renormalizations of
the $\lder{2}$ and $\sigma\cdot gF$ terms in the action are very different,
then the tree-level operator is no longer redundant and results must
change as $r$ is varied; on the other hand, if the renormalizations are
roughly the same then the net effect of the renormalizations is a
harmless shift in~$r$.\footnote{I am simplifying the discussion
here. On an anisotropic lattice, where space--time interchange is not
an exact symmetry, the temporal and spatial parts of
$\lder{2}$ and $\sigma\cdot gF$ have different radiative
corrections. Thus the $r$-test is checking on the relative
coefficients of four operators, as opposed to two in the isotropic
case.} The results of such an $r$-test for a close relative of
the D234x action are shown in Table~\ref{r-test-table}. These show that the
pion, rho, nucleon and delta masses all shift by less than a percent when
$r$~is changed
from~$2/3$ to~$1$. Since omitting the $\sigma\cdot gF$ term
shifts  masses by 20\%, our $r$-test implies that
that the relative renormalization given by tree-level tadpole
improvement is correct to within roughly 10--20\%. A 10\% change in the
coefficient of $\sigma\cdot gF$ would shift the meson masses
by only~2\%.

\begin{table}
\caption{Mass differences between two simulations, one with $r\!=\!1$
and the other with $r\!=\!2/3$, for different quark masses. The bare
quark masses in the two simulations were tuned to give the same
$m_\pi/m_\rho$.}
\label{r-test-table}
\begin{center}
\begin{tabular}{ccccc}\hline
$\quad m_\pi/m_\rho\quad$ &  $\quad\Delta m_\pi\quad$
& $\quad\Delta m_\rho\quad$
& $\quad\Delta m_N\quad$ & $\quad\Delta m_\Delta\quad$ \\ \hline
.65 & 0.6\,(2)\,\% & -0.13\,(13)\,\% & -0.3\,(2)\,\% & 0.4\,(3)\,\% \\
.57 & 0.6\,(4)\,\% & -0.6\,(1.0)\,\% \\ \hline
\end{tabular}
\end{center}
\end{table}

In~\cite{luscher96} a different test of the $\sigma\cdot gF$
coefficient is proposed. This is based on the fact that the $r$~terms
in the action break the quark's chiral symmetry even for massless
quarks. If the $\sigma\cdot gF$ coefficient is properly tuned, the
chiral symmetry breaking terms are redundant and the chiral symmetry Ward
identity should be restored. Using Monte Carlo evaluations of the
various terms involved,
the authors tune the coefficient until the identity is
satisfied. Working only with the SW action and relatively small
lattice spacings,
they find renormalizations that are larger than those above for D234:
the coefficient of $\sigma\cdot gF$ in the SW action is a little more
20\%~larger than the tadpole-improved tree-level coefficient at
$a\!\approx\!.1$\,fm. Such a shift at such small lattice spacings
has only a small effect on hadron masses; but in the same study the
authors encountered pathological behavior in the SW action, with this
correction,  at large lattice spacings. Also the correction to
tadpole improvement, though small throughout the range of their study,
grows much more
rapidly than naively expected as $a$ increases from $.03$\,fm to~$.1$\,fm.
It would be very
interesting to verify these results using the $r$-test, which provides
an independent method of testing the same correction.

Our tests of the D234~action confirm that tree-level improvement of
the quark action, using tadpole-improved operators, is quite
accurate. However the work I just described concerning the SW action
suggests that nonperturbative tuning of parameters in the action may
be important for some actions and operators. Such tuning is
straightforward for D234: the same tests I described above can be used
to tune the leading corrections in the action. In particular, the
$c_0$~parameter, which is potentially the most important, can be
easily adjusted so that $c(\pv\!=\!0)=1$.

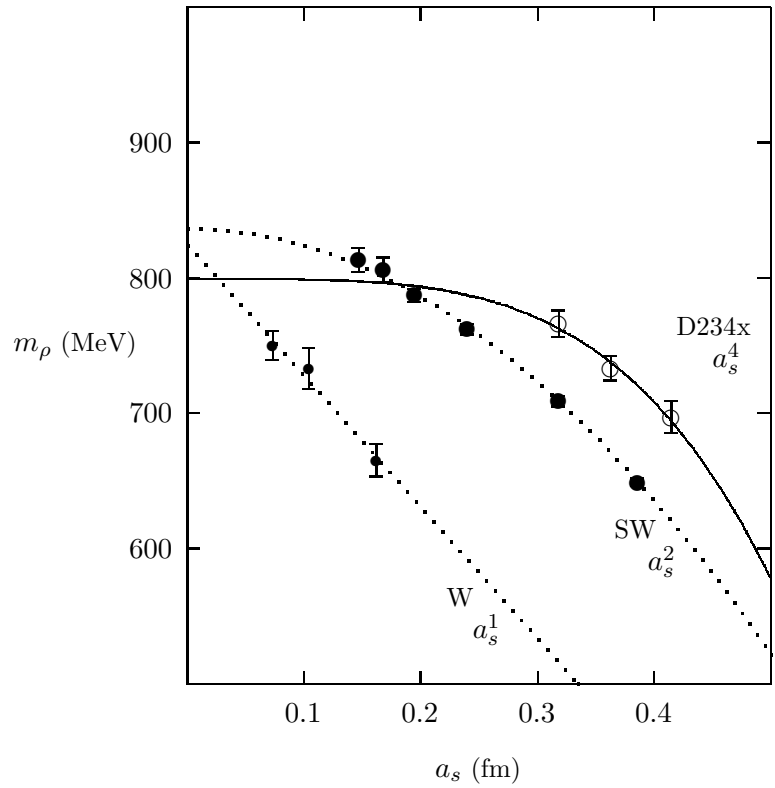
\begin{figure}
\begin{center}
\setlength{\unitlength}{0.240900pt}
\ifx\plotpoint\undefined\newsavebox{\plotpoint}\fi
\sbox{\plotpoint}{\rule[-0.200pt]{0.400pt}{0.400pt}}%
\begin{picture}(1200,1200)(0,0)
\font\gnuplot=cmr10 at 10pt
\gnuplot
\sbox{\plotpoint}{\rule[-0.200pt]{0.400pt}{0.400pt}}%
\put(220.0,326.0){\rule[-0.200pt]{4.818pt}{0.400pt}}
\put(198,326){\makebox(0,0)[r]{$600$}}
\put(1116.0,326.0){\rule[-0.200pt]{4.818pt}{0.400pt}}
\put(220.0,539.0){\rule[-0.200pt]{4.818pt}{0.400pt}}
\put(198,539){\makebox(0,0)[r]{$700$}}
\put(1116.0,539.0){\rule[-0.200pt]{4.818pt}{0.400pt}}
\put(220.0,751.0){\rule[-0.200pt]{4.818pt}{0.400pt}}
\put(198,751){\makebox(0,0)[r]{$800$}}
\put(1116.0,751.0){\rule[-0.200pt]{4.818pt}{0.400pt}}
\put(220.0,964.0){\rule[-0.200pt]{4.818pt}{0.400pt}}
\put(198,964){\makebox(0,0)[r]{$900$}}
\put(1116.0,964.0){\rule[-0.200pt]{4.818pt}{0.400pt}}
\put(403.0,113.0){\rule[-0.200pt]{0.400pt}{4.818pt}}
\put(403,68){\makebox(0,0){$0.1$}}
\put(403.0,1157.0){\rule[-0.200pt]{0.400pt}{4.818pt}}
\put(586.0,113.0){\rule[-0.200pt]{0.400pt}{4.818pt}}
\put(586,68){\makebox(0,0){$0.2$}}
\put(586.0,1157.0){\rule[-0.200pt]{0.400pt}{4.818pt}}
\put(770.0,113.0){\rule[-0.200pt]{0.400pt}{4.818pt}}
\put(770,68){\makebox(0,0){$0.3$}}
\put(770.0,1157.0){\rule[-0.200pt]{0.400pt}{4.818pt}}
\put(953.0,113.0){\rule[-0.200pt]{0.400pt}{4.818pt}}
\put(953,68){\makebox(0,0){$0.4$}}
\put(953.0,1157.0){\rule[-0.200pt]{0.400pt}{4.818pt}}
\put(220.0,113.0){\rule[-0.200pt]{220.664pt}{0.400pt}}
\put(1136.0,113.0){\rule[-0.200pt]{0.400pt}{256.318pt}}
\put(220.0,1177.0){\rule[-0.200pt]{220.664pt}{0.400pt}}
\put(45,645){\makebox(0,0){$m_\rho$ (MeV)}}
\put(678,-22){\makebox(0,0){$a_s$ (fm)}}
\put(586,219){\makebox(0,0)[l]{\shortstack{W\\$\qquad a_s^1$}}}
\put(861,326){\makebox(0,0)[l]{\shortstack{SW\\$\qquad a_s^2$}}}
\put(989,645){\makebox(0,0)[l]{\shortstack{D234x\\$\quad a_s^4$}}}
\put(220.0,113.0){\rule[-0.200pt]{0.400pt}{256.318pt}}
\put(980,532){\circle{24}}
\put(885,609){\circle{24}}
\put(803,679){\circle{24}}
\put(980.0,507.0){\rule[-0.200pt]{0.400pt}{12.286pt}}
\put(970.0,507.0){\rule[-0.200pt]{4.818pt}{0.400pt}}
\put(970.0,558.0){\rule[-0.200pt]{4.818pt}{0.400pt}}
\put(885.0,590.0){\rule[-0.200pt]{0.400pt}{9.154pt}}
\put(875.0,590.0){\rule[-0.200pt]{4.818pt}{0.400pt}}
\put(875.0,628.0){\rule[-0.200pt]{4.818pt}{0.400pt}}
\put(803.0,658.0){\rule[-0.200pt]{0.400pt}{10.118pt}}
\put(793.0,658.0){\rule[-0.200pt]{4.818pt}{0.400pt}}
\put(793.0,700.0){\rule[-0.200pt]{4.818pt}{0.400pt}}
\put(927,430){\circle*{24}}
\put(803,558){\circle*{24}}
\put(660,671){\circle*{24}}
\put(577,724){\circle*{24}}
\put(528,764){\circle*{24}}
\put(489,779){\circle*{24}}
\put(927.0,424.0){\rule[-0.200pt]{0.400pt}{2.891pt}}
\put(917.0,424.0){\rule[-0.200pt]{4.818pt}{0.400pt}}
\put(917.0,436.0){\rule[-0.200pt]{4.818pt}{0.400pt}}
\put(803.0,549.0){\rule[-0.200pt]{0.400pt}{4.095pt}}
\put(793.0,549.0){\rule[-0.200pt]{4.818pt}{0.400pt}}
\put(793.0,566.0){\rule[-0.200pt]{4.818pt}{0.400pt}}
\put(660.0,662.0){\rule[-0.200pt]{0.400pt}{4.095pt}}
\put(650.0,662.0){\rule[-0.200pt]{4.818pt}{0.400pt}}
\put(650.0,679.0){\rule[-0.200pt]{4.818pt}{0.400pt}}
\put(577.0,713.0){\rule[-0.200pt]{0.400pt}{5.059pt}}
\put(567.0,713.0){\rule[-0.200pt]{4.818pt}{0.400pt}}
\put(567.0,734.0){\rule[-0.200pt]{4.818pt}{0.400pt}}
\put(528.0,745.0){\rule[-0.200pt]{0.400pt}{9.154pt}}
\put(518.0,745.0){\rule[-0.200pt]{4.818pt}{0.400pt}}
\put(518.0,783.0){\rule[-0.200pt]{4.818pt}{0.400pt}}
\put(489.0,760.0){\rule[-0.200pt]{0.400pt}{9.154pt}}
\put(479.0,760.0){\rule[-0.200pt]{4.818pt}{0.400pt}}
\put(479.0,798.0){\rule[-0.200pt]{4.818pt}{0.400pt}}
\put(517,464){\circle*{18}}
\put(411,609){\circle*{18}}
\put(354,645){\circle*{18}}
\put(517.0,439.0){\rule[-0.200pt]{0.400pt}{12.286pt}}
\put(507.0,439.0){\rule[-0.200pt]{4.818pt}{0.400pt}}
\put(507.0,490.0){\rule[-0.200pt]{4.818pt}{0.400pt}}
\put(411.0,577.0){\rule[-0.200pt]{0.400pt}{15.418pt}}
\put(401.0,577.0){\rule[-0.200pt]{4.818pt}{0.400pt}}
\put(401.0,641.0){\rule[-0.200pt]{4.818pt}{0.400pt}}
\put(354.0,622.0){\rule[-0.200pt]{0.400pt}{11.081pt}}
\put(344.0,622.0){\rule[-0.200pt]{4.818pt}{0.400pt}}
\put(344.0,668.0){\rule[-0.200pt]{4.818pt}{0.400pt}}
\sbox{\plotpoint}{\rule[-0.500pt]{1.000pt}{1.000pt}}%
\put(220,802){\usebox{\plotpoint}}
\put(220.00,802.00){\usebox{\plotpoint}}
\put(234.16,786.84){\usebox{\plotpoint}}
\put(247.81,771.23){\usebox{\plotpoint}}
\multiput(248,771)(13.885,-15.427){0}{\usebox{\plotpoint}}
\put(261.44,755.58){\usebox{\plotpoint}}
\put(275.58,740.42){\usebox{\plotpoint}}
\multiput(276,740)(13.885,-15.427){0}{\usebox{\plotpoint}}
\put(289.25,724.81){\usebox{\plotpoint}}
\put(302.86,709.15){\usebox{\plotpoint}}
\multiput(303,709)(13.962,-15.358){0}{\usebox{\plotpoint}}
\put(316.80,693.77){\usebox{\plotpoint}}
\put(330.69,678.35){\usebox{\plotpoint}}
\multiput(331,678)(13.143,-16.064){0}{\usebox{\plotpoint}}
\put(344.30,662.70){\usebox{\plotpoint}}
\put(358.04,647.18){\usebox{\plotpoint}}
\multiput(359,646)(13.885,-15.427){0}{\usebox{\plotpoint}}
\put(371.66,631.53){\usebox{\plotpoint}}
\put(385.71,616.29){\usebox{\plotpoint}}
\multiput(387,615)(13.885,-15.427){0}{\usebox{\plotpoint}}
\put(399.47,600.76){\usebox{\plotpoint}}
\put(413.04,585.06){\usebox{\plotpoint}}
\multiput(414,584)(13.962,-15.358){0}{\usebox{\plotpoint}}
\put(426.98,569.68){\usebox{\plotpoint}}
\put(440.87,554.26){\usebox{\plotpoint}}
\multiput(442,553)(13.143,-16.064){0}{\usebox{\plotpoint}}
\put(454.43,538.57){\usebox{\plotpoint}}
\put(468.26,523.13){\usebox{\plotpoint}}
\multiput(470,521)(13.885,-15.427){0}{\usebox{\plotpoint}}
\put(481.88,507.48){\usebox{\plotpoint}}
\put(495.85,492.15){\usebox{\plotpoint}}
\multiput(498,490)(13.885,-15.427){0}{\usebox{\plotpoint}}
\put(509.70,476.71){\usebox{\plotpoint}}
\put(523.22,460.97){\usebox{\plotpoint}}
\multiput(525,459)(13.962,-15.358){0}{\usebox{\plotpoint}}
\put(537.16,445.60){\usebox{\plotpoint}}
\put(551.05,430.17){\usebox{\plotpoint}}
\multiput(553,428)(13.143,-16.064){0}{\usebox{\plotpoint}}
\put(564.56,414.44){\usebox{\plotpoint}}
\put(578.48,399.07){\usebox{\plotpoint}}
\multiput(581,396)(13.885,-15.427){0}{\usebox{\plotpoint}}
\put(592.11,383.42){\usebox{\plotpoint}}
\put(605.98,368.02){\usebox{\plotpoint}}
\multiput(609,365)(13.885,-15.427){0}{\usebox{\plotpoint}}
\put(619.92,352.65){\usebox{\plotpoint}}
\put(633.40,336.88){\usebox{\plotpoint}}
\multiput(636,334)(13.962,-15.358){0}{\usebox{\plotpoint}}
\put(647.34,321.51){\usebox{\plotpoint}}
\put(661.23,306.08){\usebox{\plotpoint}}
\multiput(664,303)(13.143,-16.064){0}{\usebox{\plotpoint}}
\put(674.70,290.30){\usebox{\plotpoint}}
\put(688.71,275.02){\usebox{\plotpoint}}
\multiput(692,271)(13.885,-15.427){0}{\usebox{\plotpoint}}
\put(702.41,259.44){\usebox{\plotpoint}}
\put(716.33,244.04){\usebox{\plotpoint}}
\multiput(720,240)(13.885,-15.427){0}{\usebox{\plotpoint}}
\put(730.17,228.57){\usebox{\plotpoint}}
\put(743.61,212.77){\usebox{\plotpoint}}
\multiput(747,209)(13.962,-15.358){0}{\usebox{\plotpoint}}
\put(757.55,197.39){\usebox{\plotpoint}}
\put(771.43,181.96){\usebox{\plotpoint}}
\multiput(775,178)(13.143,-16.064){0}{\usebox{\plotpoint}}
\put(784.86,166.14){\usebox{\plotpoint}}
\put(798.95,150.94){\usebox{\plotpoint}}
\multiput(803,146)(13.885,-15.427){0}{\usebox{\plotpoint}}
\put(812.61,135.32){\usebox{\plotpoint}}
\put(826.53,119.92){\usebox{\plotpoint}}
\multiput(831,115)(14.676,-14.676){0}{\usebox{\plotpoint}}
\put(833,113){\usebox{\plotpoint}}
\put(220,828){\usebox{\plotpoint}}
\put(220.00,828.00){\usebox{\plotpoint}}
\multiput(229,828)(20.756,0.000){0}{\usebox{\plotpoint}}
\put(240.74,827.81){\usebox{\plotpoint}}
\multiput(248,827)(20.756,0.000){0}{\usebox{\plotpoint}}
\put(261.43,826.51){\usebox{\plotpoint}}
\multiput(266,826)(20.756,0.000){0}{\usebox{\plotpoint}}
\put(282.12,825.32){\usebox{\plotpoint}}
\multiput(285,825)(20.629,-2.292){0}{\usebox{\plotpoint}}
\put(302.59,822.09){\usebox{\plotpoint}}
\multiput(303,822)(20.652,-2.065){0}{\usebox{\plotpoint}}
\multiput(313,821)(20.629,-2.292){0}{\usebox{\plotpoint}}
\put(323.20,819.73){\usebox{\plotpoint}}
\multiput(331,818)(20.261,-4.503){0}{\usebox{\plotpoint}}
\put(343.53,815.65){\usebox{\plotpoint}}
\multiput(350,815)(20.261,-4.503){0}{\usebox{\plotpoint}}
\put(363.91,811.91){\usebox{\plotpoint}}
\multiput(368,811)(19.690,-6.563){0}{\usebox{\plotpoint}}
\put(383.95,806.61){\usebox{\plotpoint}}
\multiput(387,806)(19.690,-6.563){0}{\usebox{\plotpoint}}
\put(403.96,801.23){\usebox{\plotpoint}}
\multiput(405,801)(19.690,-6.563){0}{\usebox{\plotpoint}}
\put(423.77,795.07){\usebox{\plotpoint}}
\multiput(424,795)(19.690,-6.563){0}{\usebox{\plotpoint}}
\multiput(433,792)(19.690,-6.563){0}{\usebox{\plotpoint}}
\put(443.41,788.37){\usebox{\plotpoint}}
\multiput(451,785)(19.880,-5.964){0}{\usebox{\plotpoint}}
\put(462.84,781.18){\usebox{\plotpoint}}
\multiput(470,778)(19.690,-6.563){0}{\usebox{\plotpoint}}
\put(482.13,773.61){\usebox{\plotpoint}}
\multiput(488,771)(19.271,-7.708){0}{\usebox{\plotpoint}}
\put(501.12,765.27){\usebox{\plotpoint}}
\multiput(507,762)(18.967,-8.430){0}{\usebox{\plotpoint}}
\put(519.82,756.30){\usebox{\plotpoint}}
\multiput(525,754)(18.564,-9.282){0}{\usebox{\plotpoint}}
\put(538.41,747.10){\usebox{\plotpoint}}
\multiput(544,744)(18.967,-8.430){0}{\usebox{\plotpoint}}
\put(556.95,737.81){\usebox{\plotpoint}}
\multiput(562,735)(17.798,-10.679){0}{\usebox{\plotpoint}}
\put(574.90,727.39){\usebox{\plotpoint}}
\multiput(581,724)(18.144,-10.080){0}{\usebox{\plotpoint}}
\put(592.89,717.07){\usebox{\plotpoint}}
\multiput(599,713)(18.564,-9.282){0}{\usebox{\plotpoint}}
\put(610.86,706.76){\usebox{\plotpoint}}
\multiput(618,702)(17.270,-11.513){0}{\usebox{\plotpoint}}
\put(628.13,695.25){\usebox{\plotpoint}}
\put(645.69,684.19){\usebox{\plotpoint}}
\multiput(646,684)(16.383,-12.743){0}{\usebox{\plotpoint}}
\put(662.48,672.01){\usebox{\plotpoint}}
\multiput(664,671)(16.383,-12.743){0}{\usebox{\plotpoint}}
\put(679.17,659.68){\usebox{\plotpoint}}
\multiput(683,657)(17.270,-11.513){0}{\usebox{\plotpoint}}
\put(696.15,647.77){\usebox{\plotpoint}}
\multiput(701,644)(15.513,-13.789){0}{\usebox{\plotpoint}}
\put(712.11,634.53){\usebox{\plotpoint}}
\put(728.78,622.17){\usebox{\plotpoint}}
\multiput(729,622)(15.513,-13.789){0}{\usebox{\plotpoint}}
\put(744.30,608.40){\usebox{\plotpoint}}
\multiput(747,606)(16.207,-12.966){0}{\usebox{\plotpoint}}
\put(760.24,595.12){\usebox{\plotpoint}}
\multiput(766,590)(15.513,-13.789){0}{\usebox{\plotpoint}}
\put(775.76,581.33){\usebox{\plotpoint}}
\put(791.59,567.92){\usebox{\plotpoint}}
\multiput(794,566)(14.676,-14.676){0}{\usebox{\plotpoint}}
\put(806.70,553.71){\usebox{\plotpoint}}
\multiput(812,549)(14.676,-14.676){0}{\usebox{\plotpoint}}
\put(821.69,539.38){\usebox{\plotpoint}}
\put(836.82,525.18){\usebox{\plotpoint}}
\multiput(840,522)(14.676,-14.676){0}{\usebox{\plotpoint}}
\put(851.36,510.37){\usebox{\plotpoint}}
\put(866.05,495.75){\usebox{\plotpoint}}
\multiput(868,494)(13.885,-15.427){0}{\usebox{\plotpoint}}
\put(880.13,480.52){\usebox{\plotpoint}}
\put(894.02,465.09){\usebox{\plotpoint}}
\multiput(895,464)(14.676,-14.676){0}{\usebox{\plotpoint}}
\put(908.44,450.18){\usebox{\plotpoint}}
\put(922.33,434.75){\usebox{\plotpoint}}
\multiput(923,434)(13.143,-16.064){0}{\usebox{\plotpoint}}
\put(935.91,419.09){\usebox{\plotpoint}}
\put(949.69,403.60){\usebox{\plotpoint}}
\multiput(951,402)(13.143,-16.064){0}{\usebox{\plotpoint}}
\put(962.84,387.53){\usebox{\plotpoint}}
\put(976.41,371.84){\usebox{\plotpoint}}
\multiput(979,369)(13.143,-16.064){0}{\usebox{\plotpoint}}
\put(989.71,355.91){\usebox{\plotpoint}}
\put(1002.55,339.61){\usebox{\plotpoint}}
\put(1015.60,323.48){\usebox{\plotpoint}}
\multiput(1016,323)(13.143,-16.064){0}{\usebox{\plotpoint}}
\put(1028.55,307.26){\usebox{\plotpoint}}
\put(1041.01,290.66){\usebox{\plotpoint}}
\multiput(1043,288)(12.655,-16.451){0}{\usebox{\plotpoint}}
\put(1053.62,274.18){\usebox{\plotpoint}}
\put(1066.07,257.57){\usebox{\plotpoint}}
\put(1078.14,240.69){\usebox{\plotpoint}}
\multiput(1080,238)(12.655,-16.451){0}{\usebox{\plotpoint}}
\put(1090.62,224.11){\usebox{\plotpoint}}
\put(1102.43,207.04){\usebox{\plotpoint}}
\put(1114.25,189.98){\usebox{\plotpoint}}
\put(1126.70,173.38){\usebox{\plotpoint}}
\multiput(1127,173)(11.224,-17.459){0}{\usebox{\plotpoint}}
\put(1136,159){\usebox{\plotpoint}}
\sbox{\plotpoint}{\rule[-0.200pt]{0.400pt}{0.400pt}}%
\put(220,749){\usebox{\plotpoint}}
\put(396,747.67){\rule{2.168pt}{0.400pt}}
\multiput(396.00,748.17)(4.500,-1.000){2}{\rule{1.084pt}{0.400pt}}
\put(220.0,749.0){\rule[-0.200pt]{42.398pt}{0.400pt}}
\put(442,746.67){\rule{2.168pt}{0.400pt}}
\multiput(442.00,747.17)(4.500,-1.000){2}{\rule{1.084pt}{0.400pt}}
\put(405.0,748.0){\rule[-0.200pt]{8.913pt}{0.400pt}}
\put(470,745.67){\rule{2.168pt}{0.400pt}}
\multiput(470.00,746.17)(4.500,-1.000){2}{\rule{1.084pt}{0.400pt}}
\put(451.0,747.0){\rule[-0.200pt]{4.577pt}{0.400pt}}
\put(488,744.67){\rule{2.409pt}{0.400pt}}
\multiput(488.00,745.17)(5.000,-1.000){2}{\rule{1.204pt}{0.400pt}}
\put(479.0,746.0){\rule[-0.200pt]{2.168pt}{0.400pt}}
\put(507,743.67){\rule{2.168pt}{0.400pt}}
\multiput(507.00,744.17)(4.500,-1.000){2}{\rule{1.084pt}{0.400pt}}
\put(516,742.67){\rule{2.168pt}{0.400pt}}
\multiput(516.00,743.17)(4.500,-1.000){2}{\rule{1.084pt}{0.400pt}}
\put(498.0,745.0){\rule[-0.200pt]{2.168pt}{0.400pt}}
\put(535,741.67){\rule{2.168pt}{0.400pt}}
\multiput(535.00,742.17)(4.500,-1.000){2}{\rule{1.084pt}{0.400pt}}
\put(544,740.67){\rule{2.168pt}{0.400pt}}
\multiput(544.00,741.17)(4.500,-1.000){2}{\rule{1.084pt}{0.400pt}}
\put(553,739.67){\rule{2.168pt}{0.400pt}}
\multiput(553.00,740.17)(4.500,-1.000){2}{\rule{1.084pt}{0.400pt}}
\put(562,738.67){\rule{2.409pt}{0.400pt}}
\multiput(562.00,739.17)(5.000,-1.000){2}{\rule{1.204pt}{0.400pt}}
\put(572,737.67){\rule{2.168pt}{0.400pt}}
\multiput(572.00,738.17)(4.500,-1.000){2}{\rule{1.084pt}{0.400pt}}
\put(581,736.67){\rule{2.168pt}{0.400pt}}
\multiput(581.00,737.17)(4.500,-1.000){2}{\rule{1.084pt}{0.400pt}}
\put(590,735.17){\rule{1.900pt}{0.400pt}}
\multiput(590.00,736.17)(5.056,-2.000){2}{\rule{0.950pt}{0.400pt}}
\put(599,733.67){\rule{2.409pt}{0.400pt}}
\multiput(599.00,734.17)(5.000,-1.000){2}{\rule{1.204pt}{0.400pt}}
\put(609,732.67){\rule{2.168pt}{0.400pt}}
\multiput(609.00,733.17)(4.500,-1.000){2}{\rule{1.084pt}{0.400pt}}
\put(618,731.17){\rule{1.900pt}{0.400pt}}
\multiput(618.00,732.17)(5.056,-2.000){2}{\rule{0.950pt}{0.400pt}}
\put(627,729.17){\rule{1.900pt}{0.400pt}}
\multiput(627.00,730.17)(5.056,-2.000){2}{\rule{0.950pt}{0.400pt}}
\put(636,727.17){\rule{2.100pt}{0.400pt}}
\multiput(636.00,728.17)(5.641,-2.000){2}{\rule{1.050pt}{0.400pt}}
\put(646,725.17){\rule{1.900pt}{0.400pt}}
\multiput(646.00,726.17)(5.056,-2.000){2}{\rule{0.950pt}{0.400pt}}
\put(655,723.17){\rule{1.900pt}{0.400pt}}
\multiput(655.00,724.17)(5.056,-2.000){2}{\rule{0.950pt}{0.400pt}}
\put(664,721.17){\rule{1.900pt}{0.400pt}}
\multiput(664.00,722.17)(5.056,-2.000){2}{\rule{0.950pt}{0.400pt}}
\put(673,719.17){\rule{2.100pt}{0.400pt}}
\multiput(673.00,720.17)(5.641,-2.000){2}{\rule{1.050pt}{0.400pt}}
\multiput(683.00,717.95)(1.802,-0.447){3}{\rule{1.300pt}{0.108pt}}
\multiput(683.00,718.17)(6.302,-3.000){2}{\rule{0.650pt}{0.400pt}}
\put(692,714.17){\rule{1.900pt}{0.400pt}}
\multiput(692.00,715.17)(5.056,-2.000){2}{\rule{0.950pt}{0.400pt}}
\multiput(701.00,712.95)(1.802,-0.447){3}{\rule{1.300pt}{0.108pt}}
\multiput(701.00,713.17)(6.302,-3.000){2}{\rule{0.650pt}{0.400pt}}
\multiput(710.00,709.95)(2.025,-0.447){3}{\rule{1.433pt}{0.108pt}}
\multiput(710.00,710.17)(7.025,-3.000){2}{\rule{0.717pt}{0.400pt}}
\multiput(720.00,706.95)(1.802,-0.447){3}{\rule{1.300pt}{0.108pt}}
\multiput(720.00,707.17)(6.302,-3.000){2}{\rule{0.650pt}{0.400pt}}
\multiput(729.00,703.94)(1.212,-0.468){5}{\rule{1.000pt}{0.113pt}}
\multiput(729.00,704.17)(6.924,-4.000){2}{\rule{0.500pt}{0.400pt}}
\multiput(738.00,699.95)(1.802,-0.447){3}{\rule{1.300pt}{0.108pt}}
\multiput(738.00,700.17)(6.302,-3.000){2}{\rule{0.650pt}{0.400pt}}
\multiput(747.00,696.94)(1.358,-0.468){5}{\rule{1.100pt}{0.113pt}}
\multiput(747.00,697.17)(7.717,-4.000){2}{\rule{0.550pt}{0.400pt}}
\multiput(757.00,692.94)(1.212,-0.468){5}{\rule{1.000pt}{0.113pt}}
\multiput(757.00,693.17)(6.924,-4.000){2}{\rule{0.500pt}{0.400pt}}
\multiput(766.00,688.94)(1.212,-0.468){5}{\rule{1.000pt}{0.113pt}}
\multiput(766.00,689.17)(6.924,-4.000){2}{\rule{0.500pt}{0.400pt}}
\multiput(775.00,684.94)(1.212,-0.468){5}{\rule{1.000pt}{0.113pt}}
\multiput(775.00,685.17)(6.924,-4.000){2}{\rule{0.500pt}{0.400pt}}
\multiput(784.00,680.93)(1.044,-0.477){7}{\rule{0.900pt}{0.115pt}}
\multiput(784.00,681.17)(8.132,-5.000){2}{\rule{0.450pt}{0.400pt}}
\multiput(794.00,675.93)(0.933,-0.477){7}{\rule{0.820pt}{0.115pt}}
\multiput(794.00,676.17)(7.298,-5.000){2}{\rule{0.410pt}{0.400pt}}
\multiput(803.00,670.93)(0.933,-0.477){7}{\rule{0.820pt}{0.115pt}}
\multiput(803.00,671.17)(7.298,-5.000){2}{\rule{0.410pt}{0.400pt}}
\multiput(812.00,665.93)(0.933,-0.477){7}{\rule{0.820pt}{0.115pt}}
\multiput(812.00,666.17)(7.298,-5.000){2}{\rule{0.410pt}{0.400pt}}
\multiput(821.00,660.93)(1.044,-0.477){7}{\rule{0.900pt}{0.115pt}}
\multiput(821.00,661.17)(8.132,-5.000){2}{\rule{0.450pt}{0.400pt}}
\multiput(831.00,655.93)(0.762,-0.482){9}{\rule{0.700pt}{0.116pt}}
\multiput(831.00,656.17)(7.547,-6.000){2}{\rule{0.350pt}{0.400pt}}
\multiput(840.00,649.93)(0.762,-0.482){9}{\rule{0.700pt}{0.116pt}}
\multiput(840.00,650.17)(7.547,-6.000){2}{\rule{0.350pt}{0.400pt}}
\multiput(849.00,643.93)(0.762,-0.482){9}{\rule{0.700pt}{0.116pt}}
\multiput(849.00,644.17)(7.547,-6.000){2}{\rule{0.350pt}{0.400pt}}
\multiput(858.00,637.93)(0.721,-0.485){11}{\rule{0.671pt}{0.117pt}}
\multiput(858.00,638.17)(8.606,-7.000){2}{\rule{0.336pt}{0.400pt}}
\multiput(868.00,630.93)(0.645,-0.485){11}{\rule{0.614pt}{0.117pt}}
\multiput(868.00,631.17)(7.725,-7.000){2}{\rule{0.307pt}{0.400pt}}
\multiput(877.00,623.93)(0.645,-0.485){11}{\rule{0.614pt}{0.117pt}}
\multiput(877.00,624.17)(7.725,-7.000){2}{\rule{0.307pt}{0.400pt}}
\multiput(886.00,616.93)(0.645,-0.485){11}{\rule{0.614pt}{0.117pt}}
\multiput(886.00,617.17)(7.725,-7.000){2}{\rule{0.307pt}{0.400pt}}
\multiput(895.00,609.93)(0.626,-0.488){13}{\rule{0.600pt}{0.117pt}}
\multiput(895.00,610.17)(8.755,-8.000){2}{\rule{0.300pt}{0.400pt}}
\multiput(905.00,601.93)(0.560,-0.488){13}{\rule{0.550pt}{0.117pt}}
\multiput(905.00,602.17)(7.858,-8.000){2}{\rule{0.275pt}{0.400pt}}
\multiput(914.00,593.93)(0.560,-0.488){13}{\rule{0.550pt}{0.117pt}}
\multiput(914.00,594.17)(7.858,-8.000){2}{\rule{0.275pt}{0.400pt}}
\multiput(923.00,585.93)(0.495,-0.489){15}{\rule{0.500pt}{0.118pt}}
\multiput(923.00,586.17)(7.962,-9.000){2}{\rule{0.250pt}{0.400pt}}
\multiput(932.00,576.93)(0.553,-0.489){15}{\rule{0.544pt}{0.118pt}}
\multiput(932.00,577.17)(8.870,-9.000){2}{\rule{0.272pt}{0.400pt}}
\multiput(942.59,566.74)(0.489,-0.553){15}{\rule{0.118pt}{0.544pt}}
\multiput(941.17,567.87)(9.000,-8.870){2}{\rule{0.400pt}{0.272pt}}
\multiput(951.00,557.93)(0.495,-0.489){15}{\rule{0.500pt}{0.118pt}}
\multiput(951.00,558.17)(7.962,-9.000){2}{\rule{0.250pt}{0.400pt}}
\multiput(960.59,547.56)(0.489,-0.611){15}{\rule{0.118pt}{0.589pt}}
\multiput(959.17,548.78)(9.000,-9.778){2}{\rule{0.400pt}{0.294pt}}
\multiput(969.00,537.92)(0.495,-0.491){17}{\rule{0.500pt}{0.118pt}}
\multiput(969.00,538.17)(8.962,-10.000){2}{\rule{0.250pt}{0.400pt}}
\multiput(979.59,526.56)(0.489,-0.611){15}{\rule{0.118pt}{0.589pt}}
\multiput(978.17,527.78)(9.000,-9.778){2}{\rule{0.400pt}{0.294pt}}
\multiput(988.59,515.56)(0.489,-0.611){15}{\rule{0.118pt}{0.589pt}}
\multiput(987.17,516.78)(9.000,-9.778){2}{\rule{0.400pt}{0.294pt}}
\multiput(997.59,504.37)(0.489,-0.669){15}{\rule{0.118pt}{0.633pt}}
\multiput(996.17,505.69)(9.000,-10.685){2}{\rule{0.400pt}{0.317pt}}
\multiput(1006.58,492.59)(0.491,-0.600){17}{\rule{0.118pt}{0.580pt}}
\multiput(1005.17,493.80)(10.000,-10.796){2}{\rule{0.400pt}{0.290pt}}
\multiput(1016.59,480.19)(0.489,-0.728){15}{\rule{0.118pt}{0.678pt}}
\multiput(1015.17,481.59)(9.000,-11.593){2}{\rule{0.400pt}{0.339pt}}
\multiput(1025.59,467.19)(0.489,-0.728){15}{\rule{0.118pt}{0.678pt}}
\multiput(1024.17,468.59)(9.000,-11.593){2}{\rule{0.400pt}{0.339pt}}
\multiput(1034.59,454.00)(0.489,-0.786){15}{\rule{0.118pt}{0.722pt}}
\multiput(1033.17,455.50)(9.000,-12.501){2}{\rule{0.400pt}{0.361pt}}
\multiput(1043.58,440.26)(0.491,-0.704){17}{\rule{0.118pt}{0.660pt}}
\multiput(1042.17,441.63)(10.000,-12.630){2}{\rule{0.400pt}{0.330pt}}
\multiput(1053.59,426.00)(0.489,-0.786){15}{\rule{0.118pt}{0.722pt}}
\multiput(1052.17,427.50)(9.000,-12.501){2}{\rule{0.400pt}{0.361pt}}
\multiput(1062.59,411.82)(0.489,-0.844){15}{\rule{0.118pt}{0.767pt}}
\multiput(1061.17,413.41)(9.000,-13.409){2}{\rule{0.400pt}{0.383pt}}
\multiput(1071.59,396.82)(0.489,-0.844){15}{\rule{0.118pt}{0.767pt}}
\multiput(1070.17,398.41)(9.000,-13.409){2}{\rule{0.400pt}{0.383pt}}
\multiput(1080.58,381.93)(0.491,-0.808){17}{\rule{0.118pt}{0.740pt}}
\multiput(1079.17,383.46)(10.000,-14.464){2}{\rule{0.400pt}{0.370pt}}
\multiput(1090.59,365.45)(0.489,-0.961){15}{\rule{0.118pt}{0.856pt}}
\multiput(1089.17,367.22)(9.000,-15.224){2}{\rule{0.400pt}{0.428pt}}
\multiput(1099.59,348.45)(0.489,-0.961){15}{\rule{0.118pt}{0.856pt}}
\multiput(1098.17,350.22)(9.000,-15.224){2}{\rule{0.400pt}{0.428pt}}
\multiput(1108.59,331.45)(0.489,-0.961){15}{\rule{0.118pt}{0.856pt}}
\multiput(1107.17,333.22)(9.000,-15.224){2}{\rule{0.400pt}{0.428pt}}
\multiput(1117.58,314.60)(0.491,-0.912){17}{\rule{0.118pt}{0.820pt}}
\multiput(1116.17,316.30)(10.000,-16.298){2}{\rule{0.400pt}{0.410pt}}
\multiput(1127.59,296.08)(0.489,-1.077){15}{\rule{0.118pt}{0.944pt}}
\multiput(1126.17,298.04)(9.000,-17.040){2}{\rule{0.400pt}{0.472pt}}
\put(525.0,743.0){\rule[-0.200pt]{2.409pt}{0.400pt}}
\end{picture}
\end{center}
\caption{Rho mass versus lattice spacing computed using the Wilson
quark action (W), the Sheikholeslami-Wohlert action (SW) and the D234x
action. The charmonium S-P splitting was used to determine
$a_s$. These simulations have no quark vacuum polarization and
therefore the rho mass is expected to differ from its experimental value.}
\label{mrho-a-fig}
\end{figure}

Finally, in Fig.~\ref{mrho-a-fig}, I show simulation results for the rho mass
as a function of lattice spacing. I give results for the
D234x~action\,\cite{alford96}, which
should have $a_s^4$~errors, for the SW~action\,\cite{edwards96}, which
should have $a^2$~errors,
and for the Wilson action\,\cite{weingarten94}, which should have
$a^1$~errors. The D234x
and SW simulations use an improved gluon action; the Wilson results are
for an uncorrected gluon action. Assuming that the
$a$~dependence of the errors is as expected, all three actions agree quite well
on the continuum limit ($a\!\to\!0$) mass of the rho, but the more improved the
quark action, the sooner it gets there. This figure demonstrates very clearly
that we  know how to systematically improve quark actions for use on very
coarse lattices. Furthermore it is clear that simulations that are accurate to
within a few percent are possible on lattices with spatial lattice spacings
between .3~and .4\,fm.

\subsection{$t$-Staggered Quarks}
A second approach to removing doublers is to ``stagger'' the quark degrees of
freedom on the lattice. If we rewrite the Dirac field in terms of two
two-component fields, $\phi$ and $\chi$,
\be
\psi = \left(\gamma_t\right)^{t/a_t}\,
\left[ \begin{array}{c}\phi+\chi\\ \phi-\chi\end{array}\right]
\ee
the naive lattice Dirac equation~\eq{naive-dirac} falls apart into two
decoupled two-component equations:
\bearray\label{staggered-t}
\left(m_c + \Delta_t + (-1)^{t/a_t}\Deltav\cdot\sigmav\right) \phi &=& 0\\
\left(m_c + \Delta_t + (-1)^{1+t/a_t}\Deltav\cdot\sigmav\right) \chi &=& 0
\eearray
The second equation is identical to the first equation, but shifted
one lattice spacing in the $t$~direction; it leads to identical quark
physics. One of these equations describes the quark while the other
describes the degenerate doubler. We get rid of the (temporal)
doubler simply by
dropping one of the two equations, say the second one. The remaining
equation describes two quark polarizations and two antiquark
polarizations, but no doublers. This approach is called ``staggered''
because dropping the $\chi$~field is equivalent to specifying only
half the spinor components of the original Dirac field on
time slices of the lattice with even~$t/a_t$, while only the other half are
specified on time slices with odd~$t/a_t$. The spinor components are
staggered between time slices.

This discretization still has $a^2$~errors and the spatial derivatives
have poor behavior at high momentum. These problems are easily remedied by
replacing
\bearray
\Deltav &\to& \Deltav - \frac{a_s^2}{6}\,\Deltav^{(3)}\\
m &\to& m - s\,a_s^5\,\left(\Deltav^{(2)}\right)^3
\eearray
in~\eq{staggered-t},
where the correction terms involve spatial derivatives only.
The parameter~$s$ is set at some small value ($\approx .02$) that is large
enough to keep the high-momentum energy large but small enough that is has
little effect on the low-momentum physics. The corrected equations still have
$a_t^2$~errors, but these are negligible on anisotropic lattices with
$a_t \le a_s/2$.

This action has never been tried. It should be as accurate as the
D234-type actions. It has half as many fields, no field
transformation, no ghosts, and is somewhat simpler than the
D234~actions; it may well be competitive.

A more conventional action, the Kogut-Susskind or staggered-quark action, is
obtained by staggering the quark degrees of freedom in spatial as well as
temporal directions. This reduces the number of doublers to only three
(from fifteen in the original naive action). Rather
than removing the remaining doublers,
in this approach one treats them as three extra
flavors of quark. If these extra flavors were degenerate with the
original quark, this approach might be computationally very efficient, perhaps
giving four times the statistics for the same computational cost as other
discretizations. Unfortunately the gluonic interactions lift the degeneracy
between the flavors in the simplest version of this theory, resulting in
significant mass differences. It would be very interesting to develop improved
versions of this theory where the degeneracy is more accurately preserved.

\subsection{Summary}
Temporal derivatives usually cause problems when improving the
discretization of an action. This is because successive improvements
to the temporal derivatives
involve operators that are increasingly nonlocal in time
and therefore that almost inevitably
result in
ghosts of one sort or another. This problem is unusually serious
for quark actions where even the leading-order discretization leads to
doublers.

As illustrated in the previous sections, there are two
standard techniques for removing the doublers. One is to introduce
redundant operators that break the doubling symmetry; this leads to
the SW and D234-type actions. The second is to stagger the quark
spinor components on the lattice; this leads to actions for
$t$-staggered quarks and Kogut-Susskind quarks.

Having repaired the
leading-order operator, we face an even more acute problem in
the $a_t^2$ correction to $\Delta_t\gamma_t$. This can be dealt with,
as in the D234 actions, by using anisotropic lattices, with reduced
$a_t$'s, to push the ghosts to high energies,
and by adding irrelevant operators to cancel some of the ghosts.
Alternatively, using anisotropic lattices, one
might leave the $a_t^2$~errors uncorrected; these will be negligible
compared with $a_s^4$ errors provided $a_t$ is small enough
compared with $a_s$.
(Numerical experiments suggest that the $a_t^2$ errors are only a few
percent when $a_t$ is less than .2\,fm.) A more complicated
alternative, which I have not discussed, is to add irrelevant
operators that replace the $a_t^2 \lder{3}_t\gamma_t$ correction term
with less troublesome operators: for example, one
can replace
\bearray
\lder{3}_t\gamma_t &\,\,\to\,\,&
-\half\left[\left(\Deltav\cdot\gamma+m_{\rm c}\right)
(\Delta_t\gamma_t)^2 + (\Delta_t\gamma_t)^2
\left(\Deltav\cdot\gamma+m_{\rm c}\right) \right]
\\
&\to& \left(\Deltav\cdot\gamma+m_{\rm c}\right) \Delta_t\gamma_t
\left(\Deltav\cdot\gamma+m_{\rm c}\right)
\\
&\to& -\half\left\{\Deltav\cdot\gamma+m_{\rm c}\, ,\,
m_{\rm c} - \Deltav^{(2)} - \sigma\cdot gF/2\right\}
\eearray
using successive field transformations and obtain, in the last step, an
operator that has only spatial derivatives.

The numerical results in this section suggest that these various
strategies work about as well as expected, and that accurate
simulations are possible with spatial lattice spacings between .3~and
.4\,fm. They also show that the meson dispersion relation (i.e., $c(\pv)$) at
high and low momenta, the $r$-test and the $\rho$~mass are all very
sensitive to finite-$a$ errors.
\newcommand{\Mbz}{{M_{\rm b}^0}}
\newcommand{\Ev}{{\bf E}}
\newcommand{\Bv}{{\bf B}}
\newcommand{\msb}{{\rm\overline{MS}}}
\section{Heavy Quarks}

As a final illustration of improved actions I briefly review
techniques for simulating heavy quarks like the $\rm c$~and $\rm
b$~quarks. Heavy quarks play a central role in the experimental study
of weak interactions, and consequently it is important that we have
reliable simulation techniques for their study. However heavy quark
dynamics is not easily simulated using the quark actions of the
previous section. Those actions only work well for quarks with
energies and momenta small compared with the ultraviolet
cutoff~$\pi/a$.  But heavy-quark hadrons have large rest
energies (masses) and so necessitate small lattice
spacings.
Simulating an~$\Upsilon$, for example, using such techniques
requires a lattice spacing of order $1/9.4\,{\rm GeV} = .02$\,fm or
less\,---\,much too costly with today's computers. Fortunately the
heavy quarks are generally nonrelativistic in such mesons. Most of
such a meson's energy is due to the rest mass energy of its
heavy-quark constituents, and rest mass energy plays little role in
the dynamics of nonrelativistic particles. Consequently we can develop
accurate simulation techniques that work well when the lattice spacing
is chosen according to the size, rather than the mass, of the
heavy-quark hadron: that is, for example, $a\!=\!.1$\,fm rather than
.02\,fm for upsilons.\footnote{Actually since it is only the energy
that is large,
only the temporal lattice spacing need be small. Consequently the
actions from the previous section can be used efficiently if one works
with anisotropic lattices and small~$a_t$'s. Experiments using
D234~actions are very encouraging\,\cite{alford96}.
Yet another approach to heavy-quark dynamics can be found in\,\cite{flab96}.}

Here I discuss only the NRQCD approach to heavy quark dynamics,
where one takes maximum advantage of the fact that the heavy quarks  are
nonrelativistic\,\cite{gpl88}. I illustrate the technique by reviewing
simulation results for
the upsilon and psi families of mesons. These are attractive systems to study
because so much is known already about them. The quark potential model, for
example, provides an accurate phenomenological model of the internal structure
of the mesons. Also there is  much experimental data for these mesons. The
low-lying states are largely insensitive to light-quark vacuum
polarization\,---\,for example, the
$\Upsilon$, $\Upsilon^\prime$,
$\chi_{\rm b}$\,\ldots\,are all far below the threshold for decays into
$B$~mesons\,---\,and therefore can be accurately simulated with
$n_{\rm f}\!=\!0$ light-quark flavors. Furthermore, these mesons are very
small; for
example, the~$\Upsilon$ is about five times smaller than a light hadron. This
makes them an excellent testing ground for our ideas concerning large lattice
spacings.

Given that the heavy quarks are nonrelavistic, the most
important momentum scales governing a heavy-quark meson's dynamics are smaller
than the heavy quark's mass. We can take advantage of this fact by choosing an
inverse lattice spacing of order the quark mass, thereby excluding relativistic
states from the theory. Then it is efficient to analyze the heavy-quark
dynamics
using a nonrelativistic lagrangian (NRQCD). The lagrangian used to generate the
results I show below was
\bearray \label{nrqcd-lag}
\Lag_{\rm NRQCD} &= &
\sum_x \left\{\psi^\dagger(x\!+\!a_t \hat{t})
\left(1\!-\!\frac{a_tH_0}{2n}\right)^{n}
 U^\dagger_{\hat t}
 \left(1\!-\!\frac{a_tH_0}{2n}\right)^{n}\left(1\!-\!a_t\delta H\right) \psi(x)
\right\}\nl
 &-&\sum _x \psi^\dagger(x)\psi(x),
 \eearray
where $n=2$, $H_0$ is the nonrelativistic kinetic-energy operator,
 \be
 H_0 = - {\Deltav^{(2)}\over2M_0},
 \ee
$M_0$~is the bare heavy-quark mass, and $\delta H$ is the leading relativistic
and finite-lattice-spacing correction,
 \begin{eqnarray}
\delta H
&=& - \frac{(\Deltav^{(2)})^2}{8M_0^3}\left(1+\frac{a_t M_0}{2n}\right)
    + \frac{a_s^2\Deltav^{(4)}}{24M_0} \nl
& & - \frac{g}{2M_0}\,\sigmav\cdot\Bv
            + \frac{ig}{8M_0^2}\left(\Deltav\cdot\Ev - \Ev\cdot\Deltav\right)
\nl & & - \frac{g}{8M_0^2} \sigmav\cdot(\Deltav\times\Ev - \Ev\times\Deltav) .
\label{deltaH}
\end{eqnarray}
Here $\Ev$~and~$\Bv$ are the chromoelectric and chromomagnetic fields.
As usual, the entire  action is tadpole
improved by dividing every link operator~$U_\mu$ by $u_0$.  Potential models
indicate that corrections beyond~$\delta H$ contribute only of order
5~MeV to $\Upsilon$ energies, and two or three times  this
for~$\psi$'s.

The details of this lagrangian are  unimportant to us here. What I want to
focus on is the extent to which the improvement program outlined in earlier
sections is effective for heavy-quark dynamics. The
$\delta H$~term consists of corrections just like the ones we have been
analyzing for the other parts of QCD, the only difference here being that we
are
correcting both for finite-$a$ and for the absence of relativity (ie,
order~$v^2/c^2$ errors). The coefficients of the correction terms were
determined with tree-level perturbation theory and tadpole improvement, using
precisely the techniques outlined in the earlier sections. The extent to
which~$\delta H$ improves the simulation results is a measure of the
efficacy of
all the techniques discussed in these lectures.

\begin{figure}
\begin{center}\mbox{}
\setlength{\unitlength}{.02in}
\begin{picture}(130,150)(5,920)
\put(10,935){\line(0,1){125}}
\multiput(8,950)(0,50){3}{\line(1,0){4}}
\multiput(9,950)(0,10){10}{\line(1,0){2}}
\put(7,950){\makebox(0,0)[r]{9.5}}
\put(7,1000){\makebox(0,0)[r]{10.0}}
\put(7,1050){\makebox(0,0)[r]{10.5}}
\put(7,1065){\makebox(0,0)[r]{GeV}}



\put(27,920){\makebox(0,0)[t]{${^1\rm{}S}_0$}}
\put(25,943.1){\circle*{3}}
\put(30,942){\circle{3}}

\put(52,920){\makebox(0,0)[t]{${^3\rm{}S}_1$}}
\multiput(43,946)(3,0){7}{\line(1,0){2}}
\put(50,946){\circle*{3}}
\put(55,946){\circle{3}}

\multiput(43,1002)(3,0){7}{\line(1,0){2}}
\put(50,1004.1){\circle*{3}}
\put(50,1005.1){\line(0,1){0.2}}
\put(50,1003.1){\line(0,-1){0.2}}
\put(55,1003){\circle{3}}
\put(55,1004){\line(0,1){1.4}}
\put(55,1002){\line(0,-1){1.4}}

\multiput(43,1036)(3,0){7}{\line(1,0){2}}
\put(50,1060){\circle*{3}}
\put(50,1061){\line(0,1){11}}
\put(50,1059){\line(0,-1){11}}
\put(55,1039.1){\circle{3}}
\put(55,1039.1){\line(0,1){7.2}}
\put(55,1039.1){\line(0,-1){7.2}}

\put(92,920){\makebox(0,0)[t]{${^1\rm{}P}_1$}}

\multiput(83,990)(3,0){7}{\line(1,0){2}}
\put(90,987.6){\circle*{3}}
\put(95,989){\circle{3}}
\put(95,990){\line(0,1){0.2}}
\put(95,988){\line(0,-1){0.2}}

\multiput(83,1026)(3,0){7}{\line(1,0){2}}
\put(90,1038.7){\circle*{3}}
\put(90,1039.7){\line(0,1){1.4}}
\put(90,1037.7){\line(0,-1){1.4}}
\put(95,1023){\circle{3}}
\put(95,1023){\line(0,1){7.2}}
\put(95,1023){\line(0,-1){7.2}}

\put(120,920){\makebox(0,0)[t]{${^1\rm{}D}_2$}}
\put(120,1019.2){\circle*{3}}
\put(120,1020.2){\line(0,1){6}}
\put(120,1018.2){\line(0,-1){6}}
\end{picture}
\end{center}
\caption{NRQCD simulation results for the spectrum of the
$\Upsilon (^3\rm{}S_1)$ and $h_b (^1\rm{}P_1)$ and their radial excitations.
Experimental values (dashed lines) are indicated for the S-states, and for the
spin-average of the P-states. Simulation results are for $n_{\rm f}\!=\!0$
(solid circles) and $n_{\rm f}\!=\!2$ (open circles) light-quark flavors. The
energy zero for the simulation results is adjusted to give the correct mass to
the~$\Upsilon$.}
\label{spectups}
\end{figure}
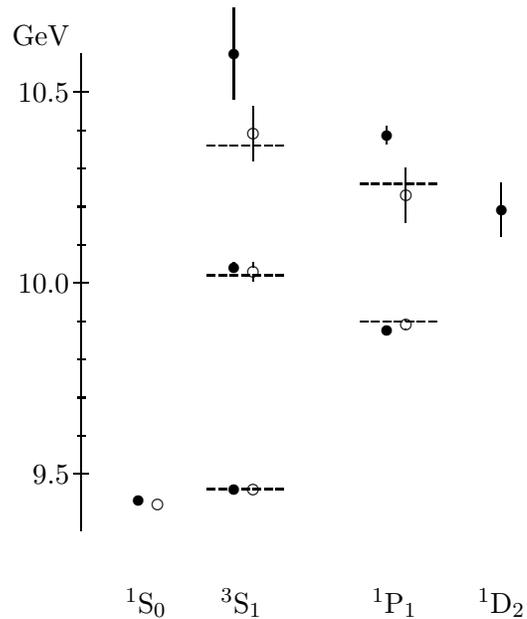

Consider first a set of simulations of the $\Upsilon$ family that used
a lattice
spacing of about $1/12$\,fm\,---\,roughly half the meson's
radius\,\cite{davies94}.
The spectrum is very well described by the simulations; results for the
low-lying spectrum in different channels are shown in
Figure~\ref{spectups}. These compare well with experimental results (the
horizontal lines), as they should since systematic errors are estimated to be
less than 10--20\,MeV. Two sets of Monte Carlo results are shown: one has no
light-quark vacuum polarization, the other includes $\rm u$ ~and $\rm
d$~quarks; as expected quark vacuum polarization has little effect on the
specturm. It is important to realize that these are calculations from first
principles. The only inputs are the lagrangians describing gluon and quark
dynamics, and the only parameters are the bare coupling constant and the bare
quark mass. In particular, these results are not based on a phenomenological
quark-potential model. These are among the most accurate lattice results to
date.

\begin{figure}
\begin{center}
\setlength{\unitlength}{.02in}
\begin{picture}(95,100)(40,-50)

\put(50,-50){\line(0,1){85}}
\multiput(48,-40)(0,20){4}{\line(1,0){4}}
\multiput(49,-40)(0,10){7}{\line(1,0){2}}
\put(47,-40){\makebox(0,0)[r]{$-40$}}
\put(47,-20){\makebox(0,0)[r]{$-20$}}
\put(47,0){\makebox(0,0)[r]{$0$}}
\put(47,20){\makebox(0,0)[r]{$20$}}
\put(47,35){\makebox(0,0)[r]{MeV}}


\put(63,-5){\makebox(0,0)[l]{$h_{\rm b}$}}

\put(70,-0.8){\circle*{3}}
\put(75,-2.9){\circle{3}}
\put(75,-2.9){\line(0,1){1.2}}
\put(75,-2.9){\line(0,-1){1.2}}
\multiput(90,-40)(3,0){7}{\line(1,0){2}}
\put(110,-40){\makebox(0,0)[l]{$\chi_{\rm b0}$}}
\put(97,-24){\circle*{3}}
\put(97,-23){\line(0,1){1}}
\put(97,-25){\line(0,-1){1}}
\put(102,-34){\circle{3}}
\put(102,-34){\line(0,1){5}}
\put(102,-34){\line(0,-1){5}}

\multiput(90,-8)(3,0){7}{\line(1,0){2}}
\put(110,-8){\makebox(0,0)[l]{$\chi_{\rm b1}$}}
\put(97,-8.6){\circle*{3}}
\put(102,-7.9){\circle{3}}
\put(102,-7.9){\line(0,1){2.4}}
\put(102,-7.9){\line(0,-1){2.4}}

\multiput(90,13)(3,0){7}{\line(1,0){2}}
\put(110,13){\makebox(0,0)[l]{$\chi_{\rm b2}$}}
\put(97,10.1){\circle*{3}}

\put(102,11.5){\circle{3}}
\put(102,11.5){\line(0,1){2.4}}
\put(102,11.5){\line(0,-1){2.4}}
\end{picture}
\end{center}
\caption{Simulation results for the spin structure of the lowest lying
P-state in the $\Upsilon$~family. The dashed lines are the experimental
values for the triplet states.Simulation results are for $n_{\rm f}\!=\!0$
(solid circles) and $n_{\rm f}\!=\!2$ (open circles) light-quark flavors.}
\label{fsups}
\end{figure}
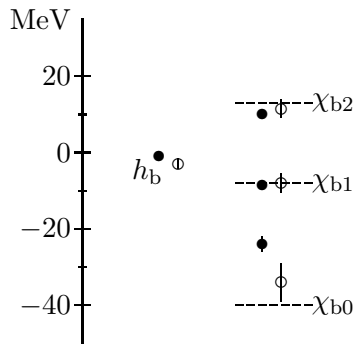

The corrections in~$\delta H$ have only a small effect on the overall
spectrum, but the spin structure is strongly affected. The lagrangian
without~$\delta H$ is spin independent, and gives no spin splittings at all.
Simulation results for the spin structure of the lowest lying P~state are
shown in Figure~\ref{fsups}. Again these compare very well with the data,
giving strong evidence that corrected lagrangians work. Systematic errors
here are estimated to be of order~5~MeV. Note that the spin terms in~$\delta
H$ all involve either chromoelectric or chromomagnetic field operators. These
operators are built from products of four-link operators and so tadpole
improvement increases their magnitude by almost a factor of two
at the lattice spacing used here. Simulations without tadpole improvement give
spin splittings that are much too small.

The correction terms also affect the extent to which the simulation
reproduces continuum symmetries. For example, the low-momentum
dispersion relation of an upsilon in the simulation must have the form
\be
E_\Upsilon(\pv) = M_1 + \frac{\pv^2}{2\,M_2} + \cdots.
\ee
In a relativistically invariant theory $M_1\!=\!M_2$ is the upsilon mass.
However in a purely nonrelativistic theory $M_2$
equals the sum of the quark masses, which differs from $M_1$ since $M_1$
also includes the quarks' binding energy. Only when
relativistic corrections are added is $M_2$ shifted to include the
binding energy. Simulations tuned to the correct mass $M_1\!=\!9.5(1)$~GeV
give $M_2\!=\!8.2(1)$~GeV when $\delta H$  is omitted. When~$\delta H$ is
included,  the simulations give $M_2\!=\!9.5(1)$~GeV, in excellent agreement
with $M_1$.
All of the spin-independent pieces of~$\delta H$ contribute to the shift
in~$M_2$; once again we have striking evidence that
corrected actions work.

This $\Upsilon$~simulation has only two parameters: the bare coupling constant
and the bare quark mass. These were tuned until the simulation agreed with
experimental data. From the bare parameters we can compute the renormalized
coupling and mass. This simulation implies that the renormalized or ``pole
mass'' of the
b~quark is\,\cite{davies94b}
\be
M_{\rm b} = 5.0\,(2)\,{\rm GeV}.
\ee
The renormalized coupling that is obtained
corresponds to\,\cite{davies95a}
\be
\alpha_\msb^{(5)}(M_{\rm Z}) = \cases{ .1175\,(25) &\mbox{from $\chi_{\rm
b}$--$\Upsilon$ splitting}\cr
.1180\,(27)&\mbox{from $\Upsilon^\prime$--$\Upsilon$ splitting}}
\ee
depending upon which $\Upsilon$~splitting in the simulation is tuned to agree
exactly with experiment. These values agree
with results from high-energy phenomenology (but  are more accurate). This
last result is striking: it shows that the QCD of hadronic structure and the
QCD of high-energy quark and gluon jets are really the same theory.

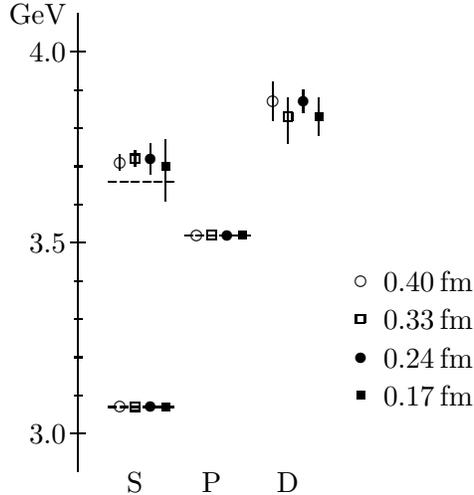
\begin{figure}
\begin{center}
\setlength{\unitlength}{.02in}
\begin{picture}(100,120)(10,290)
\put(89,340){\circle{3}}\put(95,340){\makebox(0,0)[l]{$0.40$\,fm}}
\put(88,329){\framebox(2,2){\mbox{}}}
             \put(95,330){\makebox(0,0)[l]{$0.33$\,fm}}
\put(89,320){{\circle*{3}}}\put(95,320){\makebox(0,0)[l]{$0.24$\,fm}}
\put(88,310){\rule[-\unitlength]{2\unitlength}{2\unitlength}}
         \put(95,310){\makebox(0,0)[l]{$0.17$\,fm}}

\put(15,290){\line(0,1){120}}
\multiput(13,300)(0,50){3}{\line(1,0){4}}
\multiput(14,310)(0,10){9}{\line(1,0){2}}
\put(12,300){\makebox(0,0)[r]{3.0}}
\put(12,350){\makebox(0,0)[r]{3.5}}
\put(12,400){\makebox(0,0)[r]{4.0}}
\put(12,410){\makebox(0,0)[r]{GeV}}

\put(30,290){\makebox(0,0)[t]{S}}

\multiput(23,307)(3,0){6}{\line(1,0){2}}
\put(26,307){\circle{3}}
\put(29,306){\framebox(2,2){\mbox{}}}
\put(34,307){\circle*{3}}
\put(37,306){\rule{2\unitlength}{2\unitlength}}

\multiput(23,366)(3,0){6}{\line(1,0){2}}
\put(26,371){\circle{3}}
\put(26,369){\line(0,1){4}}
\put(29,371){\framebox(2,2){\mbox{}}}
\put(30,370){\line(0,1){4}}
\put(34,372){\circle*{3}}
\put(34,368){\line(0,1){8}}
\put(37,369){\rule{2\unitlength}{2\unitlength}}
\put(38,361){\line(0,1){16}}

\put(50,290){\makebox(0,0)[t]{P}}

\multiput(43,352)(3,0){6}{\line(1,0){2}}
\put(46,352){\circle{3}}
\put(49,351){\framebox(2,2){}}
\put(54,352){\circle*{3}}
\put(54,351){\line(0,1){2}}
\put(57,351){\rule{2\unitlength}{2\unitlength}}
\put(58,351.5){\line(0,1){1}}

\put(70,290){\makebox(0,0)[t]{D}}

\put(66,387){\circle{3}}
\put(66,382){\line(0,1){10}}
\put(69,382){\framebox(2,2){}}
\put(70,376){\line(0,1){12}}
\put(74,387){\circle*{3}}
\put(74,384){\line(0,1){6}}
\put(77,382){\rule{2\unitlength}{2\unitlength}}
\put(78,378){\line(0,1){10}}
\end{picture}

\end{center}
\caption{ S, P and D states of charmonium computed on lattices with a range of
lattice spacings. Improved quark and gluon actions were used in each case
except $a\!=\!.17$\,fm where only the quark action was improved.}
\label{spectpsi}
\end{figure}

Another check on the corrected action is to show that results for physical
quantities are independent of the lattice spacing. In Fig.~\ref{spectpsi}
I show the low-lying spin-averaged spectrum of the $\psi$~mesons
computed using very coarse lattices, with $a$~ranging from .17\,fm
to~.40\,fm\,\cite{alford95}.
The 1S~and 1P~masses are tuned; the 2S~and 1D~masses are predicted. The latter
masses are independent of lattice spacing to within statistical errors of a few
percent.

\begin{figure}
\begin{center}
\setlength{\unitlength}{0.240900pt}
\ifx\plotpoint\undefined\newsavebox{\plotpoint}\fi
\sbox{\plotpoint}{\rule[-0.175pt]{0.350pt}{0.350pt}}%
\begin{picture}(1200,900)(0,0)
\tenrm
\sbox{\plotpoint}{\rule[-0.175pt]{0.350pt}{0.350pt}}%
\put(264,158){\rule[-0.175pt]{210.065pt}{0.350pt}}
\put(264,158){\rule[-0.175pt]{0.350pt}{151.526pt}}
\put(264,420){\rule[-0.175pt]{4.818pt}{0.350pt}}
\put(242,420){\makebox(0,0)[r]{$5$}}
\put(1116,420){\rule[-0.175pt]{4.818pt}{0.350pt}}
\put(264,682){\rule[-0.175pt]{4.818pt}{0.350pt}}
\put(242,682){\makebox(0,0)[r]{$10$}}
\put(1116,682){\rule[-0.175pt]{4.818pt}{0.350pt}}
\put(264,158){\rule[-0.175pt]{0.350pt}{4.818pt}}
\put(264,113){\makebox(0,0){$0$}}
\put(264,767){\rule[-0.175pt]{0.350pt}{4.818pt}}
\put(700,158){\rule[-0.175pt]{0.350pt}{4.818pt}}
\put(700,113){\makebox(0,0){$0.5$}}
\put(700,767){\rule[-0.175pt]{0.350pt}{4.818pt}}
\put(1136,158){\rule[-0.175pt]{0.350pt}{4.818pt}}
\put(1136,113){\makebox(0,0){$1$}}
\put(1136,767){\rule[-0.175pt]{0.350pt}{4.818pt}}
\put(264,158){\rule[-0.175pt]{210.065pt}{0.350pt}}
\put(1136,158){\rule[-0.175pt]{0.350pt}{151.526pt}}
\put(264,787){\rule[-0.175pt]{210.065pt}{0.350pt}}
\put(45,472){\makebox(0,0)[l]{\shortstack{$R_{\rm 1S}(r)$}}}
\put(700,68){\makebox(0,0){$r$ (fm)}}
\put(700,832){\makebox(0,0){1S Radial Wavefunction}}
\put(264,158){\rule[-0.175pt]{0.350pt}{151.526pt}}
\put(1006,722){\makebox(0,0)[r]{$a=0.40$\,fm}}
\put(1050,722){\circle*{18}}
\put(867,200){\circle*{18}}
\put(1116,167){\circle*{18}}
\put(264,619){\circle*{18}}
\put(612,348){\circle*{18}}
\put(960,189){\circle*{18}}
\put(756,243){\circle*{18}}
\put(1042,175){\circle*{18}}
\put(1006,677){\makebox(0,0)[r]{$a=0.24$\,fm}}
\put(1050,677){\circle{18}}
\put(861,199){\circle{18}}
\put(1025,172){\circle{18}}
\put(630,319){\circle{18}}
\put(781,226){\circle{18}}
\put(964,179){\circle{18}}
\put(897,190){\circle{18}}
\put(1054,169){\circle{18}}
\put(264,731){\circle{18}}
\put(475,512){\circle{18}}
\put(686,284){\circle{18}}
\put(897,192){\circle{18}}
\put(1108,172){\circle{18}}
\put(562,390){\circle{18}}
\put(736,249){\circle{18}}
\put(931,184){\circle{18}}
\put(1134,169){\circle{18}}
\put(995,174){\circle{18}}
\put(1134,164){\circle{18}}
\sbox{\plotpoint}{\rule[-0.250pt]{0.500pt}{0.500pt}}%
\put(1006,632){\makebox(0,0)[r]{quark model}}
\put(1028,632){\usebox{\plotpoint}}
\put(1048,632){\usebox{\plotpoint}}
\put(1069,632){\usebox{\plotpoint}}
\put(1090,632){\usebox{\plotpoint}}
\put(1094,632){\usebox{\plotpoint}}
\put(264,738){\usebox{\plotpoint}}
\put(264,738){\usebox{\plotpoint}}
\put(279,724){\usebox{\plotpoint}}
\put(295,711){\usebox{\plotpoint}}
\put(309,696){\usebox{\plotpoint}}
\put(324,680){\usebox{\plotpoint}}
\put(338,665){\usebox{\plotpoint}}
\put(352,650){\usebox{\plotpoint}}
\put(365,634){\usebox{\plotpoint}}
\put(378,617){\usebox{\plotpoint}}
\put(391,601){\usebox{\plotpoint}}
\put(404,585){\usebox{\plotpoint}}
\put(417,569){\usebox{\plotpoint}}
\put(429,553){\usebox{\plotpoint}}
\put(442,536){\usebox{\plotpoint}}
\put(455,520){\usebox{\plotpoint}}
\put(468,504){\usebox{\plotpoint}}
\put(481,487){\usebox{\plotpoint}}
\put(494,471){\usebox{\plotpoint}}
\put(507,455){\usebox{\plotpoint}}
\put(521,440){\usebox{\plotpoint}}
\put(534,424){\usebox{\plotpoint}}
\put(549,409){\usebox{\plotpoint}}
\put(563,394){\usebox{\plotpoint}}
\put(577,379){\usebox{\plotpoint}}
\put(592,364){\usebox{\plotpoint}}
\put(607,350){\usebox{\plotpoint}}
\put(622,336){\usebox{\plotpoint}}
\put(638,323){\usebox{\plotpoint}}
\put(654,309){\usebox{\plotpoint}}
\put(671,297){\usebox{\plotpoint}}
\put(688,285){\usebox{\plotpoint}}
\put(705,273){\usebox{\plotpoint}}
\put(722,262){\usebox{\plotpoint}}
\put(740,252){\usebox{\plotpoint}}
\put(759,242){\usebox{\plotpoint}}
\put(777,233){\usebox{\plotpoint}}
\put(797,225){\usebox{\plotpoint}}
\put(816,217){\usebox{\plotpoint}}
\put(835,210){\usebox{\plotpoint}}
\put(855,204){\usebox{\plotpoint}}
\put(875,198){\usebox{\plotpoint}}
\put(895,193){\usebox{\plotpoint}}
\put(915,188){\usebox{\plotpoint}}
\put(936,184){\usebox{\plotpoint}}
\put(956,180){\usebox{\plotpoint}}
\put(977,177){\usebox{\plotpoint}}
\put(997,174){\usebox{\plotpoint}}
\put(1018,172){\usebox{\plotpoint}}
\put(1038,170){\usebox{\plotpoint}}
\put(1059,168){\usebox{\plotpoint}}
\put(1080,166){\usebox{\plotpoint}}
\put(1101,165){\usebox{\plotpoint}}
\put(1121,164){\usebox{\plotpoint}}
\put(1136,164){\usebox{\plotpoint}}
\end{picture}
\setlength{\unitlength}{0.240900pt}
\ifx\plotpoint\undefined\newsavebox{\plotpoint}\fi
\sbox{\plotpoint}{\rule[-0.175pt]{0.350pt}{0.350pt}}%
\begin{picture}(1200,900)(0,0)
\tenrm
\sbox{\plotpoint}{\rule[-0.175pt]{0.350pt}{0.350pt}}%
\put(264,158){\rule[-0.175pt]{210.065pt}{0.350pt}}
\put(264,158){\rule[-0.175pt]{0.350pt}{151.526pt}}
\put(264,368){\rule[-0.175pt]{4.818pt}{0.350pt}}
\put(242,368){\makebox(0,0)[r]{$2$}}
\put(1116,368){\rule[-0.175pt]{4.818pt}{0.350pt}}
\put(264,577){\rule[-0.175pt]{4.818pt}{0.350pt}}
\put(242,577){\makebox(0,0)[r]{$4$}}
\put(1116,577){\rule[-0.175pt]{4.818pt}{0.350pt}}
\put(264,787){\rule[-0.175pt]{4.818pt}{0.350pt}}
\put(242,787){\makebox(0,0)[r]{$6$}}
\put(1116,787){\rule[-0.175pt]{4.818pt}{0.350pt}}
\put(264,158){\rule[-0.175pt]{0.350pt}{4.818pt}}
\put(264,113){\makebox(0,0){$0$}}
\put(264,767){\rule[-0.175pt]{0.350pt}{4.818pt}}
\put(700,158){\rule[-0.175pt]{0.350pt}{4.818pt}}
\put(700,113){\makebox(0,0){$0.5$}}
\put(700,767){\rule[-0.175pt]{0.350pt}{4.818pt}}
\put(1136,158){\rule[-0.175pt]{0.350pt}{4.818pt}}
\put(1136,113){\makebox(0,0){$1$}}
\put(1136,767){\rule[-0.175pt]{0.350pt}{4.818pt}}
\put(264,158){\rule[-0.175pt]{210.065pt}{0.350pt}}
\put(1136,158){\rule[-0.175pt]{0.350pt}{151.526pt}}
\put(264,787){\rule[-0.175pt]{210.065pt}{0.350pt}}
\put(45,472){\makebox(0,0)[l]{\shortstack{$R_{\rm 1P}(r)$}}}
\put(700,68){\makebox(0,0){$r$ (fm)}}
\put(700,832){\makebox(0,0){1P Radial Wavefunction}}
\put(264,158){\rule[-0.175pt]{0.350pt}{151.526pt}}
\put(1006,722){\makebox(0,0)[r]{$a=0.40$\,fm}}
\put(1050,722){\circle*{18}}
\put(612,521){\circle*{18}}
\put(756,411){\circle*{18}}
\put(1042,243){\circle*{18}}
\put(867,321){\circle*{18}}
\put(1116,214){\circle*{18}}
\put(960,299){\circle*{18}}
\put(1042,244){\circle*{18}}
\put(1116,211){\circle*{18}}
\put(1028,722){\rule[-0.175pt]{15.899pt}{0.350pt}}
\put(1028,712){\rule[-0.175pt]{0.350pt}{4.818pt}}
\put(1094,712){\rule[-0.175pt]{0.350pt}{4.818pt}}
\put(612,503){\rule[-0.175pt]{0.350pt}{8.431pt}}
\put(602,503){\rule[-0.175pt]{4.818pt}{0.350pt}}
\put(602,538){\rule[-0.175pt]{4.818pt}{0.350pt}}
\put(756,404){\rule[-0.175pt]{0.350pt}{3.613pt}}
\put(746,404){\rule[-0.175pt]{4.818pt}{0.350pt}}
\put(746,419){\rule[-0.175pt]{4.818pt}{0.350pt}}
\put(1042,238){\rule[-0.175pt]{0.350pt}{2.650pt}}
\put(1032,238){\rule[-0.175pt]{4.818pt}{0.350pt}}
\put(1032,249){\rule[-0.175pt]{4.818pt}{0.350pt}}
\put(867,314){\rule[-0.175pt]{0.350pt}{3.373pt}}
\put(857,314){\rule[-0.175pt]{4.818pt}{0.350pt}}
\put(857,328){\rule[-0.175pt]{4.818pt}{0.350pt}}
\put(1116,211){\rule[-0.175pt]{0.350pt}{1.445pt}}
\put(1106,211){\rule[-0.175pt]{4.818pt}{0.350pt}}
\put(1106,217){\rule[-0.175pt]{4.818pt}{0.350pt}}
\put(960,289){\rule[-0.175pt]{0.350pt}{4.577pt}}
\put(950,289){\rule[-0.175pt]{4.818pt}{0.350pt}}
\put(950,308){\rule[-0.175pt]{4.818pt}{0.350pt}}
\put(1042,240){\rule[-0.175pt]{0.350pt}{1.927pt}}
\put(1032,240){\rule[-0.175pt]{4.818pt}{0.350pt}}
\put(1032,248){\rule[-0.175pt]{4.818pt}{0.350pt}}
\put(1116,208){\rule[-0.175pt]{0.350pt}{1.686pt}}
\put(1106,208){\rule[-0.175pt]{4.818pt}{0.350pt}}
\put(1106,215){\rule[-0.175pt]{4.818pt}{0.350pt}}
\put(1006,677){\makebox(0,0)[r]{$a=0.24$\,fm}}
\put(1050,677){\circle{18}}
\put(1054,231){\circle{18}}
\put(897,286){\circle{18}}
\put(931,265){\circle{18}}
\put(1025,222){\circle{18}}
\put(995,251){\circle{18}}
\put(1134,205){\circle{18}}
\put(475,555){\circle{18}}
\put(562,560){\circle{18}}
\put(736,442){\circle{18}}
\put(931,295){\circle{18}}
\put(964,248){\circle{18}}
\put(1134,246){\circle{18}}
\put(1054,214){\circle{18}}
\put(630,521){\circle{18}}
\put(781,399){\circle{18}}
\put(1108,164){\circle{18}}
\put(964,275){\circle{18}}
\put(1134,162){\circle{18}}
\put(1134,194){\circle{18}}
\put(686,480){\circle{18}}
\put(736,430){\circle{18}}
\put(861,329){\circle{18}}
\put(897,312){\circle{18}}
\put(1025,242){\circle{18}}
\put(1054,238){\circle{18}}
\put(781,389){\circle{18}}
\put(897,304){\circle{18}}
\put(1028,677){\rule[-0.175pt]{15.899pt}{0.350pt}}
\put(1028,667){\rule[-0.175pt]{0.350pt}{4.818pt}}
\put(1094,667){\rule[-0.175pt]{0.350pt}{4.818pt}}
\put(1054,221){\rule[-0.175pt]{0.350pt}{4.577pt}}
\put(1044,221){\rule[-0.175pt]{4.818pt}{0.350pt}}
\put(1044,240){\rule[-0.175pt]{4.818pt}{0.350pt}}
\put(897,264){\rule[-0.175pt]{0.350pt}{10.840pt}}
\put(887,264){\rule[-0.175pt]{4.818pt}{0.350pt}}
\put(887,309){\rule[-0.175pt]{4.818pt}{0.350pt}}
\put(931,254){\rule[-0.175pt]{0.350pt}{5.300pt}}
\put(921,254){\rule[-0.175pt]{4.818pt}{0.350pt}}
\put(921,276){\rule[-0.175pt]{4.818pt}{0.350pt}}
\put(1025,212){\rule[-0.175pt]{0.350pt}{4.818pt}}
\put(1015,212){\rule[-0.175pt]{4.818pt}{0.350pt}}
\put(1015,232){\rule[-0.175pt]{4.818pt}{0.350pt}}
\put(995,236){\rule[-0.175pt]{0.350pt}{6.986pt}}
\put(985,236){\rule[-0.175pt]{4.818pt}{0.350pt}}
\put(985,265){\rule[-0.175pt]{4.818pt}{0.350pt}}
\put(1134,197){\rule[-0.175pt]{0.350pt}{3.854pt}}
\put(1124,197){\rule[-0.175pt]{4.818pt}{0.350pt}}
\put(1124,213){\rule[-0.175pt]{4.818pt}{0.350pt}}
\put(475,483){\rule[-0.175pt]{0.350pt}{34.690pt}}
\put(465,483){\rule[-0.175pt]{4.818pt}{0.350pt}}
\put(465,627){\rule[-0.175pt]{4.818pt}{0.350pt}}
\put(562,522){\rule[-0.175pt]{0.350pt}{18.308pt}}
\put(552,522){\rule[-0.175pt]{4.818pt}{0.350pt}}
\put(552,598){\rule[-0.175pt]{4.818pt}{0.350pt}}
\put(736,407){\rule[-0.175pt]{0.350pt}{17.104pt}}
\put(726,407){\rule[-0.175pt]{4.818pt}{0.350pt}}
\put(726,478){\rule[-0.175pt]{4.818pt}{0.350pt}}
\put(931,265){\rule[-0.175pt]{0.350pt}{14.213pt}}
\put(921,265){\rule[-0.175pt]{4.818pt}{0.350pt}}
\put(921,324){\rule[-0.175pt]{4.818pt}{0.350pt}}
\put(964,237){\rule[-0.175pt]{0.350pt}{5.300pt}}
\put(954,237){\rule[-0.175pt]{4.818pt}{0.350pt}}
\put(954,259){\rule[-0.175pt]{4.818pt}{0.350pt}}
\put(1134,207){\rule[-0.175pt]{0.350pt}{18.790pt}}
\put(1124,207){\rule[-0.175pt]{4.818pt}{0.350pt}}
\put(1124,285){\rule[-0.175pt]{4.818pt}{0.350pt}}
\put(1054,207){\rule[-0.175pt]{0.350pt}{3.373pt}}
\put(1044,207){\rule[-0.175pt]{4.818pt}{0.350pt}}
\put(1044,221){\rule[-0.175pt]{4.818pt}{0.350pt}}
\put(630,483){\rule[-0.175pt]{0.350pt}{18.067pt}}
\put(620,483){\rule[-0.175pt]{4.818pt}{0.350pt}}
\put(620,558){\rule[-0.175pt]{4.818pt}{0.350pt}}
\put(781,375){\rule[-0.175pt]{0.350pt}{11.563pt}}
\put(771,375){\rule[-0.175pt]{4.818pt}{0.350pt}}
\put(771,423){\rule[-0.175pt]{4.818pt}{0.350pt}}
\put(1108,158){\rule[-0.175pt]{0.350pt}{6.022pt}}
\put(1098,158){\rule[-0.175pt]{4.818pt}{0.350pt}}
\put(1098,183){\rule[-0.175pt]{4.818pt}{0.350pt}}
\put(964,255){\rule[-0.175pt]{0.350pt}{9.636pt}}
\put(954,255){\rule[-0.175pt]{4.818pt}{0.350pt}}
\put(954,295){\rule[-0.175pt]{4.818pt}{0.350pt}}
\put(1134,158){\rule[-0.175pt]{0.350pt}{3.613pt}}
\put(1124,158){\rule[-0.175pt]{4.818pt}{0.350pt}}
\put(1124,173){\rule[-0.175pt]{4.818pt}{0.350pt}}
\put(1134,185){\rule[-0.175pt]{0.350pt}{4.095pt}}
\put(1124,185){\rule[-0.175pt]{4.818pt}{0.350pt}}
\put(1124,202){\rule[-0.175pt]{4.818pt}{0.350pt}}
\put(686,440){\rule[-0.175pt]{0.350pt}{19.031pt}}
\put(676,440){\rule[-0.175pt]{4.818pt}{0.350pt}}
\put(676,519){\rule[-0.175pt]{4.818pt}{0.350pt}}
\put(736,411){\rule[-0.175pt]{0.350pt}{9.154pt}}
\put(726,411){\rule[-0.175pt]{4.818pt}{0.350pt}}
\put(726,449){\rule[-0.175pt]{4.818pt}{0.350pt}}
\put(861,312){\rule[-0.175pt]{0.350pt}{8.191pt}}
\put(851,312){\rule[-0.175pt]{4.818pt}{0.350pt}}
\put(851,346){\rule[-0.175pt]{4.818pt}{0.350pt}}
\put(897,282){\rule[-0.175pt]{0.350pt}{14.454pt}}
\put(887,282){\rule[-0.175pt]{4.818pt}{0.350pt}}
\put(887,342){\rule[-0.175pt]{4.818pt}{0.350pt}}
\put(1025,228){\rule[-0.175pt]{0.350pt}{6.745pt}}
\put(1015,228){\rule[-0.175pt]{4.818pt}{0.350pt}}
\put(1015,256){\rule[-0.175pt]{4.818pt}{0.350pt}}
\put(1054,221){\rule[-0.175pt]{0.350pt}{8.191pt}}
\put(1044,221){\rule[-0.175pt]{4.818pt}{0.350pt}}
\put(1044,255){\rule[-0.175pt]{4.818pt}{0.350pt}}
\put(781,370){\rule[-0.175pt]{0.350pt}{8.913pt}}
\put(771,370){\rule[-0.175pt]{4.818pt}{0.350pt}}
\put(771,407){\rule[-0.175pt]{4.818pt}{0.350pt}}
\put(897,292){\rule[-0.175pt]{0.350pt}{5.782pt}}
\put(887,292){\rule[-0.175pt]{4.818pt}{0.350pt}}
\put(887,316){\rule[-0.175pt]{4.818pt}{0.350pt}}
\sbox{\plotpoint}{\rule[-0.250pt]{0.500pt}{0.500pt}}%
\put(1006,632){\makebox(0,0)[r]{quark model}}
\put(1028,632){\usebox{\plotpoint}}
\put(1048,632){\usebox{\plotpoint}}
\put(1069,632){\usebox{\plotpoint}}
\put(1090,632){\usebox{\plotpoint}}
\put(1094,632){\usebox{\plotpoint}}
\put(264,158){\usebox{\plotpoint}}
\put(264,158){\usebox{\plotpoint}}
\put(271,177){\usebox{\plotpoint}}
\put(279,196){\usebox{\plotpoint}}
\put(287,215){\usebox{\plotpoint}}
\put(295,234){\usebox{\plotpoint}}
\put(303,253){\usebox{\plotpoint}}
\put(311,273){\usebox{\plotpoint}}
\put(319,292){\usebox{\plotpoint}}
\put(328,311){\usebox{\plotpoint}}
\put(337,329){\usebox{\plotpoint}}
\put(346,348){\usebox{\plotpoint}}
\put(355,366){\usebox{\plotpoint}}
\put(366,385){\usebox{\plotpoint}}
\put(376,403){\usebox{\plotpoint}}
\put(386,421){\usebox{\plotpoint}}
\put(398,438){\usebox{\plotpoint}}
\put(410,455){\usebox{\plotpoint}}
\put(422,471){\usebox{\plotpoint}}
\put(436,487){\usebox{\plotpoint}}
\put(450,501){\usebox{\plotpoint}}
\put(467,514){\usebox{\plotpoint}}
\put(485,525){\usebox{\plotpoint}}
\put(503,534){\usebox{\plotpoint}}
\put(524,538){\usebox{\plotpoint}}
\put(544,540){\usebox{\plotpoint}}
\put(565,538){\usebox{\plotpoint}}
\put(585,533){\usebox{\plotpoint}}
\put(604,526){\usebox{\plotpoint}}
\put(623,517){\usebox{\plotpoint}}
\put(641,506){\usebox{\plotpoint}}
\put(659,496){\usebox{\plotpoint}}
\put(675,483){\usebox{\plotpoint}}
\put(692,471){\usebox{\plotpoint}}
\put(708,458){\usebox{\plotpoint}}
\put(724,445){\usebox{\plotpoint}}
\put(740,432){\usebox{\plotpoint}}
\put(756,418){\usebox{\plotpoint}}
\put(772,405){\usebox{\plotpoint}}
\put(788,392){\usebox{\plotpoint}}
\put(804,378){\usebox{\plotpoint}}
\put(820,365){\usebox{\plotpoint}}
\put(836,352){\usebox{\plotpoint}}
\put(853,340){\usebox{\plotpoint}}
\put(869,327){\usebox{\plotpoint}}
\put(886,315){\usebox{\plotpoint}}
\put(903,303){\usebox{\plotpoint}}
\put(920,291){\usebox{\plotpoint}}
\put(937,280){\usebox{\plotpoint}}
\put(955,269){\usebox{\plotpoint}}
\put(973,259){\usebox{\plotpoint}}
\put(992,249){\usebox{\plotpoint}}
\put(1011,241){\usebox{\plotpoint}}
\put(1030,232){\usebox{\plotpoint}}
\put(1049,225){\usebox{\plotpoint}}
\put(1068,218){\usebox{\plotpoint}}
\put(1088,210){\usebox{\plotpoint}}
\put(1108,205){\usebox{\plotpoint}}
\put(1128,199){\usebox{\plotpoint}}
\put(1136,197){\usebox{\plotpoint}}
\end{picture}
\end{center}
\caption{The radial wavefunctions for the 1S and 1P charmonium
computed using improved actions and two different lattice spacings.
Wavefunctions from a continuum quark model are also shown. Statistical
errors are negligible for the $1S$~wavefunction.}
\label{psi-wfcns}
\end{figure}
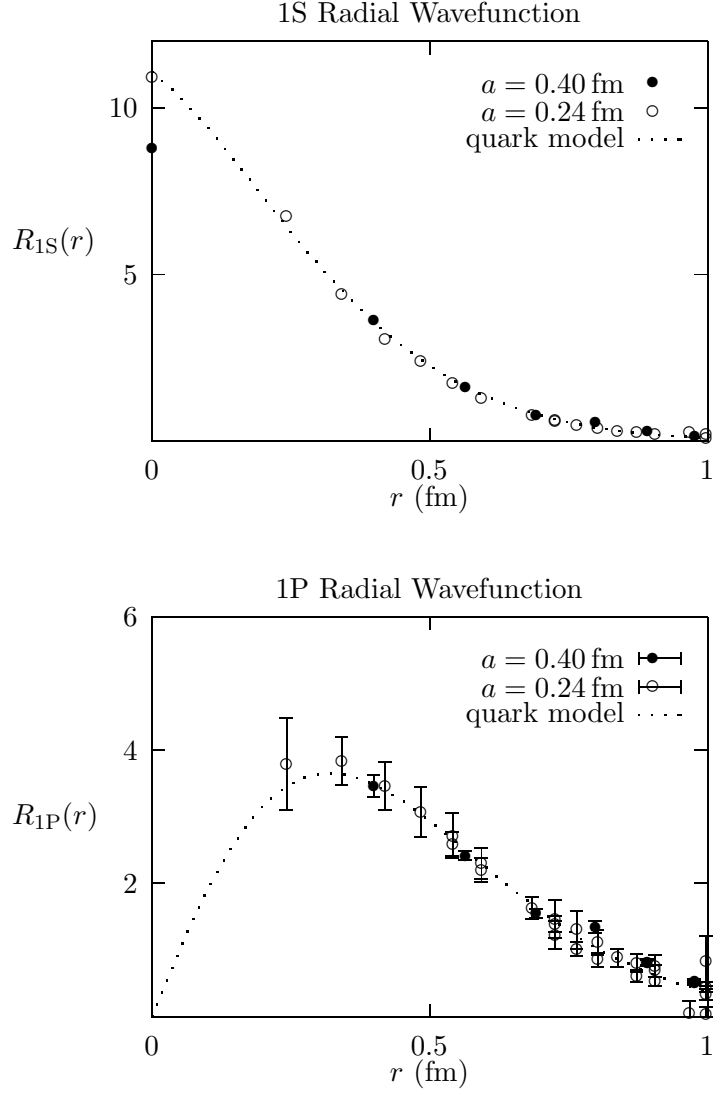

In Fig.~\ref{psi-wfcns} I show the
$1S$~and $1P$~radial wavefunctions for charmonium,
computed on lattices with two different
lattice spacings using the (improved) NRQCD actions\,\cite{alford95}.
The most striking observation from these pictures is that the charge radius of
the~$\psi$ is almost exactly equal to one lattice spacing on the coarser
lattice, and yet the results from the coarser lattice are essentially
identical to those from the finer lattice. These data confirm that in general
one needs only a few lattice points per direction within a hadron to obtain
accurate results (few percent) from an $a^2$-accurate action.
Numerical experiments with the NRQCD action, where $a^2$~corrections are
easily turned on and off, suggest that
\be \label{nrqcd-errors}
r_{\rm hadron}\! \approx\! a  \Rightarrow
\cases{ \order(a^2)~\mbox{corrections}\approx \mbox{15--20\%} \cr
        \mbox{} \cr
        \order(\alpha_s a^2,a^4) \approx \mbox{2--4\%} }
\ee

NRQCD simulations of heavy quarks were among the very first in which
tadpole improvement played a crucial role. They remain among the best
demonstrations of the technique.

\begin{exercise}
Verify that the continuum limit of the lattice NRQCD lagrangian~\eq{nrqcd-lag}
is a Schr\"odinger action with the standard relativistic corrections
(chromomagnetic moment, Darwin term, spin-orbit coupling\,\ldots).
\end{exercise}
\section{Conclusions}
Improved discretizations of QCD dynamics and large lattice spacings
work. Tree-level (i.e., classical) improvement of the quark and gluon
actions give results that are accurate to within a few to several percent for
lattice spacings as large as .3--.4\,fm, provided the all operators
are tadpole-improved. Tree-level improvement is {\em not} surprisingly
effective; it isn't perfect.
Rather, it works about as well as one would naively expect: there
are certainly $\order(\alpha_s)$ corrections (and probably also
nonperturbative corrections), and these will shift couplings by~10
or~20\% and physical quantities by a few percent. That QCD simulations
now conform to naive expectations and intuition is a major advance,
perhaps {\em the} major advance.
The current level of precision is more
than sufficient to have a large impact on QCD phenomenology.
And these techniques provide a firm foundation, a starting
point, for future
high-precision work that will rely upon some mixture of systematic
perturbative calculations of the corrections, nonperturbative tuning,
nontrivial transformations of the fields like those that lead to
``perfect actions'', and perhaps analytic techniques like
strong-coupling expansions.

The potential impact on QCD simulations of coarse lattices is
enormous. This potential will be realized if lots of people become
involved in the design, implementation and application this technology.
\newcommand{\lambibitem}[1]{\bibitem{#1}}

\end{document}